\documentclass[preprints,article,accept,moreauthors,pdftex]{Definitions/mdpi} 

\firstpage{1} 
\pubvolume{1}
\issuenum{1}
\articlenumber{0}
\pubyear{2021}
\copyrightyear{2021}
\externaleditor{{Academic Editor}: Veronica Dexheimer}

\datereceived{1 October 2021} 
\dateaccepted{25 November 2021} 
\datepublished{} 
\hreflink{https://doi.org/} % Please use \linebreak if need to do line break
\usepackage{soul}
\usepackage{physics}
\usepackage{mathrsfs}

\newcommand{\Eyq}{\widetilde{\mathbb{E}}_y^{(q)}}
\newcommand{\Tnorm}{\widetilde{T}_{q}}
\newcommand{\Vnorm}{\widetilde{\mathbb{V}}_{q}}

\newcommand{\h}{h_q}
\newcommand{\D}{\Delta_q}
\newcommand{\Drefq}{\hat{\Delta}_q}
\newcommand{\Dref}{\hat{\Delta}}
\newcommand{\beqy}{\begin{eqnarray}}
\newcommand{\eeqy}{\end{eqnarray}}

\newcommand{\Deltabar}{\bar{\Delta}_q}
\newcommand{\Vbar}{\bar{\mathbb{V}}_q}
\newcommand{\Tbar}{\bar{T}_q}
\newcommand{\Ex}{\mathbb{E}_x^{(q)}}
\newcommand{\ExLandau}{\breve{\mathbb{E}}_x^{(q)}}

\newcommand{\Fermi}{\varepsilon_{Fq}}

\newcommand{\Cutoff}{\varepsilon_{\Lambda}}
\newcommand{\asinh}{\text{asinh}}

\pdfoutput=1

\Title{$^1$S$_0$ Pairing Gaps, Chemical Potentials and Entrainment Matrix in Superfluid Neutron-Star Cores for the \mbox{Brussels–Montreal Functionals}}

\TitleCitation{$^1$S$_0$ Pairing Gaps, Chemical Potential and Entrainment Matrix in Superfluid Neutron-Star Cores for the Brussels–Montreal Functionals}

\Author{Valentin Allard \orcidA{}, Nicolas Chamel *\orcidB{} }
\AuthorNames{Valentin Allard, Nicolas Chamel}

\AuthorCitation{Allard, V.; Chamel, N.}
\address[1]{Institute of Astronomy and Astrophysics, Universit\'e Libre de Bruxelles, CP 226, Boulevard du Triomphe,\linebreak B-1050 Brussels, Belgium; valentin.allard@ulb.be}

\corres{\hangafter=1 \hangindent=1.05em \hspace{-0.82em}Correspondence: nicolas.chamel@ulb.be}

\abstract{Temperature and velocity-dependent $^1$S$_0$ pairing gaps, chemical potentials and entrainment matrix in dense homogeneous neutron–proton superfluid mixtures constituting the outer core of neutron stars, are determined  fully self-consistently by solving numerically the time-dependent Hartree–Fock–Bogoliubov equations over the whole range of temperatures and flow velocities for which superfluidity can exist. Calculations have been made for $npe\mu$ in beta-equilibrium using the Brussels–Montreal functional BSk24. The accuracy of various approximations is assessed and the physical meaning of the different velocities and momentum densities appearing in the theory is clarified. Together with the unified equation of state published earlier, the present results provide consistent microscopic inputs for modeling superfluid neutron-star cores. 
}


\def\apj{Astrophys. J.}  % Astrophysical Journal
\def\apjl{Astrophys. J. Lett.} % Astrophysical Journal, Letters
\def\prl{Phys.~Rev.~Lett.} % Physical Review Letters
\def\prc{Phys.~Rev.~C} % Physical Review C
\def\prd{Phys.~Rev.~D} % Physical Review C
\def\aap{Astron. Astrophys.}   % Astronomy and Astrophysics
\def\nphysa{Nucl.~Phys.~A}   % Nuclear Physics A

\def\mnras{Mon. Not. R. Astron. Soc.}             % Monthly Notices of the RAS
\begin{document}
%%%%%%%%%%%%%%%%%%%%%%%%%%%%%%%%%%%%%%%%%%
%\setcounter{section}{-1} %% Remove this when starting to work on the template.

\section{Introduction}

Different superfluid and superconducting phases are predicted to exist in neutron stars (see, e.g., \cite{chamel2017} for a review). In particular, the (electrically charge neutral) outer core is expected to be made of a neutron–proton superfluid mixture in beta-equilibrium with a normal gas of leptons. More speculative superfluid phases involving other particles such as hyperons or quarks might occur in the innermost region of the core of a neutron star but will not be considered here. Although the interior of the star is highly degenerate, thermal effects may still play an important role in the rotational evolution of the star~\cite{ho2012,gusakov2014,ho2015}. This stems from the fact that the critical temperatures $T_{cq}$ above which superfluidity is destroyed ($q=n,p$ for neutrons and protons respectively) are  much smaller than the Fermi temperatures $T_{Fq}$, and may thus be comparable to the actual temperature $T$ of the star. Superfluidity leads to a very complicated dynamics, characterized by the coexistence of different flows (see, e.g., \cite{andersson2021} for a recent review). The core of a neutron star involves at least three distinct fluids: 
the neutron and proton superfluids, as well as a normal fluid made of leptons and excitations. Due to nuclear interactions, the neutron and proton superfluids in the core do not flow freely. They are mutually coupled by entrainment effects of the same kind as the ones discussed by Andreev and Bashkin~\cite{andreev1976} in the context of superfluid $^4$He-$^3$He mixtures: the mass currents $\pmb{\rho_q}$ are expressible as linear combinations of the velocity $\pmb{v_N}$ of the normal fluid and of the so-called ``superfluid velocities'' $\pmb{V_q}$ as

\beqy\label{eq:NeutronMassCurrent}
    \pmb{\rho_n}=\left(\rho_n - \rho_{nn} -\rho_{np}\right)\pmb{v_N} + \rho_{nn}\pmb{V_n} + \rho_{np}\pmb{V_p} \, ,
\eeqy
\beqy\label{eq:ProtonMassCurrent}
    \pmb{\rho_p}=\left(\rho_p - \rho_{pp} -\rho_{pn}\right)\pmb{v_N} + \rho_{pp}\pmb{V_p} + \rho_{pn}\pmb{V_n}\, , 
\eeqy
where $\rho_q=m n_q$ is the mass density of nucleon of charge type $q$ with associated nucleon  number density $n_q$ (ignoring the small difference between neutron and proton masses, which is denoted simply by $m$). As shown by Carter and Khalatnikov in the context of Landau's canonical two-fluid model of superfluid $^4$He~\cite{carter1994} (see also \cite{prix2004variational}),  $\pmb{V_q}$ are not true velocities but physically represent average nucleon momenta per unit mass. Indeed, it can be easily seen that the true velocities should be defined in terms of the mass currents as $\pmb{v_q}=\pmb{\rho_q}/\rho_q$ and in general these velocities do not coincide with $\pmb{V_q}$. To avoid any misinterpretation, the superfluid ``velocities'' are thus written with a capital letter. Please note that $\pmb{v_N}$ is a true velocity. The importance of entrainment effects is measured by the (symmetric) matrix $\rho_{qq^\prime}$, which is a key microscopic input for modeling the dynamics of neutron stars, see, e.g.,~\cite{andersson2021} and references therein. The entrainment matrix is expected to depend not only on the composition and the baryon density $n=n_n+n_p$, but also on the superfluid velocities as well as the temperature, and this may have an impact on  neutron-star oscillations~\cite{gusakov2013,dommes2019,kantor2020,kantor2021}. The influence of temperature and velocity on  entrainment effects has been previously studied within Landau's theory~\cite{gusakov2005,leinson2017,leinson2018}. However, the Landau parameters, the critical temperatures, and the composition had to be given. Recently, we have derived the entrainment matrix self-consistently for arbitrary superfluid velocities and temperatures within the nuclear-energy-density functional theory by solving exactly the time-dependent Hartree–Fock–Bogoliubov (TDHFB)  equations~\cite{ChamelAllard2019,ChamelAllard2020}. 
The expressions we have obtained are quite general and applicable to a wide variety of functionals. 

In this paper, we have calculated various properties of homogeneous neutron–proton superfluid mixtures in the outer core of neutron stars by solving numerically the self-consistent TDHFB equations using the Brussels–Montreal functional BSk24~\cite{goriely2013} for which unified equations of state are already available~\cite{pearson2018,shelley2020,pearson2020,mutafchieva2019} as well as gravitoelectric and gravitomagnetic tidal Love numbers up to $\ell=5$~\cite{perot2019,perot2021}. More importantly, unlike most available functionals, BSk24 has been accurately calibrated to realistic microscopic calculations of $^1$S$_0$ pairing gaps in infinite homogeneous nuclear matter (at zero temperature and in the absence of currents). 
We have determined within the same microscopic framework not only the entrainment matrix but also the  $^1$S$_0$ pairing gaps and chemical potentials of the superfluid mixture without any approximation, varying the superfluid velocities, the temperature and the baryon density. In this way, we have also been able to assess the accuracy of various approximations. We have focused on $^1$S$_0$ nuclear superfluid phases, which are reliably predicted to exist in the outer core of neutron stars (see, e.g., \cite{sedrakian2019} and references therein). We have not considered $^3$PF$_2$ neutron superfluidity. As a matter of fact, it has been recently shown that in regions where both types of pairing can potentially exist, the $^3$PF$_2$ superfluid phase is completely excluded by the $^1$S$_0$ phase unless strong magnetic fields are present~\cite{yasui2020}.

The formalism to describe nuclear superfluidity is presented in Section~\ref{sec:theory}. After briefly recapitulating the general principles of the TDHFB theory in Section~\ref{sec:TDHFB} and the functionals in Section~\ref{sec:functionals}, the exact solution in homogeneous nuclear matter is given in Section~\ref{sec:TDHFB-neutron-star-cores}, where explicit expressions for various quantities entering the calculations of 
superfluid properties are derived. The physical interpretation of the different  velocities and momentum densities are clarified in Section~\ref{sec:interpretation}. In Section~\ref{sec:LandauApprox}, it is shown how the TDHFB theory can be recast into Landau's theory after introducing a series of approximations. 
Applications to neutron stars are presented in Section~\ref{sec:results}. The main features of the Brussels–Montreal functionals are recapitulated in Section~\ref{sec:BSk}. After describing our numerical implementation of the TDHFB equations in Section~\ref{sec:implementation}, detailed numerical results for various properties of the superfluid mixture are presented and analyzed in\linebreak Sections~\ref{sec:gaps}--\ref{sec:chemical}. The accuracy of various approximations and interpolations are also discussed. Our conclusions are given in Section~\ref{sec:conclusions}.

\section{Nuclear Superfluidity within  the Time-Dependent Hartree–Fock–Bogoliubov Theory}
\label{sec:theory}

\subsection{General Principles}
\label{sec:TDHFB}

The TDHFB theory~\cite{blaizot1986} provides 
a unified microscopic framework for studying the dynamics of various nuclear systems, ranging from atomic nuclei to the dense nuclear matter present in neutron stars---the main focus of this work. 

Introducing the one-body density matrix $n_q^{ij}=\langle c_q^{j\dagger} c_q^i \rangle= n^{ji*}_q$  and pairing tensor $\kappa_q^{ij}=\langle c_q^j c_q^i \rangle=-\kappa_q^{ji}$ defined in terms of thermal averages of products of creation ($c_q^{i\dagger}$) and destruction ($c_q^{i}$) operators for nucleons of charge type $q$ in a quantum state $i$  (using the symbol $\dagger$ for Hermitian conjugation and $*$ for complex conjugation), and assuming that the energy $E$ of a nucleon-matter element of volume $V$
is a function of $n_q^{ij}$, $\kappa_q^{ij}$ and $\kappa_q^{ij*}$ lead to the following equations of motion~\cite{blaizot1986}:  
\begin{align}
\label{eq:TDHFB1}
i\hbar \frac{\partial n_q^{ij}}{\partial t}=\sum_k \left(\h^{ik}n_q^{kj}-n_q^{ik}\h^{kj}+\kappa_q^{ik}\D^{kj*} -\D^{ik}\kappa_q^{kj*}\right) \, ,
\end{align}
\begin{align}
\label{eq:TDHFB2}
i \hbar \frac{\partial \kappa_q^{ij}}{\partial t}=\sum_k  \left[(\h^{ik}-\lambda_q\delta^{ik})\kappa_q^{kj}+\kappa_q^{ik}(\h^{kj*}-\lambda_q \delta^{kj})-\D^{ik}n_q^{kj*} - n_q^{ik}\D^{kj}\right]+ \D^{ij}\, ,
\end{align}
where $\lambda_q$ denote the chemical potentials. The matrices $h_q^{ij}$ and  $\D^{ij}$, defined by 
\beqy\label{eq:Hamiltonian-matrix}
\h^{ij}=\frac{\partial E}{\partial n_q^{ji}}=\h^{ji*}
\, , 
\eeqy
\beqy\label{eq:pairing-matrix}
\D^{ij}=\frac{\partial  E}{\partial \kappa_q^{ij*}}=-\D^{ji}
\, ,
\eeqy
generally depend themselves on $n_q^{ij}$, $\kappa_q^{ij}$ and $\kappa_q^{ij*}$ and must therefore be determined\linebreak self-consistently. 

\subsection{Functionals of Local Densities and Currents}
\label{sec:functionals}

The class of energy functionals $E(n_q^{ij},\kappa_q^{ij},\kappa_q^{ij*})$ that we consider here depend on the density matrices and pair tensors only through the following local densities and currents: 

\begin{enumerate}
\item[(i)]	the nucleon densities at position $\pmb{r}$ at time $t$ ($\sigma=\pm1$ distinguishing the two spin states),
\begin{align}\label{eq:Density}
    n_q(\pmb{r},t)=\sum_{\sigma=\pm 1} n_q(\pmb{r},\sigma;\pmb{r},\sigma;t)\, ,
\end{align}
\item[(ii)]	the kinetic-energy densities (in units of $\hbar^2/2m)$ at position $\pmb{r}$ and time $t$,
\begin{align}\label{eq:KineticDensity}
    \tau_q(\pmb{r},t)=\sum_{\sigma=\pm 1}\int\text{d}^3\pmb{r'}\; \delta (\pmb{r}-\pmb{r'}) \grad\cdot\grad\pmb{'} n_q(\pmb{r},\sigma;\pmb{r'},\sigma;t)\, ,
\end{align}
\item[(iii)]	the momentum densities (in units of $\hbar$) at position $\pmb{r}$ and time $t$,
\begin{align}\label{eq:total-momentum-density}
\pmb{j_q}(\pmb{r},t)=-\frac{ i}{2}\sum_{\sigma=\pm 1}\int\,{\rm d}^3\pmb{r^\prime}\,\delta(\pmb{r}-\pmb{r^\prime}) (\pmb{\nabla} -\pmb{\nabla^\prime})n_q(\pmb{r}, \sigma; \pmb{r^\prime}, \sigma;t)\, ,
\end{align}
\item[(iv)]	 the abnormal densities at position $\pmb{r}$ and time $t$
\begin{align}
    \widetilde{n}_q(\pmb{r},t)=\sum_{\sigma = \pm 1}\widetilde{n}_q (\pmb{r},\sigma ; \pmb{r},\sigma ;t)\, , 
\end{align}
\end{enumerate}

\noindent where
\begin{equation}\label{eq:DensityMatrixCoordinateSpaceDef}
n_q(\pmb{r}, \sigma; \pmb{r^\prime}, \sigma^\prime; t)=\sum_{i,j}n^{ij}_{q}\varphi^{(q)}_i (\pmb{r},\sigma)\varphi^{(q)}_j (\pmb{r^\prime},\sigma^\prime)^{*}\, ,
\end{equation}

\begin{equation}\label{eq:AbnormalDensityMatrixCoordinateSpaceDef}
\widetilde{n}_q(\pmb{r}, \sigma; \pmb{r^\prime}, \sigma^\prime; t)=-\sigma^\prime\sum_{i,j}\kappa^{ij}_{q}\varphi^{(q)}_i(\pmb{r},\sigma)\varphi^{(q)}_j(\pmb{r^\prime},-\sigma^\prime)\, ,
\end{equation}
are the so-called normal and abnormal density matrices respectively~\cite{dobaczewski1984}, 
and $\varphi^{(q)}_i(\pmb{r},\sigma)$ is the single-particle wavefunction associated with the state $i$. The abnormal densities are the local order parameters of the neutron and proton superfluid phases~\cite{ChamelAllard2020}. These densities are complex and the gradient of their respective phase $\phi_q(\pmb{r},t)$ defines the superfluid velocities as follows: 
\begin{align}\label{eq:SuperfluidVelocity-def}
\pmb{V_q}(\pmb{r},t)=\frac{\hbar}{2m} \pmb{\nabla}\phi_q(\pmb{r},t)\, .
\end{align}
The matrices (\ref{eq:Hamiltonian-matrix}) and (\ref{eq:pairing-matrix}) can be alternatively expressed as~\cite{ChamelAllard2020} 
\beqy\label{eq:Hamiltonian-matrix2}
\h^{ij}(t)=\sum_{\sigma}\int {\rm d}^3\pmb{r} \,  \varphi^{(q)}_i(\pmb{r},\sigma)^*\h(\pmb{r},t)\varphi^{(q)}_j(\pmb{r},\sigma)\, , 
\eeqy
\beqy\label{eq:pairing-matrix2}
\D^{ij}(t)=-\sum_{\sigma}\sigma \int {\rm d}^3\pmb{r} \,  \varphi^{(q)}_i(\pmb{r},\sigma)^*\D(\pmb{r},t)\varphi^{(q)}_j(\pmb{r},-\sigma)^*\, , 
\eeqy
where
\beqy
\label{eq:Hamiltonian}
\h(\pmb{r},t)&=&-\pmb{\nabla}\cdot \frac{\hbar^2}{2 m_q^\oplus(\pmb{r},t)}\pmb{\nabla} + U_q(\pmb{r},t)-
\frac{i}{2}\biggl[\pmb{I_q}(\pmb{r},t)\cdot\pmb{\nabla}+\pmb{\nabla}\cdot\pmb{I_q}(\pmb{r},t)\biggr]\, ,
\eeqy
\beqy
\label{eq:def-fields}
\frac{\hbar^2}{2 m_q^\oplus(\pmb{r},t)}=\frac{\delta E}{\delta \tau_q(\pmb{r},t)}, \qquad U_q(\pmb{r},t)=\frac{\delta E}{\delta n_q(\pmb{r},t)},\qquad 
\pmb{I_q}(\pmb{r},t)=  \frac{\delta E}{\delta \pmb{j_q}(\pmb{r},t)}\, ,
\eeqy
\beqy
\label{eq:pair-pot}
\D(\pmb{r},t) =2\frac{\delta E}{\delta \widetilde{n}_q(\pmb{r},t)^*}= 2\frac{\delta E}{\delta\vert \widetilde{n}_q(\pmb{r},t)\vert^2} \widetilde{n}_q(\pmb{r},t)\, .
\eeqy

The last equality in Equation~\eqref{eq:pair-pot} arises from the requirement that the energy must be real. Please note that the pair potential $\Delta_q(\pmb{r},t)$ is a complex field sharing the same phase $\phi_q(\pmb{r},t)$ as the abnormal density $\widetilde{n}_q(\pmb{r},t)$. Let us recall that this phase enters through the definition of the superfluid velocities~\eqref{eq:SuperfluidVelocity-def}. 

\subsection{Application to Homogeneous Systems}
\label{sec:TDHFB-neutron-star-cores}

Considering a homogeneous neutron–proton superfluid mixture with stationary flows in the normal-fluid rest frame where $\pmb{v_N}=\pmb{0}$, the TDHFB equations can be solved exactly~\cite{ChamelAllard2020}. In particular, the entrainment matrix reads ($\delta_{qq'}$ denotes the Kronecker symbol and $\rho = \rho_n + \rho_p$ is the total mass density)
\begin{align}\label{eq:EntrainmentMatrix}
    \rho_{qq'}=\rho_q\left(1-\mathcal{Y}_q\right)\left(\frac{m}{m_q^{\oplus}}\delta_{qq'} + \frac{\mathcal{I}_{qq'}}{\hbar}\right)\, , 
\end{align}
with
\beqy\label{eq:def-Bq}
\frac{m}{m_q^\oplus} =1+ \frac{2 \rho}{\hbar^2}\Biggl(\frac{\delta E^j_{\rm nuc}}{\delta X_0}-\frac{\delta E^j_{\rm nuc}}{\delta X_1}\Biggr)  + \frac{4 \rho_q}{\hbar^2} \frac{\delta E^j_{\rm nuc}}{\delta X_1}
\, ,
\eeqy
\begin{align}\label{eq:Inn-def}
\mathcal{I}_{nn}&=\frac{2}{\hbar}\rho_{n}  \left(1-\mathcal{Y}_{n}\right)\Theta\left[\frac{\delta E^j_{\rm nuc}}{\delta X_1}\left(\frac{8}{\hbar^2}\frac{\delta E^j_{\rm nuc}}{\delta X_0}m_p^{\oplus}n_p\mathcal{Y}_p-1\right)-\frac{\delta E^j_{\rm nuc}}{\delta X_0}\right]\, ,
\end{align}
\begin{align}\label{eq:Ipp-def}
\mathcal{I}_{pp}&=\frac{2}{\hbar}\rho_{p}  \left(1-\mathcal{Y}_{p}\right)\Theta\left[\frac{\delta E^j_{\rm nuc}}{\delta X_1}\left(\frac{8}{\hbar^2}\frac{\delta E^j_{\rm nuc}}{\delta X_0}m_n^{\oplus}n_n\mathcal{Y}_n-1\right)-\frac{\delta E^j_{\rm nuc}}{\delta X_0}\right]\, ,
\end{align}
\begin{align}\label{eq:Inp-def}
\mathcal{I}_{np}&=\frac{2}{\hbar}\rho_{p}  \left(1-\mathcal{Y}_{p}\right)\Theta\left(\frac{\delta E^j_{\rm nuc}}{\delta X_1}-\frac{\delta E^j_{\rm nuc}}{\delta X_0}\right)\, ,
\end{align}
\begin{align}\label{eq:Ipn-def}
\mathcal{I}_{pn}&=\frac{2}{\hbar}\rho_{n}  \left(1-\mathcal{Y}_{n}\right)\Theta\left(\frac{\delta E^j_{\rm nuc}}{\delta X_1}-\frac{\delta E^j_{\rm nuc}}{\delta X_0}\right)\, ,
\end{align} 
\begin{align}\label{eq:Theta-def}
\Theta &\equiv\left[ 1- \frac{2}{\hbar^2}\left(\frac{\delta E^j_{\rm nuc}}{\delta X_0} +\frac{\delta E^j_{\rm nuc}}{\delta X_1}\right)\left(m_n^{\oplus}n_{n}\mathcal{Y}_{n} + m_p^{\oplus}n_{p}\mathcal{Y}_{p} \right)\right.\nonumber \\
&\qquad\;\; \left. +\left(\frac{4}{\hbar^2}\right)^2\frac{\delta E^j_{\rm nuc}}{\delta X_0}\frac{\delta E^j_{\rm nuc}}{\delta X_1} m_n^{\oplus}n_{n}m_p^{\oplus}n_{p}\mathcal{Y}_{n} \mathcal{Y}_{p} \right]^{-1}\, .
\end{align}
Here $E^j_{\rm nuc}$ represents the nuclear-energy terms contributing to the mass currents. Galilean invariance requires  that these terms depend on the following combinations:  
\begin{equation}
X_0(\pmb{r},t)=n_0(\pmb{r},t)\tau_0(\pmb{r},t) -  \pmb{j_0}(\pmb{r},t)^2\, ,     
\end{equation}

\begin{equation}
X_1(\pmb{r},t)=n_1(\pmb{r},t)\tau_1(\pmb{r},t) -  \pmb{j_1}(\pmb{r},t)^2 \, .
\end{equation} 
The subscripts $0$ and $1$ denote isoscalar and isovector quantities, respectively, namely sums over neutrons and protons for the former (e.g., $n_0\equiv n=n_n+n_p$) and differences between neutrons and protons for the latter (e.g., $n_1=n_n-n_p$). The temperature and velocity-dependent functions $\mathcal{Y}_q$ are defined by ($k_\text{B}$ being the Boltzmann constant) 
\begin{align}\label{eq:YqFunction}
\mathcal{Y}_q(T,\pmb{\mathbb{V}_q})\equiv \frac{\hbar}{m_q^{\oplus}n_q \mathbb{V}_q^2}\frac{1}{V}\sum_{\pmb{k}} \pmb{k}\cdot\pmb{\mathbb{V}_q}\tanh\left(\frac{\mathfrak{E}^{(q)}_{\pmb{k}}}{2 k_\text{B} T}\right)\, ,
\end{align}
where we have introduced the effective superfluid velocities
\begin{align}\label{eq:EffectiveSuperfluidVelocity}
    \pmb{\mathbb{V}_q}\equiv \frac{m}{m_q^{\oplus}}\pmb{V_q}+\frac{\pmb{I_q}}{\hbar}\, ,  
\end{align}
and $\mathfrak{E}_{\pmb{k}}^{(q)}$ represent the energies of quasiparticle excitations, given by
\begin{align}\label{eq:QuasiparticleEnergy}
    \mathfrak{E}_{\pmb{k}}^{(q)}=\hbar\pmb{k}\cdot\pmb{\mathbb{V}_q} + \sqrt{\varepsilon^{(q)2}_{\pmb{k}} +  \Delta_q^2}\, ,
\end{align}
with
\begin{align}\label{eq:varepsilon}
\varepsilon^{(q)}_{\pmb{k}} =\frac{\hbar^2\pmb{k}^2}{2m^{\oplus}_q} +\frac{1}{2} m_q^\oplus\left(\pmb{\mathbb{V}_q}+\frac{\pmb{I_q}}{\hbar}\right)\cdot\left(\pmb{\mathbb{V}_q}-\frac{\pmb{I_q}}{\hbar}\right)  + U_q - \lambda_q \, .
\end{align}
In turn, the vector $\pmb{I_q}$ is expressible in terms of the superfluid velocities as follows
\beqy \label{eq:Iq}
\pmb{I_q}=\sum_{q'} \mathcal{I}_{qq'} \pmb{V_{q'}} \, .
\eeqy 

The pairing gaps (as defined as the nonvanishing matrix elements of the pair potential, 
see \cite{ChamelAllard2020}) are obtained from the self-consistent equations
\begin{align}
\label{eq:GapEquation}
    \Delta_q(T,\pmb{\mathbb{V}_q}) = -\frac{2}{V}\frac{\delta E}{\delta\vert \widetilde{n}_q\vert^2} \sum_{\pmb{k}} \frac{\Delta_q(T,\pmb{\mathbb{V}_q})}{\sqrt{\varepsilon_{\pmb{k}}^{(q)2} +\Delta_q(T,\pmb{\mathbb{V}_q})^2}}\tanh\left(\frac{\mathfrak{E}_{\pmb{k}}^{(q)}}{2 k_\text{B} T}\right)\, ,
\end{align}
where 
it  is understood that the summation must be regularized to remove ultraviolet divergences, as will be discussed below. 
The gap equations must be solved together with the particle number conservation conditions  
\begin{align}
\label{eq:DensityHomogeneous}
n_q=
\frac{1}{V}\sum_{\pmb{k}}\left[1-\frac{\varepsilon^{(q)}_{\pmb{k}}}{\sqrt{\varepsilon^{(q)2}_{\pmb{k}} +  \Delta_q^2}}\tanh\left(\frac{\mathfrak{E}_{\pmb{k}}^{(q)}}{2 k_\text{B} T}\right)\right]\, . 
\end{align}

As can be seen from Equation~\eqref{eq:varepsilon}, Equations~\eqref{eq:YqFunction}, \eqref{eq:GapEquation} and \eqref{eq:DensityHomogeneous} all depend on the \emph{reduced} chemical potentials defined by 
\beqy\label{eq:ReducedChemicalPotential}
\mu_q=\lambda_q - U_q -\frac{1}{2} m_q^\oplus\left(\pmb{\mathbb{V}_q}+\frac{\pmb{I_q}}{\hbar}\right)\cdot\left(\pmb{\mathbb{V}_q}-\frac{\pmb{I_q}}{\hbar}\right) 
\eeqy 
so that neither the pairing gaps nor the entrainment matrix require the explicit form of the potentials $U_q$.

From now on, we will take the continuum limit, i.e., we will replace discrete summations over wave vectors $\pmb{k}$ by integrations as follows: 
\beqy\label{eq:continuum}
\frac{1}{V} \sum_{\pmb{k}} \dotsi \rightarrow   \int \frac{\text{d}^3\pmb{k}}{(2\pi)^3} \dotsi =\int \frac{\text{d}\Omega_{\pmb{k}}}{4 \pi}\int_{-\mu_q}^{+\infty} \text{d}\varepsilon\, \mathcal{D}(\varepsilon) \dotsi
\eeqy 
with $\Omega_{\pmb{k}}$ the solid angle in $\pmb{k}$-space and $\mathcal{D}(\varepsilon)$ the density of single-particle states per one spin state given by
\beqy\label{eq:DoS}
\mathcal{D}(\varepsilon)=\frac{m_q^\oplus}{2 \pi^2 \hbar^3}\sqrt{2 m_q^\oplus(\varepsilon+\mu_q)}\, .
\eeqy 

Integrating over solid angle and changing variables, Equation~\eqref{eq:YqFunction} can thus be expressed as 
\begin{align}\label{eq:Yq}
\mathcal{Y}_q&=&\frac{3}{8}\frac{\Tbar}{\Vbar^2}\int_0^{+\infty}\text{d}x\;\sqrt{x} \log\left\{\left[1+\text{e}^{-\left(\Ex - 2\Vbar\sqrt{x}\right)/\Tbar}\right]\left[1+\text{e}^{-\left(\Ex + 2\Vbar\sqrt{x}\right)/\Tbar}\right]\right\}\nonumber\\
&& +\frac{3}{16}\frac{\Tbar^2}{\Vbar^3}\int_0^{+\infty}\text{d}x\; \left\{\text{Li}_2\left[-\text{e}^{-\left(\Ex - 2\Vbar\sqrt{x}\right)/\Tbar}\right]-\text{Li}_2\left[-\text{e}^{-\left(\Ex + 2\Vbar\sqrt{x}\right)/\Tbar}\right]\right\}
\end{align}
where 
\begin{align}
    \Ex=\sqrt{\left(x-\bar{\mu}_{q}\right)^2+ \bar{\Delta}_q^2} \, ,
\end{align}
$\displaystyle \text{Li}_2(x) = \int_1^{1-x}\frac{\log{u}}{1-u}\text{d}u$ is the dilogarithm function, and we have  introduced the dimensionless ratios
\begin{align}
\Tbar \equiv \frac{T}{T_{Fq}}\, , \hskip0.5cm \Vbar \equiv \frac{\mathbb{V}_q}{V_{Fq}}\, , \hskip0.5cm \bar{\mu}_{q}\equiv \frac{\mu_q}{\Fermi}\, , \hskip0.5cm  \bar{\Delta}_q\equiv \frac{\Delta_q}{\Fermi}\, .
\end{align}
The Fermi temperature is defined by $T_{Fq}=\Fermi/k_\text{B}$ with the Fermi energy
\begin{equation}\label{eq:eF}
\Fermi=\frac{\hbar^2k_{Fq}^2}{2m_q^{\oplus}}
\end{equation}
and Fermi wave number $k_{Fq}=(3\pi^2 n_q)^{1/3}$; the Fermi velocity is given by 
\begin{equation}
V_{Fq}=\frac{\hbar k_{Fq}}{m_q^{\oplus}} \, .
\end{equation}
Similarly, the gap Equation~\eqref{eq:GapEquation} and the particle number conservation  Equation~\eqref{eq:DensityHomogeneous} become, respectively
\beqy\label{eq:DimensionlessGapEquation}
\Delta_q&=& -\frac{m_q^{\oplus}k_{Fq}}{2\pi^2\hbar^2}\frac{\Tbar}{\Vbar}\frac{\delta E}{\delta\vert \widetilde{n}_q\vert^2} \Delta_q \displaystyle \int_0^{(\mu_q + \Cutoff)/\Fermi}\frac{\text{d}x}{\Ex} \nonumber \\
&&\times \log\left[\cosh\left(\frac{\Ex}{2\Tbar} +\frac{\Vbar}{\Tbar}\sqrt{x}\right) \sech\left(\frac{\Ex}{2\Tbar} -\frac{\Vbar}{\Tbar}\sqrt{x}\right)  \right]\, ,
\eeqy

\beqy\label{eq:DimensionlessChemicalPotentialEquation}
\frac{4}{3}&=&\int_0^{+\infty}\text{d}x\;\left\{\sqrt{x}-\frac{\Tbar}{\Vbar}\frac{x-\bar{\mu}_q}{2\Ex} \right. \nonumber \\ 
&&\left.\times \log\left[\cosh\left(\frac{\Ex}{2\Tbar} +\frac{\Vbar}{\Tbar}\sqrt{x}\right) \sech\left(\frac{\Ex}{2\Tbar} -\frac{\Vbar}{\Tbar}\sqrt{x}\right)  \right]\right\}\, ,
\eeqy
and $\Cutoff$ is a cutoff above the Fermi level to regularize the ultraviolet divergences\linebreak (see Section~\ref{sec:BSk}). 
Expressing the hyperbolic functions in terms of the exponential function, 
we can alternatively rewrite~\eqref{eq:DimensionlessGapEquation} and~\eqref{eq:DimensionlessChemicalPotentialEquation} as 

\beqy
\Delta_q&=& -\frac{m_q^{\oplus}k_{Fq}}{\pi^2\hbar^2}\frac{\delta E}{\delta\vert \widetilde{n}_q\vert^2} \Delta_q \displaystyle \int_0^{(\mu_q + \Cutoff)/\Fermi}\text{d}x\;\frac{\sqrt{x}}{\Ex} \nonumber \\
&&\times \left\{ 1 + \frac{\Tbar}{2\Vbar\sqrt{x}}\log\left[\frac{1+\text{e}^{-\left(\Ex + 2\Vbar\sqrt{x}\right)/\Tbar} }{1+\text{e}^{-\left(\Ex - 2\Vbar\sqrt{x}\right)/\Tbar} }\right]\right\}\, ,
\eeqy 
\beqy
\frac{4}{3}&=&\int_0^{+\infty}\text{d}x\;\sqrt{x}\left\{1-\frac{x-\bar{\mu}_q}{\Ex} \right. \nonumber \\ 
&&\left.\times \left[1+ \frac{\Tbar}{2\Vbar\sqrt{x}} \log\left(\frac{1+\text{e}^{-\left(\Ex + 2\Vbar\sqrt{x}\right)/\Tbar} }{1+\text{e}^{-\left(\Ex - 2\Vbar\sqrt{x}\right)/\Tbar} }\right)  \right]\right\}\, .
\eeqy
We have made use of the identity $\log\left[\cosh\left(a+b\right)\sech\left(a-b\right)\right]=2b + \log\left[1 + e^{-2\left(a+b\right)}\right]-\log\left[1+e^{-2(a-b)}\right]$.

It is worth remarking that although the pairing gaps and the entrainment matrix depend in general on the directions of the superfluid velocities $\pmb{V_q}$, this dependence is entirely contained in the norm of the effective superfluid velocities $\pmb{\mathbb{V}_q}$. Using Equations~\eqref{eq:EffectiveSuperfluidVelocity} and \eqref{eq:Iq}, it can be seen that the two kinds of velocities are related by 
    \begin{align}\label{eq:EffectiveSuperfluidVelocity-def}
    \pmb{\mathbb{V}_q}=\sum_{q'=n,p}\left(\frac{m}{m_{q}^{\oplus}}\delta_{qq'}+\frac{\mathcal{I}_{qq'}}{\hbar}\right)\pmb{V_{q'}}.
    \end{align}
    
It is important to realize that this relation is highly non-linear because the matrix elements $\mathcal{I}_{qq'}$, defined by Equations~\eqref{eq:Inn-def}--\eqref{eq:Theta-def}, depend themselves on the effective superfluid velocities through the functions $\mathcal{Y}_q$. For this reason, 
the mapping between $\pmb{\mathbb{V}_n}$, $\pmb{\mathbb{V}_p}$ and $\pmb{V_n}, \pmb{V_p}$ is quite complicated. It is, therefore, much more convenient to express the results in terms of $\pmb{\mathbb{V}_q}$ instead of  $\pmb{V_q}$. In particular, it can be seen that the neutron (proton) pairing gaps depend only the norms of neutron (proton) effective superfluid velocity. It is only when the chemical potentials $\lambda_q$ are needed rather than the reduced ones $\mu_q$ that the directions of the superflows become important since $\lambda_q$ are obtained from Equation~\eqref{eq:ReducedChemicalPotential} using \mbox{Equation~\eqref{eq:Iq}, namely} 
\begin{align}\label{eq:LambdaN}
    \lambda_n= \mu_n + \left(\frac{1}{2}\frac{m}{m_n^{\oplus}} + \frac{\mathcal{I}_{nn}}{\hbar}\right)m\pmb{V_n}^2 + \frac{\mathcal{I}_{np}}{\hbar}m\pmb{V_n}\cdot\pmb{V_p} +U_n\, ,
\end{align}
\begin{align}\label{eq:LambdaP}
    \lambda_p= \mu_p + \left(\frac{1}{2}\frac{m}{m_p^{\oplus}} + \frac{\mathcal{I}_{pp}}{\hbar}\right)m\pmb{V_p}^2 + \frac{\mathcal{I}_{pn}}{\hbar}m\pmb{V_p}\cdot\pmb{V_n} +U_p\, .
\end{align}

The potentials $U_q$ are functions of the nucleon densities $n_q$, the momentum densities $\pmb{j_q}$ and the kinetic densities $\tau_q$. The momentum density $\pmb{j_q}$ can be expressed as~\cite{ChamelAllard2020}
\beqy\label{eq:MomentumDensityHomogeneous}
\pmb{j_q} = 
\frac{\rho_q}{\hbar}\left[(1-\mathcal{Y}_q)\pmb{V_q} -\frac{m_q^{\oplus}}{m}\mathcal{Y}_q\sum_{q'=n,p}\frac{\mathcal{I}_{qq'}}{\hbar}\pmb{V_{q'}}\right] \, .
\eeqy
Using Equation~\eqref{eq:KineticDensity}, we find for the kinetic-energy density: 
\begin{align}\label{eq:KineticDensityHomogeneous}
    \tau_q =& \frac{3}{4}(3\pi^2)^{2/3}n_q^{5/3}\int_0^{(\mu_q + \Cutoff)/\Fermi}\text{d}x \;x\left\{\sqrt{x}-\frac{\Tbar}{\Vbar}\frac{x-\bar{\mu}_q}{2\Ex} \right. \nonumber \\ 
&\qquad\qquad\left.\times \log\left[\cosh\left(\frac{\Ex}{2\Tbar} +\frac{\Vbar}{\Tbar}\sqrt{x}\right) \sech\left(\frac{\Ex}{2\Tbar} -\frac{\Vbar}{\Tbar}\sqrt{x}\right)  \right]\right\} \nonumber \\
&\qquad +\frac{1}{2}\left(\frac{2m_q^{\oplus}}{\hbar^2}\right)\rho_q\pmb{V_q}\cdot\left(\pmb{V_q}-2\mathcal{Y}_q \pmb{\mathbb{V}_q}\right)\, .
\end{align}

In the regime $\Tbar \ll 1$, $\Vbar\ll 1$, $\Deltabar \ll 1$ and $\bar \mu_q \approx 1$, the second term in the right-hand side of Equation~\eqref{eq:KineticDensityHomogeneous} becomes negligible and the integral reduces to the Thomas-Fermi expression $\tau_q \approx \frac{3}{5}(3\pi^2)^{2/3}n_q^{5/3}$. The kinetic-energy density can be equivalently expressed in terms of the exponential function as 
\begin{align}
\tau_q =& \frac{3}{4}(3\pi^2)^{2/3}n_q^{5/3}\int_0^{(\mu_q + \Cutoff)/\Fermi}\text{d}x \;x^{3/2}\left\{1-\frac{x-\bar{\mu}_q}{\Ex} \right. \nonumber \\ 
&\qquad\qquad\left.\times \left[1+ \frac{\Tbar}{2\Vbar\sqrt{x}} \log\left(\frac{1+\text{e}^{-\left(\Ex + 2\Vbar\sqrt{x}\right)/\Tbar} }{1+\text{e}^{-\left(\Ex - 2\Vbar\sqrt{x}\right)/\Tbar} }\right)  \right]\right\}\nonumber \\
&\qquad +\frac{1}{2}\left(\frac{2m_q^{\oplus}}{\hbar^2}\right)\rho_q\pmb{V_q}\cdot\left(\pmb{V_q}-2\mathcal{Y}_q \pmb{\mathbb{V}_q}\right)\, .
\end{align}

\subsection{Physical Interpretation of the Different Velocities and Momentum Densities}
\label{sec:interpretation}

Using Equation~(66) of \cite{ChamelAllard2020}, it can be immediately seen that the true velocities associated with the transport of nucleons (mass) are related to the effective superfluid \linebreak velocities~\eqref{eq:EffectiveSuperfluidVelocity} through the relation
\beqy\label{eq:true-velocity}
\pmb{v_q}\equiv\frac{\pmb{\rho_q}}{\rho_q}=(1-\mathcal{Y}_q)\pmb{\mathbb{V}_q}\, .
\eeqy 
Let us recall that these velocities are measured relative to the normal-fluid rest frame. 
At zero temperature and subcritical superflow of nucleons of type $q$, the functions $\mathcal{Y}_q$ will be shown to vanish in Section~\ref{sec:Yq}: in this case, the effective superfluid velocity thus actually represents the true velocity $\pmb{v_q}=\pmb{\mathbb{V}_q}$. At finite temperatures, the excitation of quasiparticles entails a finite fraction $\mathcal{Y}_q>0$: nucleons thus move with a lower speed at $T>0$ than at $T=0$. If nucleons of type $q$ are nonsuperfluid, $\mathcal{Y}_q=1$ as we will see in Section~\ref{sec:Yq}, therefore their true velocity vanishes $v_q=0$: nucleons move with the normal fluid (however, the other nucleon species can flow with a different velocity if it is 
superfluid). The function $\mathcal{Y}_q$ thus measures the relative importance of quasiparticle excitations for the transport of nucleons of type $q$.

As already mentioned earlier, the superfluid ``velocity'' $\pmb{V_q}$ defined by the gradient of the phase of the condensate through Equation~\eqref{eq:SuperfluidVelocity-def} represents the momentum per unit mass of the superfluid. The superfluid momentum density of the nucleon species $q$, given by $\rho_q\pmb{V_q}$, does not coincide with the momentum density $\hbar \pmb{j_q}$ introduced in Equation~\eqref{eq:total-momentum-density}. This stems from the fact that the latter not only accounts for the superfluid momentum density but also includes the contribution from quasiparticles. This can be directly seen from Equation~\eqref{eq:MomentumDensityHomogeneous}, which can be equivalently written as
\beqy\label{eq:total-momentum-density}
\hbar \pmb{j_q} = 
\rho_q\pmb{V_q} -\mathcal{Y}_q \rho_q \frac{m_q^\oplus}{m}\pmb{\mathbb{V}_q}\, .
\eeqy

The second term can be interpreted as the momentum density of quasiparticles. Indeed, this contribution vanishes if $\mathcal{Y}_q=0$, i.e., in the absence of quasiparticle 
excitations. It is only in this limiting case that the total momentum density 
$\hbar \pmb{j_q}$ coincides with the superfluid momentum density $\rho_q \pmb{V_q}$. In
general, it can be shown using the self-consistent solutions of the TDHFB
equations presented in the previous section that the total mass current is equal to the total momentum density 
\beqy
\pmb{\rho_n}+\pmb{\rho_p}=\hbar (\pmb{j_n}+\pmb{j_p})\, , 
\eeqy 
as required by Galilean invariance (this identity can be more easily demonstrated using the general expression of the mass currents~\cite{ChamelAllard2019}).

The distinction between the different velocities and momentum densities becomes 
irrelevant if both nucleon species are nonsuperfluid since $\mathcal{Y}_n=\mathcal{Y}_p=1$
implies that $\pmb{v_n}$, $\pmb{v_p}$, $\hbar \pmb{j_n}/\rho_n$ and 
$\hbar \pmb{j_p}/\rho_p$ all vanish in the fluid rest frame, i.e., all nucleons move 
with the normal fluid, as expected. Likewise, in the limiting case of a single superfluid constituent at zero temperature and subcritical superflow, we have $\pmb{v_q}=\pmb{V_q}=\pmb{\mathbb{V}_q}=\hbar \pmb{j_q}/\rho_q$.

\subsection{Landau's Approximations}
\label{sec:LandauApprox}

The neutron–proton superfluid mixture can be alternatively described using Landau's theory~\cite{gusakov2005,leinson2017,leinson2018}. The TDHFB theory can be reduced  
to a similar form after introducing a series of approximations.
Specifically, assuming that the critical temperatures 
and the critical superfluid velocities are small compared to their Fermi counterpart, 

\begin{itemize}
    \item  instead of solving Equation~\eqref{eq:DimensionlessChemicalPotentialEquation},   the  reduced chemical potentials~\eqref{eq:ReducedChemicalPotential} are approximated by their associated Fermi energies ($\mu_q\approx \Fermi$), thus ignoring any dependence on temperature, currents, and pairing gaps; 
    \item the single-particle energies~\eqref{eq:varepsilon} are calculated at zero temperature, in the absence of currents ignoring any dependence on the pairing gaps, and expanding linearly around the Fermi surface (denoting by $\breve{Q}$ the approximate expression for a \mbox{quantity $Q$})
    \beqy\label{eq:Landau-energy}
    \varepsilon^{(q)}_{\pmb{k}} \approx \breve{\varepsilon}^{(q)}_{\pmb{k}} \equiv \hbar V_{Fq}(k-k_{Fq})
    \, ; 
    \eeqy
    \item the quasiparticle energies~\eqref{eq:QuasiparticleEnergy} are similarly expanded as 
    \beqy
    \mathfrak{E}_{\pmb{k}}^{(q)}\approx \breve{\mathfrak{E}}_{\pmb{k}}^{(q)} + \hbar k_{Fq} \mathbb{V}_q \cos \theta_{\pmb{k}}\, , \qquad  \breve{\mathfrak{E}}_{\pmb{k}}^{(q)}=\sqrt{\breve{\varepsilon}_{\pmb{k}}^{(q)2}+\breve{\Delta}_q^2}\ ;
    \eeqy

    \item the density of single-particle states $\mathcal{D}(\varepsilon)$ in $\pmb{k}$-space integrations~\eqref{eq:continuum} is approximated by its value on the Fermi surface, $\mathcal{D}(\varepsilon)\approx \breve{\mathcal{D}}(0)$ with 
    \beqy\label{eq:DoS-Fermi}
    \breve{\mathcal{D}}(0)=\frac{k_{Fq} \breve{m}_q^\oplus}{2\pi^2 \hbar^2}\,  ;
    \eeqy 
    
    \item the derivatives of the nuclear-energy terms $E_{\rm nuc}^{j}$ entering  Equations~\eqref{eq:def-Bq}--\eqref{eq:Theta-def}, are evaluated in the absence of currents; 
    
\end{itemize}   

     In previous studies~\cite{gusakov2005,leinson2017,leinson2018}, the pairing gaps   $\breve{\Delta}_q$ were obtained  in the weak-coupling approximation $\breve{\Delta}_q\ll \Fermi, \Cutoff$ at zeroth order from the following approximate equation 
         (see Appendix~\ref{app:weak-coupling})
    \begin{align}\label{eq:LandauGapWeak}
        \log \left(\frac{\breve{\Delta}_q}{\breve{\Delta}_q^{(0)}}\right)\approx \int_0^{+\infty}\frac{\text{d}x}{\ExLandau}\left\{\frac{\Tbar}{2\Vbar}\log\left[\cosh\left(\frac{\ExLandau}{2\Tbar}+\frac{\Vbar}{\Tbar}\right)\sech\left(\frac{\ExLandau}{2\Tbar}-\frac{\Vbar}{\Tbar}\right)\right]-1\right\}\, ,
    \end{align}    
   or in terms of the exponential function
     \begin{align}\label{eq:LandauGapWeakExpForm}
        \log \left(\frac{\breve{\Delta}_q}{\breve{\Delta}_q^{(0)}}\right)\approx \frac{\Tbar}{2\Vbar}\int_0^{+\infty}\frac{\text{d}x}{\ExLandau}\log\left[\frac{1+\text{e}^{-\left(\ExLandau + 2\Vbar\right)/\Tbar} }{1+\text{e}^{-\left(\ExLandau - 2\Vbar\right)/\Tbar} }\right]\, ,
    \end{align}
    where  $\breve{\Delta}_q^{(0)}$ denotes the pairing gaps at $T=0 $ in the absence of currents. The latter were 
    determined using the BCS relation~ \cite{Bardeen1957} (introducing the Euler–Mascheroni constant $\gamma \simeq 0.577216$)
    \beqy\label{eq:Tc0Landau}
    \breve{\Delta}_q^{(0)}=\frac{k_{\text{B}}\pi}{\text{e}^{\gamma}}\breve{T}_{cq}^{(0)}\, ,
    \eeqy
     by fixing arbitrarily the associated critical temperatures $\breve{T}_{cq}^{(0)}$.
    
    Moreover, the functions $\mathcal{Y}_q$ were replaced by the functions $\Phi_q$ of \cite{leinson2018}, which can be expressed as 
    \beqy\label{eq:LandauYq}
\Phi_q&=& \frac{3}{4}\frac{\Tbar}{\Vbar^2}\int_0^{+\infty}\text{d}x\; \log\left\{\left[1+\text{e}^{-\left(\ExLandau - 2\Vbar\right)/\Tbar}\right]\left[1+\text{e}^{-\left(\ExLandau + 2\Vbar\right)/\Tbar}\right]\right\}\nonumber\\
&+& \frac{3}{8}\frac{\Tbar^2}{\Vbar^3}\int_0^{+\infty}\text{d}x\; \left\{\text{Li}_2\left[-\text{e}^{-\left(\ExLandau - 2\Vbar\right)/\Tbar}\right]-\text{Li}_2\left[-\text{e}^{-\left(\ExLandau + 2\Vbar\right)/\Tbar}\right]\right\}\, ,
\eeqy
where 
\beqy\label{eq:ExLandau}
\ExLandau
=\sqrt{x^2 + \left(\frac{\breve{\Delta}_q}{\Fermi}\right)^2}\, .
\eeqy

Introducing the critical effective superfluid velocities~\cite{Alexandrov2003}
\beqy\label{eq:Vc0Landau}
\breve{\mathbb{V}}_{cq}^{(0)}=\frac{\text{e}}{2}\frac{\breve{\Delta}_q^{(0)}}{\hbar k_{Fq}} ,
\eeqy
the approximate pairing gap Equation~\eqref{eq:LandauGapWeak} and the functions~\eqref{eq:LandauYq} can be equivalently expressed in terms of the reduced temperature $\Tnorm\equiv T/\breve{T}_{cq}^{(0)}$ and the reduced effective superfluid velocity $\Vnorm\equiv \mathbb{V}_q/\breve{\mathbb{V}}_{cq}^{(0)}$ as follows: 
\begin{align}\label{eq:LandauGapUniversal}
        \log \left(\frac{\breve{\Delta}_q}{\breve{\Delta}_q^{(0)}}\right)\approx&\int_0^{+\infty}\frac{\text{d}y}{\Eyq}\left\{\frac{2}{\pi}\text{e}^{\gamma-1}\frac{\Tnorm}{\Vnorm}\log\left[\cosh\left(\frac{\pi}{2}\text{e}^{-\gamma}\left(\frac{\Eyq}{\Tnorm}+\frac{\text{e}}{2}\frac{\Vnorm}{\Tnorm}\right)\right)\right.\right. \nonumber\\
        &\qquad\times\left.\left.\sech\left(\frac{\pi}{2}\text{e}^{-\gamma}\left(\frac{\Eyq}{\Tnorm}-\frac{\text{e}}{2}\frac{\Vnorm}{\Tnorm}\right)\right)\right]-1\right\}\, ,
\end{align}
or using Equation~\eqref{eq:LandauGapWeakExpForm}
\begin{align}
\log \left(\frac{\breve{\Delta}_q}{\breve{\Delta}_q^{(0)}}\right)\approx&\frac{2}{\pi}\text{e}^{\gamma-1}\frac{\Tnorm}{\Vnorm}\int_0^{+\infty}\frac{\text{d}y}{\Eyq}\log\left[\frac{1+\text{e}^{-\pi\text{e}^{-\gamma}\left(\Eyq + \frac{\text{e}}{2}\Vnorm\right)/\Tnorm} }{1+\text{e}^{-\pi\text{e}^{-\gamma}\left(\Eyq - \frac{\text{e}}{2}\Vnorm\right)/\Tnorm} }\right]\, ,
\end{align}
and
% start a new page without indent 4.6cm
%\clearpage
%\end{paracol}
%\nointerlineskip
\begin{align}\label{eq:LandauPhiUniversal}
\Phi_q&= \frac{12}{\pi}\text{e}^{\gamma-2}\frac{\Tnorm}{\Vnorm^2}\int_0^{+\infty}\text{d}y\; \log\left\{\left[1+\exp\left(-\pi\text{e}^{-\gamma}\left(\frac{\Eyq}{\Tnorm}-\frac{\text{e}}{2}\frac{\Vnorm}{\Tnorm}\right)\right)\right]\right.\notag\\
&\times\left.\left[1+\exp\left(-\pi\text{e}^{-\gamma}\left(\frac{\Eyq}{\Tnorm}+\frac{\text{e}}{2}\frac{\Vnorm}{\Tnorm}\right)\right)\right]\right\}\nonumber\\
&+ \frac{24}{\pi^2}\text{e}^{2\gamma-3}\frac{\Tnorm^2}{\Vnorm^3}\int_0^{+\infty}\text{d}y\; \left\{\text{Li}_2\left[-\exp\left(-\pi\text{e}^{-\gamma}\left(\frac{\Eyq}{\Tnorm}-\frac{\text{e}}{2}\frac{\Vnorm}{\Tnorm}\right)\right)\right]\right.\notag\\&-\left.\text{Li}_2\left[-\exp\left(-\pi\text{e}^{-\gamma}\left(\frac{\Eyq}{\Tnorm}+\frac{\text{e}}{2}\frac{\Vnorm}{\Tnorm}\right)\right)\right]\right\}\, ,
\end{align}
%\begin{paracol}{2}
%\linenumbers
%\switchcolumn
with
\beqy
\Eyq = \sqrt{y^2 + \left(\frac{\breve{\Delta}_q}{\breve{\Delta}_q^{(0)}}\right)^2} \, .
\eeqy

These alternative formulations show that $\breve{\Delta}_q/\breve{\Delta}_q^{(0)}$ and $\Phi_q$ are universal functions of suitably rescaled temperature and effective superfluid velocity, independently of the nucleon species under consideration, the composition, and the details of the adopted nuclear-energy-density functional.

\section{Application to Neutron Stars}
\label{sec:results}

Although the entrainment matrix can be written in the deceptively simple analytical form~\eqref{eq:EntrainmentMatrix}, its dependencies on 
the temperature and on the superfluid velocities remain 
implicit and highly nontrivial. To obtain actual values, numerical solutions of \linebreak Equations~\eqref{eq:DimensionlessGapEquation} and~\eqref{eq:DimensionlessChemicalPotentialEquation} are needed. In this work, we have considered the Brussels–Montreal functionals, whose main features are described in Section~\ref{sec:BSk}. Results are presented in the subsequent sections. 

\subsection{Brussels–Montreal Functionals}
\label{sec:BSk}

The Brussels–Montreal functionals from BSk16 and beyond (see \cite{chamel2015,goriely2016} for a brief overview) were constructed from extended Skyrme effective nucleon-nucleon interactions, whose parameters were precision-fitted to essentially all experimental nuclear data on atomic masses and charge radii while ensuring realistic properties of homogeneous nuclear matter (neutron-matter equation of state, effective masses, symmetry energy, incompressibility coefficient, pairing gaps). 

The functional derivatives of the energy $E_{\rm nuc}^j$ with respect to $X_0$ and $X_1$ appearing in the effective masses, the matrix $\mathcal{I}_{qq'}$ and the entrainment matrix are expressible in terms of the parameters of the effective interaction as~\cite{ChamelAllard2019}
\beqy\label{eq:dEdX0}
\frac{\delta E^j_{\rm nuc}}{\delta X_0}=\frac{3}{16}t_1+\frac{1}{4} t_2\left( \frac{5}{4}+x_2\right)+\frac{3}{16}t_4 n^\beta+\frac{1}{4} t_5\left( \frac{5}{4}+x_5\right) n^\gamma
\eeqy
\beqy \label{eq:dEdX1}
\frac{\delta E^j_{\rm nuc}}{\delta X_1}=-\frac{1}{8} t_1\left(\frac{1}{2}+x_1\right)+\frac{1}{8}
t_2\left(\frac{1}{2}+x_2\right)-\frac{1}{8} t_4\left(\frac{1}{2}+x_4\right)n^\beta+\frac{1}{8}
t_5\left(\frac{1}{2}+x_5\right) n^\gamma \, .
\eeqy

The potentials in homogeneous matter read 
(recalling the shorthand notations $n\equiv n_0$, $\pmb{j}\equiv\pmb{j_0}$ and $\tau\equiv\tau_0$)
\beqy\label{eq:UqPotential}
U_q&=& t_0\Biggl[\left(1+ \frac{1}{2} x_0\right)n - \left(\frac{1}{2} + x_0\right)n_q\Biggr] +\frac{1}{4} t_1 \Biggl[\left(1+ \frac{1}{2} x_1\right) \tau - \left(\frac{1}{2} +x_1\right) \tau_q \Biggr]  \nonumber \\
&+&\frac{1}{4} t_2 \Biggl[\left(1+ \frac{1}{2} x_2\right) \tau  + \left(\frac{1}{2} +x_2\right) \tau_q  \Biggr]  \nonumber \\
&+&\frac{1}{12} t_3n^{\alpha-1} \Biggl[\left(1+ \frac{1}{2} x_3\right) (2+\alpha)
n^{2} - \left(\frac{1}{2} +x_3\right)\left(2n n_q
+\alpha \sum_{q^\prime=n,p} n_{q^\prime}^2 \right) \Biggr]  
\nonumber \\
&+&\frac{1}{4} t_4n^{\beta-1}\Bigg[\left(1+\frac{1}{2}x_4\right) (1+\beta)n\tau -\left(\frac{1}{2}+x_4\right)
\left(n\tau_q +\beta \sum_{q^\prime=n,p}
n_{q^\prime}\tau_{q^\prime}\right)\Bigg]\nonumber\\
&+&\frac{1}{4}t_5n^{\gamma-1}\Bigg[\left(1+\frac{1}{2}x_5\right)(1+\gamma)n\tau +\left(\frac{1}{2}+x_5\right)\left(n\tau_q +\gamma\sum_{q^\prime=n,p}n_{q^\prime}\tau_{q^\prime} \right)\Bigg] \nonumber\\
&+&\frac{1}{8}t_4 \beta n^{\beta-1}\Bigg[(x_4-1) \pmb{j}^2 +4\left(\frac{1}{2}+x_4\right)\pmb{j_q}\cdot(\pmb{j_q}-\pmb{j}) \Bigg] \nonumber\\
&-&\frac{1}{8}t_5\gamma n^{\gamma-1}\Bigg[3\left(x_5+1\right)\pmb{j}^2 + 4\left(\frac{1}{2}+x_5\right)\pmb{j_q}\cdot\left(\pmb{j_q}-\pmb{j}\right)\Biggr]   \nonumber\\
&+&\frac{1}{4} \sum_{q^\prime=n,p}\frac{\partial v^{\pi q^\prime}}{\partial n_q}\, \widetilde{n}_{q^\prime}^2 \, .
\eeqy

The functional derivative of the energy $E$ with respect to the square modulus of the abnormal density $\widetilde{n}_q$ is related to the strength $v^{\pi q}$ of the effective pairing interaction as 
\begin{align}
\frac{\delta E}{\delta\vert \widetilde{n}_q\vert^2} = \frac{1}{4}v^{\pi q}\, .
\end{align}

In most existing functionals, $v^{\pi q}$ is expressed~\cite{bertsch1991} as the sum of a constant ``volume'' term and a ``surface term'' proportional to the density $n$ to some power with parameters adjusted empirically to reproduce the  average pairing gaps in some finite nuclei~\cite{dobaczewski1995b}. Such functionals may thus lead to unreliable predictions when applied to homogeneous nuclear matter~\cite{chamel2008}. On the contrary, the pairing strengths $v^{\pi\, q}[n_n,n_p]<0$ of the Brussels–Montreal functionals were determined so as to reproduce the $^1S_0$ pairing gaps in infinite homogeneous neutron matter and in symmetric nuclear matter at $T=0$ and in the absence of currents (these reference gaps will be denoted by $\Dref_{NM}$ and  $\Dref_{SM}$ respectively), as obtained from many-body calculations using realistic potentials (see \cite{chamel2008,goriely2009,goriely2009b} for details).  Very accurate analytical expressions for the pairing strengths were obtained in \cite{Chamel2010Pairing}:   
\beqy\label{eq:Vpi}
    v^{\pi q}\left[n_n,n_p\right]&=&-\frac{4\pi^2 \hbar^2 \Sigma_q}{mk_{Fq}}\left[ \log\left(\frac{64 m_q^{\oplus}\Sigma_q}{m}\frac{\Fermi\Cutoff}{ \Drefq^2 }\right) \right. \nonumber \\
    &+& \left.  2\sqrt{1+\frac{\Cutoff}{\Fermi}\frac{m}{m_q^{\oplus}\Sigma_q}} - 2 \log \left( 1+\sqrt{1+\frac{\Cutoff}{\Fermi}\frac{m}{m_q^{\oplus}\Sigma_q}}\right) - 4    
    \right]^{-1}\, .
\eeqy

The parameters $\Sigma_q$ are used here to distinguish Brussels–Montreal functionals\linebreak BSk17-29~\cite{chamel2009,goriely2009,goriely2010,goriely2013,goriely2015} which neglect self-energy corrections ($\Sigma_q=1$) from the most recent series BSk30-32~\cite{goriely2016} which include them ($\Sigma_q=m/m_q^{\oplus}$). 
Since reference pairing gaps $\Dref(n_n,n_p)$ for arbitrary composition are needed, the following interpolation ansatz was adopted in \cite{goriely2009} for BSk17 and subsequent functionals: 
\begin{equation}\label{eq:reference-gaps}
\Drefq(n_n,n_p)=\Dref_{SM}(n)(1-|\eta|) \pm \Dref_{NM}(n_q)
\,\eta\,\frac{n_q}{n}   \, ,
\end{equation}
where $\eta = (n_n-n_p)/n$ and the upper (lower) sign is to be taken 
for $q = n (p)$. Because this parametrization is empirical, we have found that $\Drefq(n_n,n_p)$
may become negative depending on the composition and density $n$. In such cases, we merely set $\Drefq(n_n,n_p)=0$. As for the nucleon mass, it is defined as $m=2(1/m_n+1/m_p)^{-1}$.

For numerical calculations, we will adopt the Brussels–Montreal functional BSk24~\cite{goriely2013}. The reference pairing gaps were taken from the extended Brueckner–Hartree–Fock calculations of \cite{cao2006}. The associated parameters are indicated in Tables~\ref{tab:BSk24} and \ref{tab:RefGaps}. The reference gaps can be conveniently represented as 
\begin{equation}
\Dref_{SM}(n)=H(k_{\rm max}-k_F)\, \Delta_0 \frac{k_F^3}{k_F^2+k_1^2}\frac{(k_F-k_2)^2}{(k_F-k_2)^2+k_3^2}\, ,
\end{equation}
\begin{equation}
\Dref_{NM}(n_q)=H(k_{\rm max}-k_{Fq})\, \Delta_0 \frac{k_{Fq}^2}{k_{Fq}^2+k_1^2}\frac{(k_{Fq}-k_2)^2}{(k_{Fq}-k_2)^2+k_3^2}\, ,
\end{equation}
where $k_{F}=(3\pi^2 n/2)^{1/3}$, $H$ is the Heaviside unit-step function, and $k_1$, $k_2$, $k_3$ and $k_{\rm max}$ are fitted parameters. 
The functional BSk24 has been recently employed for determining the composition and the equation of state of dense matter throughout all regions of a neutron star~\cite{pearson2018,shelley2020} including the pasta mantle~\cite{pearson2020} and allowing for strong magnetic fields~\cite{mutafchieva2019}. More importantly, as shown in \cite{perot2019,gulminelli2021,thi2021}, this functional turns out to be in very good agreement with existing astrophysical observations including those from the binary neutron-star merger GW170817\cite{LIGO2018} as well as from PSR~J0740+6620 and PSR~J0030+0451 by 
the Neutron star Interior Composition Explorer (NICER)~\cite{riley2019,miller2019,riley2021,miller2021}. Results for the entrainment matrix at finite temperatures but in the absence of superflows have been recently published in \cite{kantor2020} within Landau's theory using values for the Landau parameters calculated for the Brussels–Montreal functionals including BSk24 and setting arbitrarily the critical temperatures. We will present here consistent numerical results for the pairing gaps, chemical potentials and entrainment matrix for arbitrary temperatures and superfluid velocities in different regions of neutron-star cores.

 \begin{specialtable}[H]
    \tablesize{\small}
    \caption{Parameters of the functional BSk24~\cite{goriely2013}. The unit of length is femtometer and the unit of energy is megaelectronvolt.}
    \label{tab:BSk24}
\setlength{\cellWidtha}{\columnwidth/2-2\tabcolsep-1in}
\setlength{\cellWidthb}{\columnwidth/2-2\tabcolsep+1in}

\scalebox{1}[1]{\begin{tabularx}{\columnwidth}{>{\PreserveBackslash\raggedright}m{\cellWidtha}>{\PreserveBackslash\raggedright}m{\cellWidthb}}
\toprule
$t_0$ %{\scriptsize [MeV fm$^3$]}       
& $-$3970.29  \\
$t_1$ %{\scriptsize [MeV fm$^5$]}         
& 395.766   \\
$t_2$% {\scriptsize [MeV fm$^5$]}         
& $10^{-5} $      \\
$t_3$% {\scriptsize [MeV fm$^{3+3\alpha}$]} 
& 22648.6   \\
$t_4$ % {\scriptsize [MeV fm$^{5+3\beta}$]}
&$-$100.000   \\
$t_5$ %{\scriptsize [MeV fm$^{5+3\gamma}$]} 
&$-$150.000   \\
$x_0$                                      & 0.894371  \\
$x_1$                                      &0.0563535  \\
$x_2$% {\scriptsize [MeV fm$^5$]}       
&$-$0.138961 $\times$ $10^9$  \\
$x_3$                                      &1.05119   \\
$x_4$                                      &2.00000   \\
$x_5$                                      &$-$11.0000  \\
$\alpha$                                   &1/12      \\
$\beta$                                    &1/2       \\
$\gamma$                                   &1/12      \\
$\varepsilon_{\Lambda}$% {\scriptsize [MeV]}
&16.0      \\            
$\Sigma_q$                                  & 1 \\ 
\bottomrule
\end{tabularx}}

\end{specialtable}

\vspace{-6pt}

 \begin{specialtable}[H]
    \tablesize{\small}
    \caption{Parameters of the reference gaps from \cite{goriely2009b}. The unit of length is femtometer and the unit of energy is megaelectronvolt. With kind permission of The European Physical
Journal (EPJ).}
    \label{tab:RefGaps}
\setlength{\cellWidtha}{\columnwidth/6-2\tabcolsep+0.0in}
\setlength{\cellWidthb}{\columnwidth/6-2\tabcolsep+0.0in}
\setlength{\cellWidthc}{\columnwidth/6-2\tabcolsep+0.0in}
\setlength{\cellWidthd}{\columnwidth/6-2\tabcolsep+0.0in}
\setlength{\cellWidthe}{\columnwidth/6-2\tabcolsep+0.0in}
\setlength{\cellWidthf}{\columnwidth/6-2\tabcolsep+0.0in}
\scalebox{1}[1]{\begin{tabularx}{\columnwidth}{>{\PreserveBackslash\raggedright}m{\cellWidtha}>{\PreserveBackslash\raggedright}m{\cellWidthb}>{\PreserveBackslash\raggedright}m{\cellWidthc}>{\PreserveBackslash\raggedright}m{\cellWidthd}>{\PreserveBackslash\raggedright}m{\cellWidthe}>{\PreserveBackslash\raggedright}m{\cellWidthf}}
\toprule
 & $\boldsymbol{\Delta_0}$ & $\boldsymbol{k_1}$ & $\boldsymbol{k_2}$ & $\boldsymbol{k_3}$ & $\boldsymbol{k_{\rm max}}$ \\ 
 \midrule
$\Dref_{SM}$ & 133.779 & 0.943146 & 1.52786 & 2.11577 & 1.51 \\ 
$\Dref_{NM}$ & 14.9003 & 1.18847 & 1.51854 & 0.639489 & 1.52 \\ 
\bottomrule
\end{tabularx}}

\end{specialtable}

\subsection{Numerical Implementation}
\label{sec:implementation}

\textls[-15]{The TDHFB equations are solved as follows. We first compute the pairing  gaps $\Delta_q^{(0)}$ at zero temperature and in the absence of currents by solving Equations~\eqref{eq:DimensionlessGapEquation} and~\eqref{eq:DimensionlessChemicalPotentialEquation} for $T=0$ and $\mathbb{V}_q=0$ via a root-finding method with a precision of $10^{-8}$,  searching around the approximate solutions $ \mu_q \approx \Fermi$ and 
the following expression given by Equation (14) in \cite{Chamel2010Pairing}: }
\beqy\label{eq:ChamelApprox}\displaystyle
\Delta_q^{(0)}\approx \frac{8\sqrt{\Fermi\varepsilon_{\Lambda}}}{1+\sqrt{1+\varepsilon_{\Lambda}/\Fermi}}\exp\left[\frac{1}{v^{\pi q}\breve{\mathcal{D}}(0)} + \sqrt{1+\frac{\varepsilon_{\Lambda}}{\Fermi}}-2\right]\, .
\eeqy 

In a second stage, we use this solution to determine iteratively the pairing gaps $\Delta_q$ and the reduced chemical potentials $\mu_q$ at finite temperature $T>0$ and for given effective superfluid velocities $\pmb{\mathbb{V}_n}$ and $\pmb{\mathbb{V}_p}$. An initial guess for $\Delta_q$ is obtained by solving Equation~\eqref{eq:LandauGapWeak} using for $\breve{\Delta}_q^{(0)}$ the gap obtained previously. With this first estimate of the gap, Equation~\eqref{eq:DimensionlessChemicalPotentialEquation} is solved using the Newton-Raphson method and $\mu_q\approx \Fermi$ as the initial guess. Substituting these first estimates for $\Delta_q$ and $\mu_q$ in the right-hand side of Equation~\eqref{eq:DimensionlessGapEquation} leads to a new estimate for the pairing gap $\Delta_q$, which is injected in Equation~\eqref{eq:DimensionlessChemicalPotentialEquation} to refine the chemical potential $\mu_q$. The process is repeated until the difference in the pairing gaps between two successive iterations lies below $10^{-4} \Delta_q^{(0)}$.  
Having found $\Delta_q$ and $\mu_q$, the functions $\mathcal{Y}_q$ are calculated from Equation \eqref{eq:Yq}. The entrainment matrix can be easily inferred from Equations \eqref{eq:EntrainmentMatrix}--\eqref{eq:Theta-def} together with Equations \eqref{eq:dEdX0} and \eqref{eq:dEdX1}. 

\subsection{$^1$S$_0$ Pairing Gaps}
\label{sec:gaps}

\begin{figure}[H]

\centerline{\includegraphics[width=7.5cm]{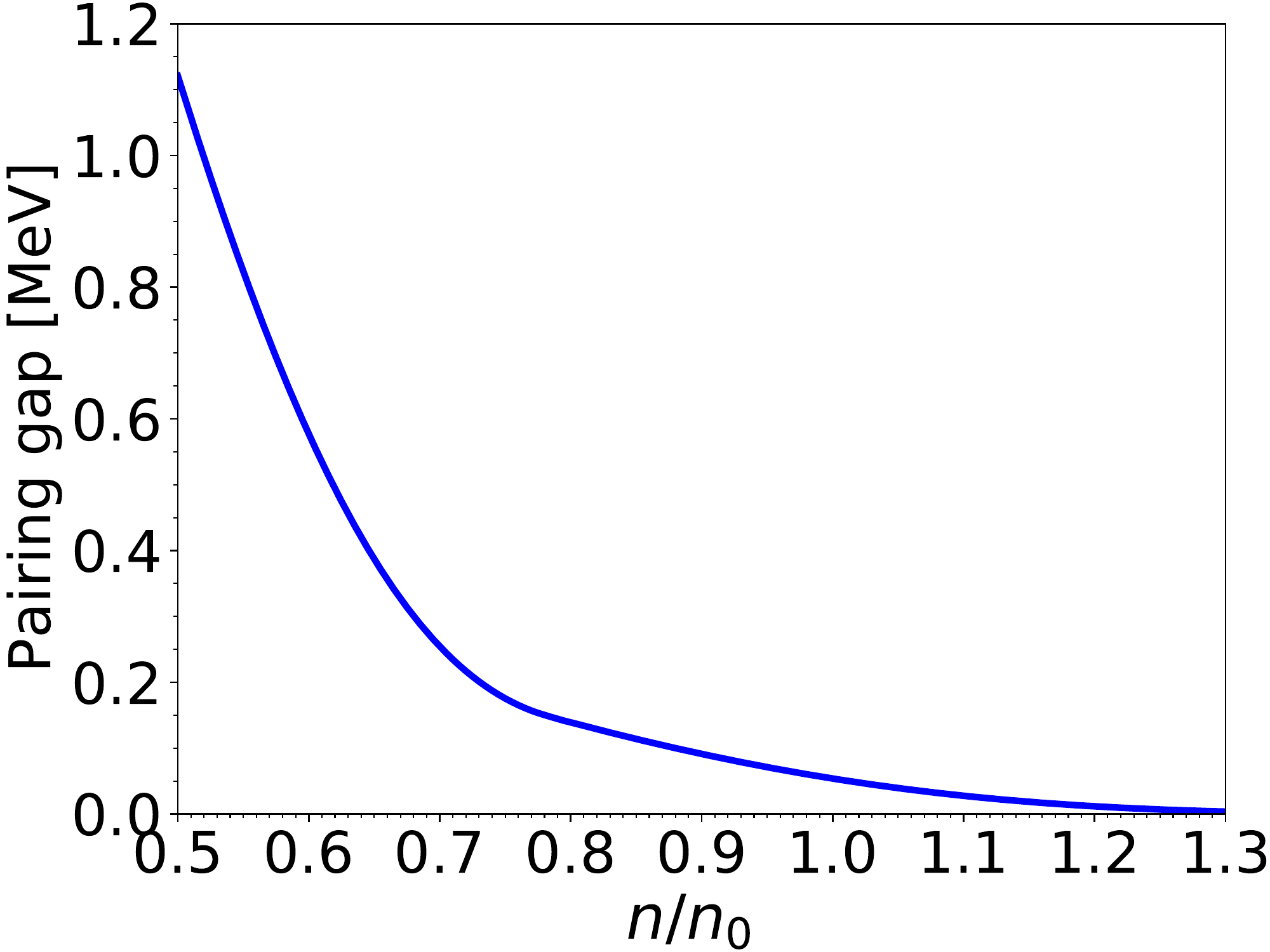}}
\caption{$^1S_0$ neutron pairing gaps (in MeV) at zero temperature and in the absence of currents for $npe\mu$ matter in beta-equilibrium as a function of the baryon density $n$ in units of saturation density $n_0$. The pairing gaps obtained from \eqref{eq:ChamelApprox} are indistinguishable from the exact ones. 
}
\label{fig:GapZeroN}
\end{figure}

\begin{figure}[H]

\centerline{\includegraphics[width=7.5cm]{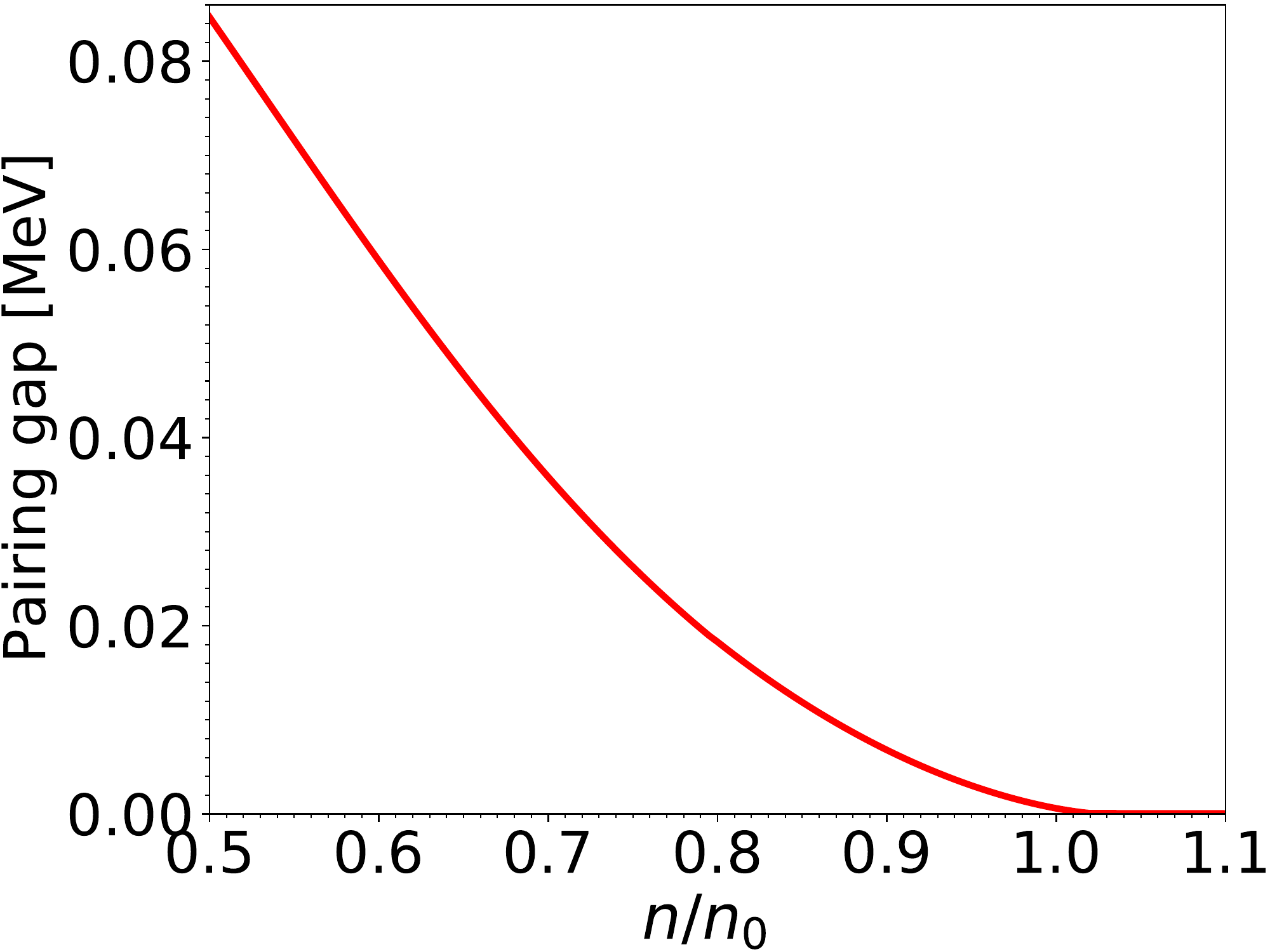}}
\caption{Same as Figure~\ref{fig:GapZeroN} for $^1S_0$ proton pairing gaps.}
\label{fig:GapZeroP}
\end{figure}

The $^1$S$_0$ neutron and proton pairing gaps $\Delta_q^{(0)}$ for $npe\mu$ matter in beta-equilibrium at $T=0$ and $\mathbb{V}_q=0$ are displayed in Figures~\ref{fig:GapZeroN} and~\ref{fig:GapZeroP} at densities relevant for the outer core of neutron stars above the crust-core transition at density $n_\text{cc}=0.08076$~fm$^{-3}\approx 0.5n_0$, where $n_0=0.1578$~fm$^{-3}$ is the nuclear saturation density with the corresponding mass density $\rho_0 = mn_0=2.654\times 10^{14}$~g~cm$^{-3}$. We have made use of the composition calculated in \cite{pearson2018}. 

The
approximate formula~\eqref{eq:ChamelApprox} is found to be in excellent agreement with the exact results, the deviations being contained within the thickness of the solid lines. With neutron-star matter containing only a few percents of protons, the reference pairing gaps for neutrons~\eqref{eq:reference-gaps} are approximately given by that in pure neutron matter $\Dref_n(n_n,n_p) \approx \Dref_{NM}(n_n)$, as obtained from the many-body calculations of \cite{cao2006} using realistic potentials. On the contrary, the reference pairing gaps for protons are mainly determined by the interpolation $\Dref_p(n_n,n_p) \approx \Dref_{NM}(n_p) n_p/n$. This explains why the proton gaps $\Delta_p$ are significantly smaller than the neutron ones $\Delta_n$ unlike 
those usually employed in neutron-star studies, as e.g., in \cite{ho2015}. This result could reveal a deficiency of the interpolation~\eqref{eq:reference-gaps}. On the other hand, 
the proton pairing gaps remain highly uncertain (see, e.g., \cite{baldo2012,lombardo2013,sedrakian2019}). Recent many-body calculations~\cite{guo2019} taking into account medium-polarization effects through self-energy and vertex corrections lead to very small proton pairing gaps in neutron-star matter of comparable magnitudes to those plotted in Figure~\ref{fig:GapZeroP}. This study also shows that the three-body interactions, especially those between two protons and one neutron, reduce considerably the domain of temperatures and densities over which protons are superfluid (see also \cite{zuo2004,zhou2004}).

\vspace{-6pt}

The variations of the neutron and proton pairing gaps with temperature and effective superfluid velocity are found to be essentially independent of density when considering the dimensionless ratios $\Delta_q/\Delta_q^{(0)}$, $T/T_{cq}^{(0)}$ and $\mathbb{V}_q/\mathbb{V}_{cq}^{(0)}$, with 
\beqy\label{eq:Tc0}
T_{cq}^{(0)}=\frac{\text{e}^{\gamma}}{k_{\text{B}}\pi}\Delta_q^{(0)} ,
\eeqy
\beqy\label{eq:Vc0}
\mathbb{V}_{cq}^{(0)}=\frac{\text{e}}{2}\frac{\Delta_q^{(0)}}{\hbar k_{Fq}} .
\eeqy

\begin{figure}[H]

\centerline{\includegraphics[width=8cm]{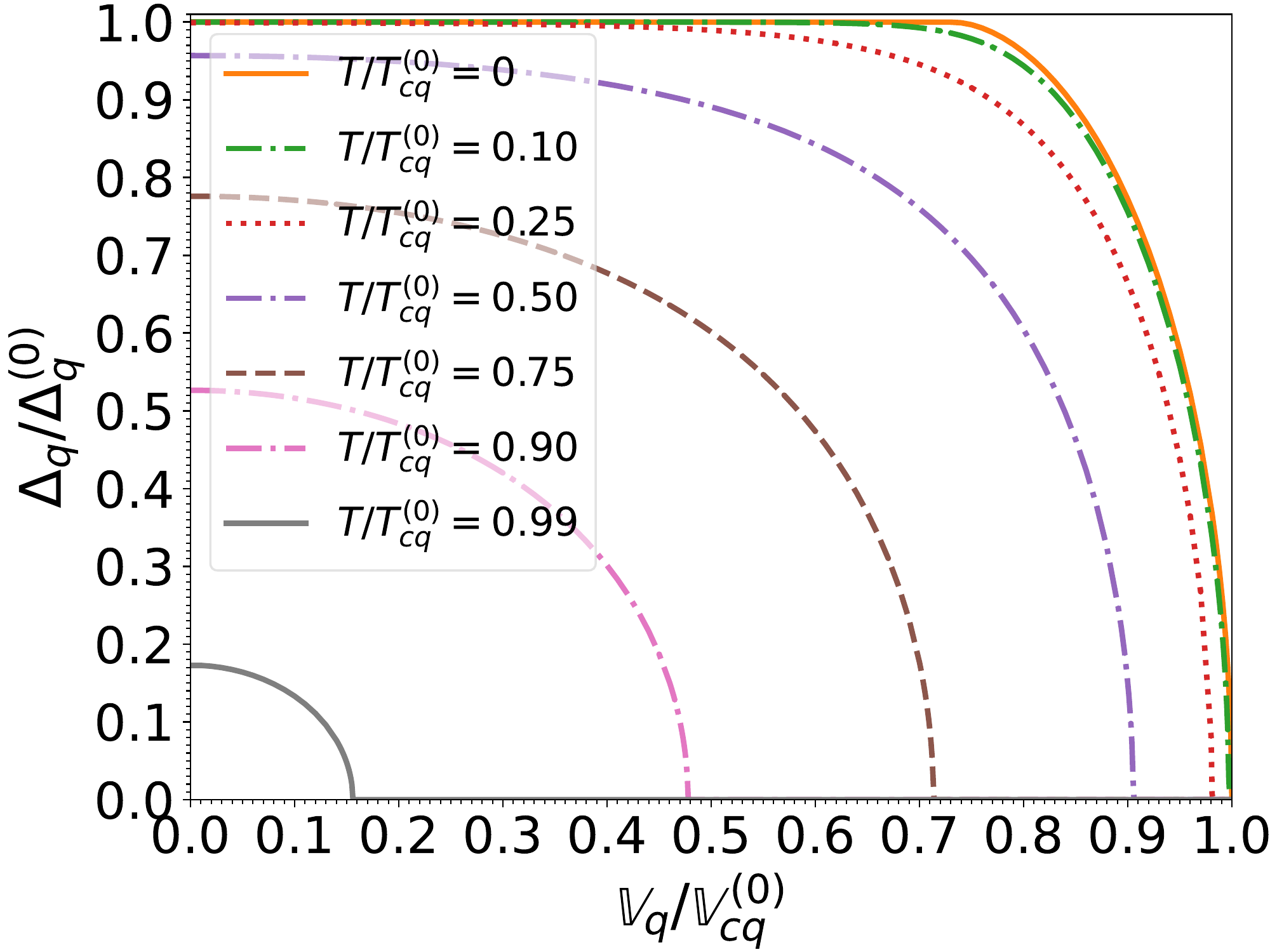}}
\caption{$^1$S$_0$ nucleon pairing gap relative to that at zero temperature and in the absence of superflow, as a function of the normalized effective superfluid velocity $\mathbb{V}_q/\mathbb{V}_{cq}^{(0)}$ for different normalized temperatures $T/T_{cq}^{(0)}$. }
\label{fig:PairingGapVitesseBSk24ncc}
\end{figure}

\begin{figure}[H]

\centerline{\includegraphics[width=8cm]{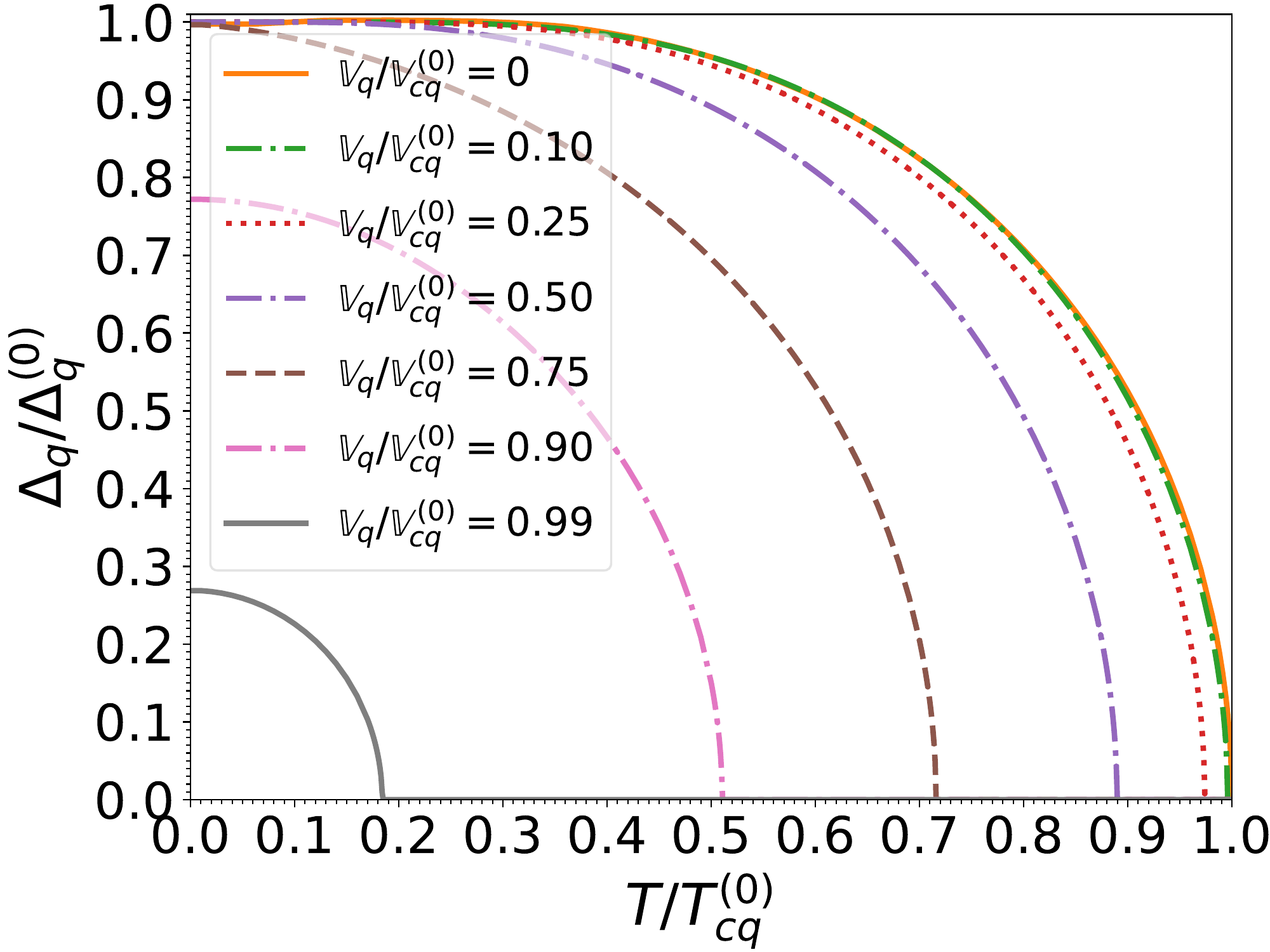}}
\caption{$^1$S$_0$ nucleon pairing gap relative to that at zero temperature and in the absence of superflow, as a  function of the normalized temperature $T/T_{cq}^{(0)}$ for different normalized effective superfluid velocities $\mathbb{V}_q/\mathbb{V}_{cq}^{(0)}$. }
\label{fig:PairingGapTemperatureBSk24ncc}
\end{figure}

As shown in Figures~\ref{fig:PairingGapVitesseBSk24ncc} and~\ref{fig:PairingGapTemperatureBSk24ncc}, the gaps for both neutrons and protons decrease monotonically with increasing temperature and effective superfluid velocity due to the excitation of quasiparticles. 
For vanishing effective superfluid velocities $\mathbb{V}_{q}=0$ (i.e., in the absence of mass flow $\pmb{\rho_q}=\pmb{0}$), the temperature dependence of the pairing gaps is well fitted by the following expression~\cite{YakovlevandLevenfish1994}:
\beqy\label{eq:YavovlevGapInterpolation}
\frac{\Delta_q (T\leq T_{cq}^{(0)},\mathbb{V}_{q}=0)}{\Delta_q^{(0)}}=\frac{\text{e}^{\gamma}}{\pi}\sqrt{1-\frac{T}{T_{cq}^{(0)}}}\left(1.456\frac{T}{T_{cq}^{(0)}} - 0.157\sqrt{\frac{T}{T_{cq}^{(0)}}} + 1.764 \right)\, .
\eeqy

This same formula was applied in \cite{gusakov2005} to evaluate the entrainment matrix. At zero temperature, the pairing gap remains equal to $\Delta_{q}^{(0)}$ until the effective superfluid velocity $\mathbb{V}_q$ reaches Landau's critical velocity  $\mathbb{V}_{Lq}$, which for BCS condensates is given by~\cite{bardeen1962}
\beqy
\mathbb{V}_{Lq} = 
\frac{\Delta_q^{(0)}}{\hbar k_{Fq}}\, .
\eeqy

Beyond this point, the pairing gap decreases with increasing effective superfluid velocity and vanishes for $\mathbb{V}_q= \mathbb{V}_{cq}^{(0)}$. We find that this behavior is well reproduced by the following interpolating formula:
% start a new page without indent 4.6cm
%\clearpage
%\end{paracol}
%\nointerlineskip
\beqy\label{eq:GapInterpolation}
\frac{\Delta_q(T=0,\mathbb{V}_{Lq}\leq\mathbb{V}_q \leq \mathbb{V}_{cq}^{(0)})}{\Delta_q^{(0)}}= 0.5081\sqrt{1-\frac{\mathbb{V}_q}{\mathbb{V}_{cq}^{(0)}}}\left(3.312\frac{\mathbb{V}_q}{\mathbb{V}_{cq}^{(0)}}-3.811\sqrt{\frac{ \mathbb{V}_{cq}^{(0)}}{\mathbb{V}_q}} +5.842\right)\, .
\eeqy
%\begin{paracol}{2}
%\linenumbers
%\switchcolumn

The maximum relative error does not exceed 0.13$\%$. 

\begin{figure}[H]

\centerline{\includegraphics[width=8cm]{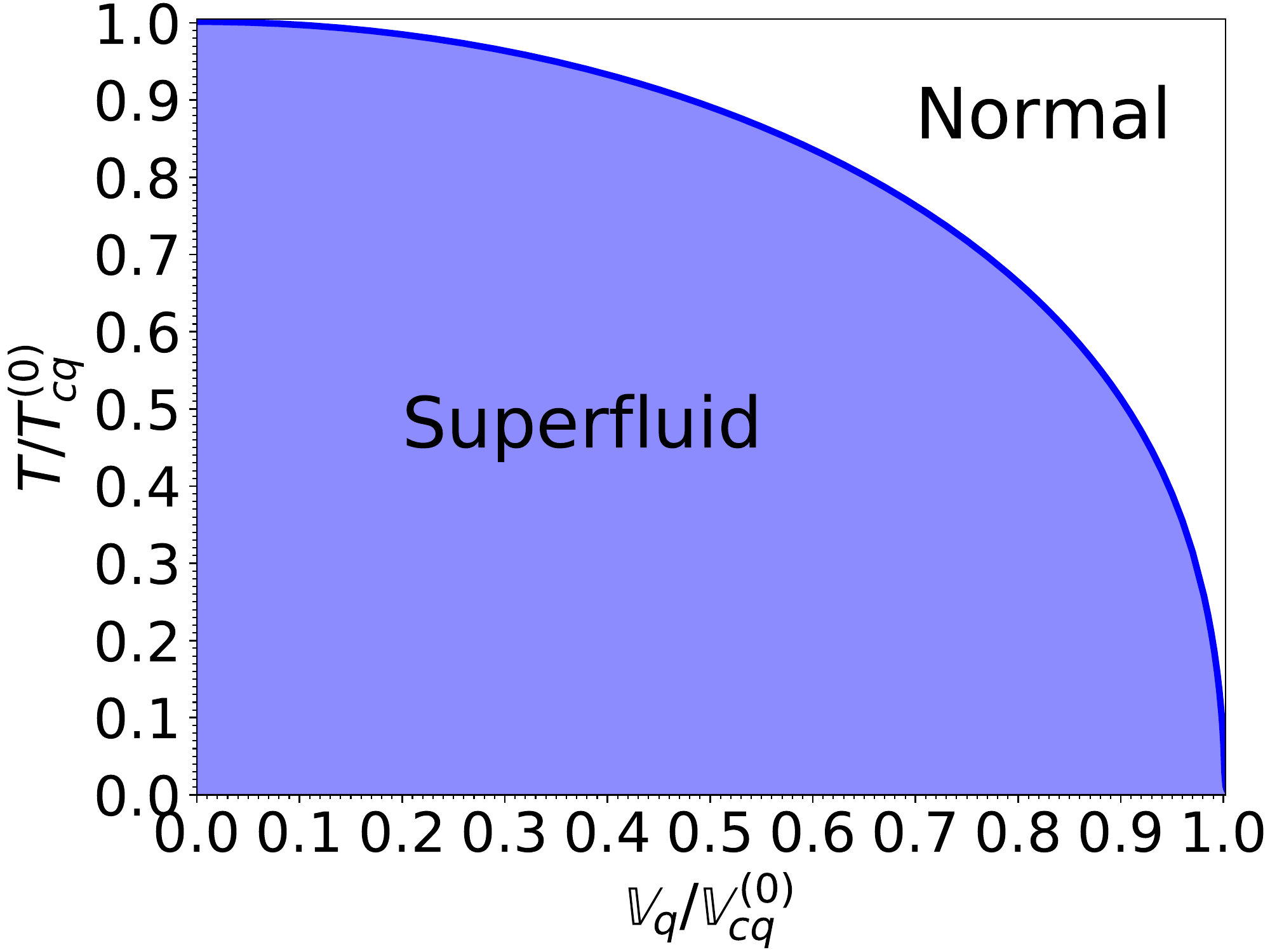}}
\caption{Phase diagram for the nucleon species $q$ in terms of the normalized temperature $T/T_{cq}^{(0)}$ and the normalized effective superfluid velocity $\mathbb{V}_q/\mathbb{V}_{cq}^{(0)}$. 
}
\label{fig:TVGraphBSk24ncc}
\end{figure}

The critical temperature and critical effective superfluid velocity delimiting the superfluid and normal phases, plotted in Figure~\ref{fig:TVGraphBSk24ncc} 
is well fitted by the following expression: 

\beqy\label{eq:InterpolationCriticalQuantities}\displaystyle
\frac{T_{cq}}{T_{cq}^{(0)}}(\mathbb{V}_q\leq \mathbb{V}_{cq}^{(0)})\simeq \left[1-\left(\frac{\mathbb{V}_{q}}{\mathbb{V}_{cq}^{(0)}}\right)^{2}\right]^{2/5}\, .
\eeqy
This interpolation is valid for both neutrons and protons. The errors are contained within the thickness of the lines in Figure~\ref{fig:TVGraphBSk24ncc}. 

The universality observed in the superfluid properties of both neutrons and protons (after a suitable choice of normalizations) is the consequence of the weak-coupling regime, as discussed in Section~\ref{sec:LandauApprox}. Indeed, as shown in Equation~\eqref{eq:LandauGapUniversal}, the normalized pairing gaps $\breve{\Delta}_q/\breve{\Delta}_q^{(0)}$ are independent of the pairing strength $v^{\pi q}$ (hence also of the associated pairing cutoff $\Cutoff$) and depend only on the rescaled temperature $T/T_{cq}^{(0)}$ and effective superfluid velocity $\mathbb{V}_q/\mathbb{V}_{cq}^{(0)}$. Estimating the exact pairing gaps $\Delta^{(0)}_q$ from Equation~\eqref{eq:ChamelApprox} and substituting in Equation~\eqref{eq:LandauGapUniversal} lead to a very good approximation for the exact pairing gaps $\Delta_q/\Delta^{(0)}_q$ at finite temperatures and arbitrary effective superfluid velocities. The largest absolute deviations are found at the crust-core interface: they are of order of $10^{-3}$ for neutrons and lie within the numerical errors for protons.

\subsection{Reduced Chemical Potentials}
\label{sec:reduced-chemical}

\begin{figure}[H]

\centerline{\includegraphics[width=8cm]{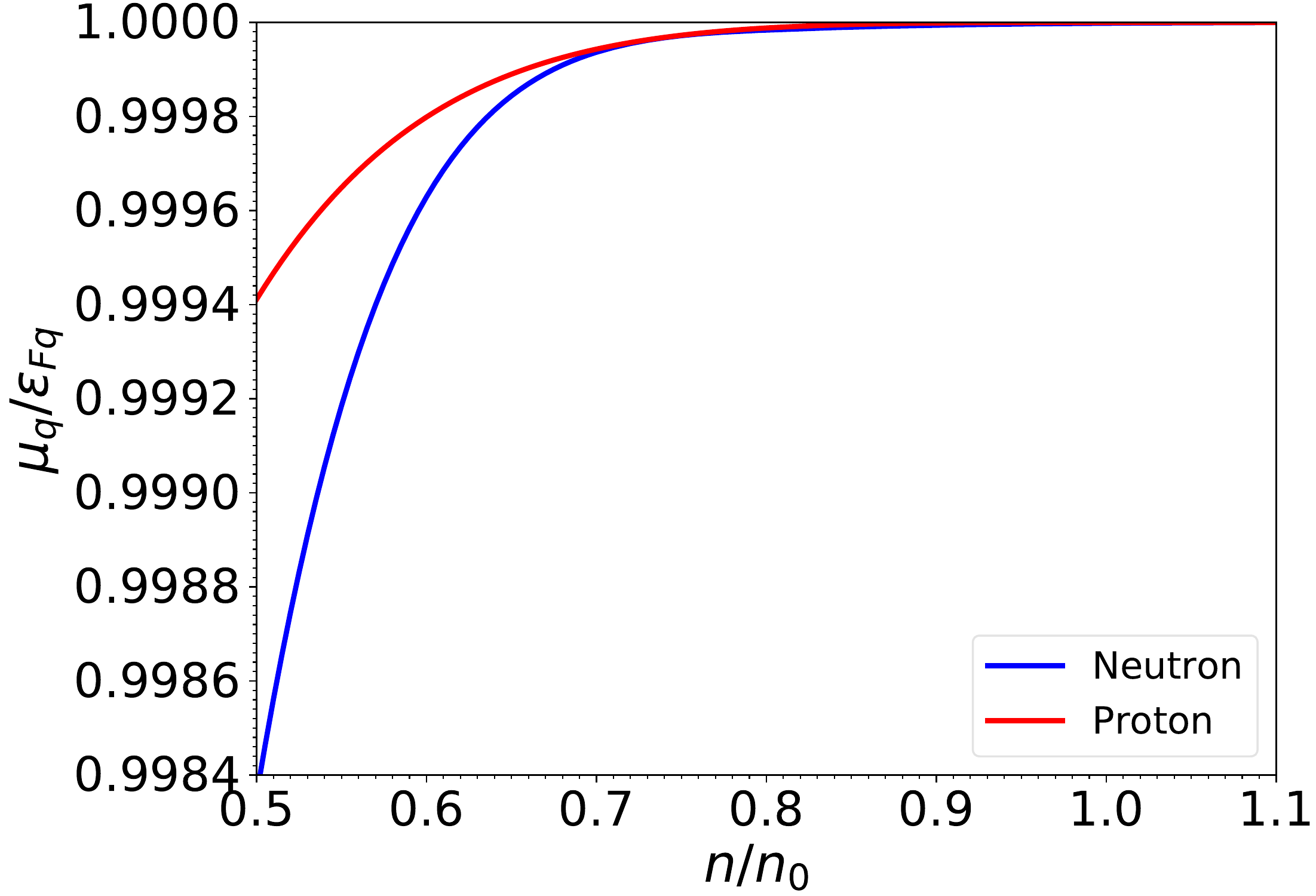}}
\caption{Reduced chemical potentials $\mu_q$ relative to the corresponding Fermi energy for $npe\mu$ matter in beta-equilibrium at baryon densities  prevailing in neutron-star cores in units of the saturation density $n_0$. 
Results obtained at $T=0$ and for $\mathbb{V}_q=0$. }
\label{fig:MuQ_TV=0_BSk24}
\end{figure}

The TDHFB theory allows the determination of the chemical potentials consistently with the pairing gaps. 
Let us recall that in Landau's theory adopted in previous studies~\cite{gusakov2005,leinson2017,leinson2018}, 
the reduced chemical potential $\mu_q$ was approximated by the corresponding Fermi energy $\Fermi$; effects 
induced by pairing, temperature, and currents were therefore ignored. To assess the precision of 
this approximation, we have computed $\mu_q$ numerically by solving simultaneously Equations~\eqref{eq:DimensionlessGapEquation} 
and \eqref{eq:DimensionlessChemicalPotentialEquation} varying the temperature and the neutron effective  superfluid velocity. 
The largest relative errors between $\mu_q$ and the Fermi energy $\Fermi$ we have found (at the crust-core interface) are 0.14$\%$ for neutrons and 0.052$\%$ for protons. Such errors have been obtained for low temperatures and small effective superfluid velocities for which pairing effects are the most important. Focusing on these conditions, 
we have plotted in Figure~\ref{fig:MuQ_TV=0_BSk24} the 
ratio $\mu_q/\Fermi$ as a function of density. As expected, the higher the density, the more precise are Landau's approximations.
To a large extent, the small deviations between $\mu_q$ and $\Fermi$ stem from the rather small pairing gaps predicted by the functional BSk24. Larger deviations cannot be excluded if another functional is adopted. In any case, let us recall that both Equations~\eqref{eq:DimensionlessGapEquation} and \eqref{eq:DimensionlessChemicalPotentialEquation} should be solved simultaneously to obtain fully consistent pairing gaps and chemical potentials. 

\subsection{Functions $\mathcal{Y}_q$}
\label{sec:Yq}

\begin{figure}[H]

\centerline{\includegraphics[width=7.5cm]{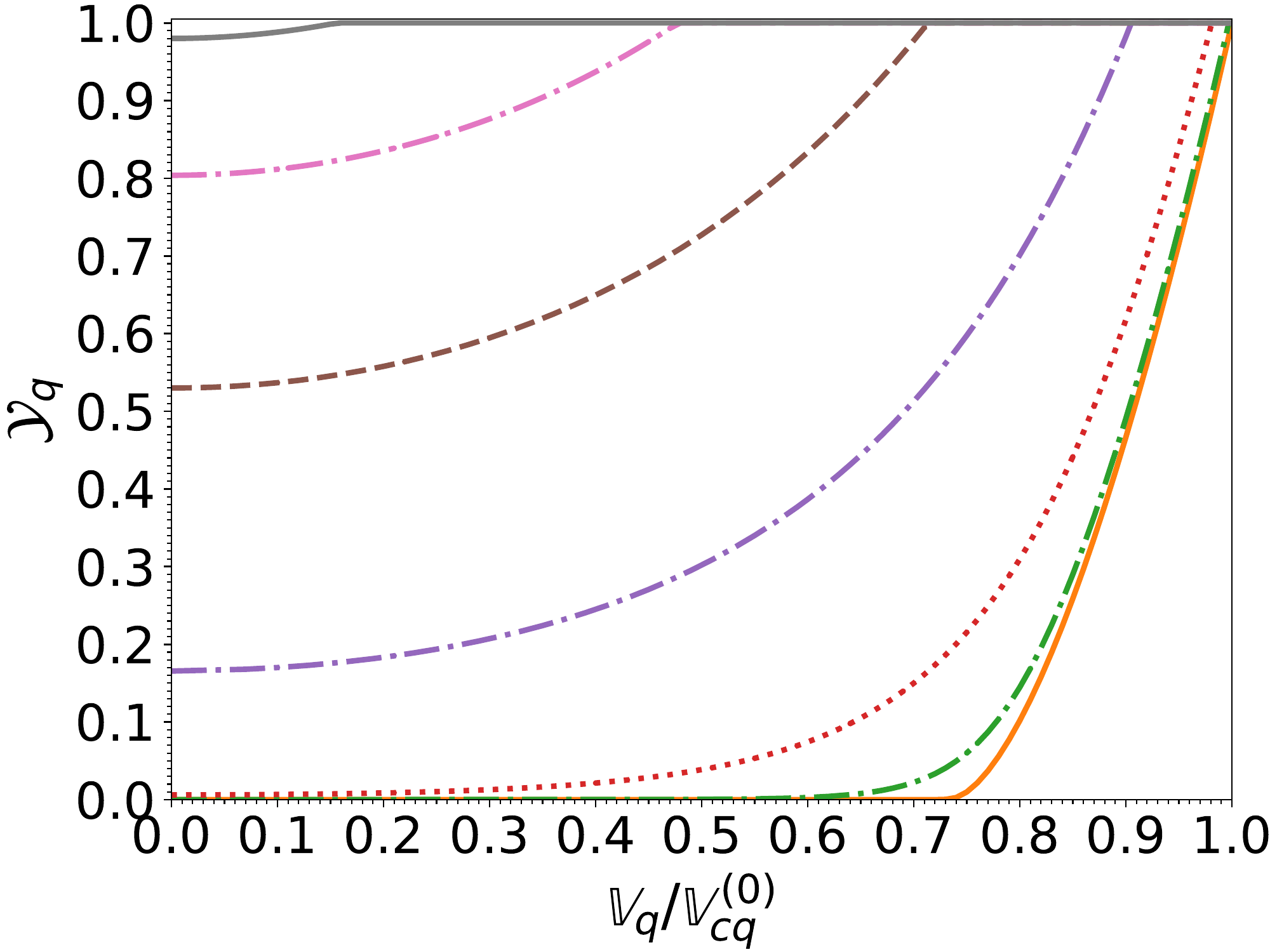}}
\caption{$\mathcal{Y}_q$ as a function of the normalized effective superfluid velocity $\mathbb{V}_q/\mathbb{V}_{cq}^{(0)}$ for different normalized temperatures $T/T_{cq}^{(0)}$. 
The legend of the curves is the same as in Figure~\ref{fig:PairingGapVitesseBSk24ncc}.}
\label{fig:YqVitesseBSk24ncc}
\end{figure}

\begin{figure}[H]

\centerline{\includegraphics[width=7.5cm]{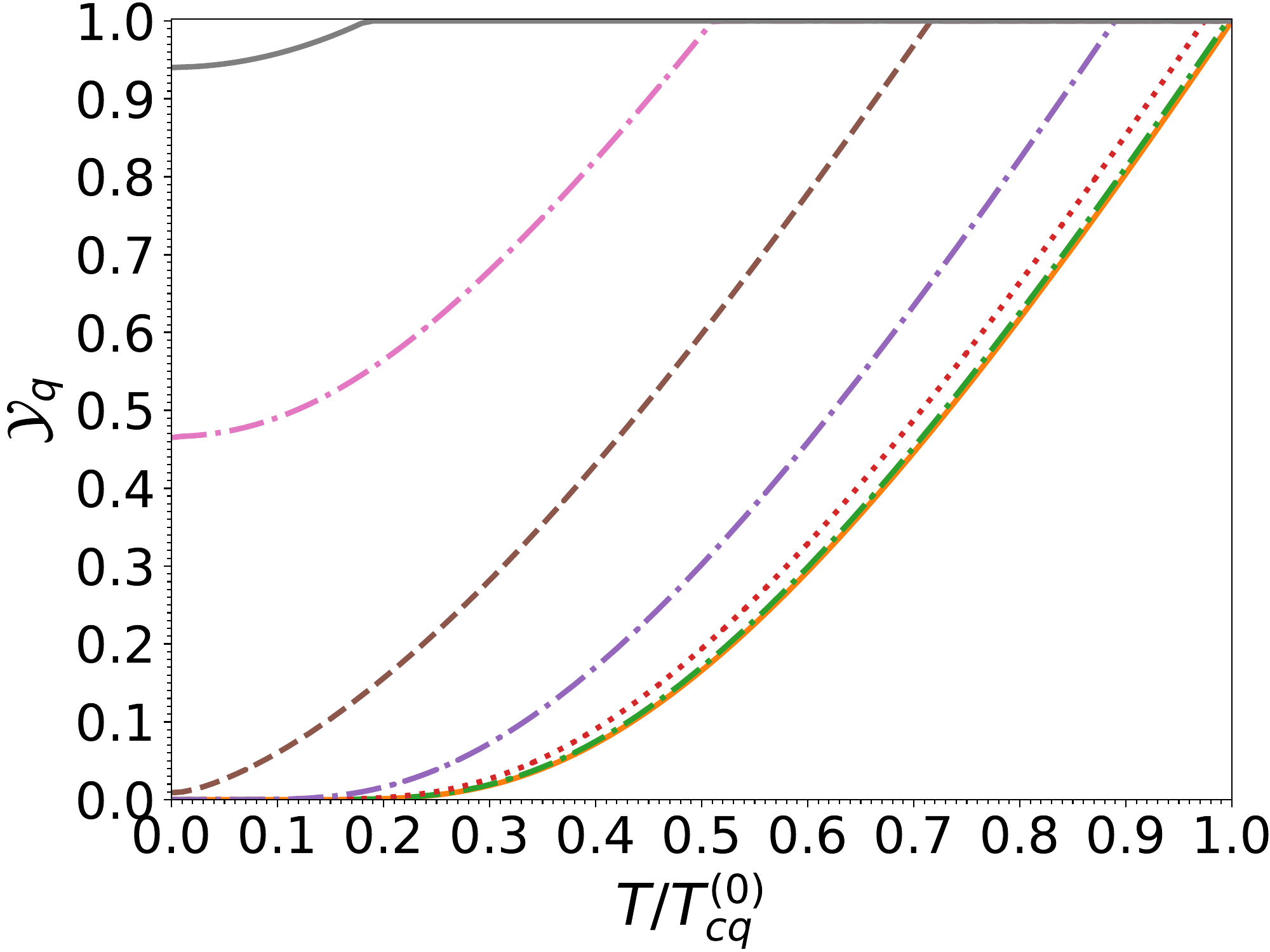}}
\caption{$\mathcal{Y}_q$ as a function of the normalized temperature $T/T_{cq}^{(0)}$ for different normalized effective superfluid velocities $\mathbb{V}_q/\mathbb{V}_{cq}^{(0)}$. 
The legend of the curves is the same as in Figure~\ref{fig:PairingGapTemperatureBSk24ncc}. }
\label{fig:YqTemperatureBSk24ncc}
\end{figure}

Having computed the pairing gaps $\Delta_q$ as well as the reduced chemical potentials $\mu_q$ at finite temperatures and for arbitrary effective superfluid velocities by solving\linebreak Equations~\eqref{eq:DimensionlessGapEquation} and~\eqref{eq:DimensionlessChemicalPotentialEquation}, we can now evaluate the functions $\mathcal{Y}_q$ from Equation~\eqref{eq:Yq}. When expressed in terms of the dimensionless ratios $T/T_{cq}^{(0)}$ and $\mathbb{V}_q/\mathbb{V}_{cq}^{(0)}$, results are found to be essentially independent of density and are summarized in Figures~\ref{fig:YqVitesseBSk24ncc} and \ref{fig:YqTemperatureBSk24ncc}. 

\vspace{-6pt}

The functions $\mathcal{Y}_q$ are well approximated by the functions $\Phi_q$ defined by Equation~\eqref{eq:LandauYq} where the pairing gaps $\breve{\Delta}_q$ are computed from Equation~\eqref{eq:LandauGapWeak} and provided $\breve{\Delta}^{(0)}_q$ are evaluated from Equation~\eqref{eq:ChamelApprox}. Since the deviations decrease with increasing density, we have focused on the crust-core interface. The absolute errors are found to be at most  of order $10^{-3}$ for $\mathcal{Y}_n$ and $10^{-4}$ for $\mathcal{Y}_p$. It follows from Equation~ \eqref{eq:LandauPhiUniversal} that the function $\mathcal{Y}_q$ is universal. 
In the absence of superflow $\mathbb{V}_q = 0$, the temperature dependence of the functions $\mathcal{Y}_q$ can be well fitted by the following expression~\cite{Gnedin1995} (errors not exceeding 2.6\%): 
\begin{align}\label{eq:GusakovPhiq}
\mathcal{Y}_q(T\leq T_{cq}^{(0)},\mathbb{V}_q=0)&\simeq\left[0.9443 + \sqrt{0.0557^2 + \left(0.1886\frac{\pi}{\text{e}^{\gamma}} \frac{\Delta_q(T)}{\Delta_q^{(0)}}\frac{T_{cq}^{(0)}}{T}\right)^2}\right]^{1/2}\nonumber\\
&\qquad\times\exp\left[1.753-\sqrt{1.753^2 + \left(\frac{\pi}{\text{e}^{\gamma}}\frac{\Delta_q(T)}{\Delta_q^{(0)}}\frac{T_{cq}^{(0)}}{T}\right)^2}\right] 
\end{align}
with $\Delta_q(T)$ computed using the interpolation~\eqref{eq:YavovlevGapInterpolation}.

\subsection{Effective versus True Superfluid Velocities}
\label{sec:velocities}

The results we have presented so far have been conveniently expressed in terms of the effective superfluid velocities $\pmb{\mathbb{V}_q}$, which are related to the original superfluid velocities $\pmb{V_q}$ by  Equations~\eqref{eq:EffectiveSuperfluidVelocity} and~\eqref{eq:Iq}. These relations are highly nontrivial, recalling that the coefficients $\mathcal{I}_{qq'}$, defined by~\eqref{eq:Inn-def}--\eqref{eq:Theta-def}, depend on $\mathbb{V}_q$ through the functions $\mathcal{Y}_q$. 

So far, we have treated the effective superfluid velocities as free parameters. In reality however, $\pmb{\mathbb{V}_n}$ and $\pmb{\mathbb{V}_p}$ are determined by the dynamics of the star, as pointed out in the previous analysis of entrainment effects in  \cite{leinson2018}. In particular, in the study of low-frequency oscillations, it is a very good approximation to assume that the electric current in the normal frame vanishes, as shown in the classical work of \cite{mendell1991}. Considering that leptons are co-moving with quasiparticle excitations, the previous condition reads $\pmb{v_p}=\pmb{0}$ (in the normal frame). It immediately follows from Equation~\eqref{eq:true-velocity} that $\pmb{\mathbb{V}_p}=\pmb{0}$. In the following, we will restrict to this case as in  \cite{leinson2018} since it is of most physical interest. Under such condition, the vectors $\pmb{V_n}$ and $\pmb{V_p}$ are aligned, and are given by 
% start a new page without indent 4.6cm
%\clearpage
%\end{paracol}
%\nointerlineskip
\beqy
\label{eq:VitesseNeutron}
\pmb{V_n}& = &\left[ 1-\dfrac{2}{\hbar^2}\left(\dfrac{\delta E^j_{\rm nuc}}{\delta X_0}+\dfrac{\delta E^j_{\rm nuc}}{\delta X_1}\right)\left(m_p^{\oplus}n_p + m_n^{\oplus}n_n\mathcal{Y}_n \right)+\dfrac{16}{\hbar^4}\dfrac{\delta E^j_{\rm nuc}}{\delta X_0}\dfrac{\delta E^j_{\rm nuc}}{\delta X_1}m_p^{\oplus}n_p m_n^{\oplus}n_n\mathcal{Y}_n \right] \nonumber \\ 
&&\times \Xi \dfrac{m_n^{\oplus}}{m}\pmb{\mathbb{V}_n}  ,
\eeqy
%\begin{paracol}{2}
%\linenumbers
%\switchcolumn
\beqy\label{eq:VitesseProton}
\pmb{V_p} = \left[\frac{2}{\hbar^2}\left(\frac{\delta E^j_{\rm nuc}}{\delta X_0}-\frac{\delta E^j_{\rm nuc}}{\delta X_1}\right)\left(1-\mathcal{Y}_n\right)m_n^{\oplus}n_n\right] \Xi \dfrac{m_p^{\oplus}}{m}\pmb{\mathbb{V}_n}  ,
\eeqy

\beqy
\Xi=\left[1-\dfrac{2}{\hbar^2}\left(\frac{\delta E^j_{\rm nuc}}{\delta X_0}+\frac{\delta E^j_{\rm nuc}}{\delta X_1}\right)\left( m_p^{\oplus}n_p + m_n^{\oplus}n_n\right)+\frac{16}{\hbar^4}\frac{\delta E^j_{\rm nuc}}{\delta X_0}\frac{\delta E^j_{\rm nuc}}{\delta X_1}m_p^{\oplus}n_p m_n^{\oplus}n_n\right]^{-1}.
\eeqy 

These superfluid velocities depend on the baryon density $n$, the temperature $T$ and the neutron effective superfluid velocity $\pmb{\mathbb{V}_n}$. Please note that under Landau's approximations, the norm of~\eqref{eq:VitesseNeutron} reduces to Equation (79) of~\cite{leinson2018} (these authors adopted the notation $\tilde{\Phi}_q$ for $\mathcal{Y}_q$, $m_n^{*}$ for the neutron effective mass $m_n^{\oplus}$, $\tilde{V}_n$ for the neutron effective superfluid velocity $\mathbb{V}_n$ and $v_n$ for the neutron true superfluid velocity $V_n$; the Landau parameters $F_1^{qq'}$ are given by Equation (100) of \cite{ChamelAllard2020}). 

\begin{figure}[H]
\raggedright
\centerline{\includegraphics[width=7cm]{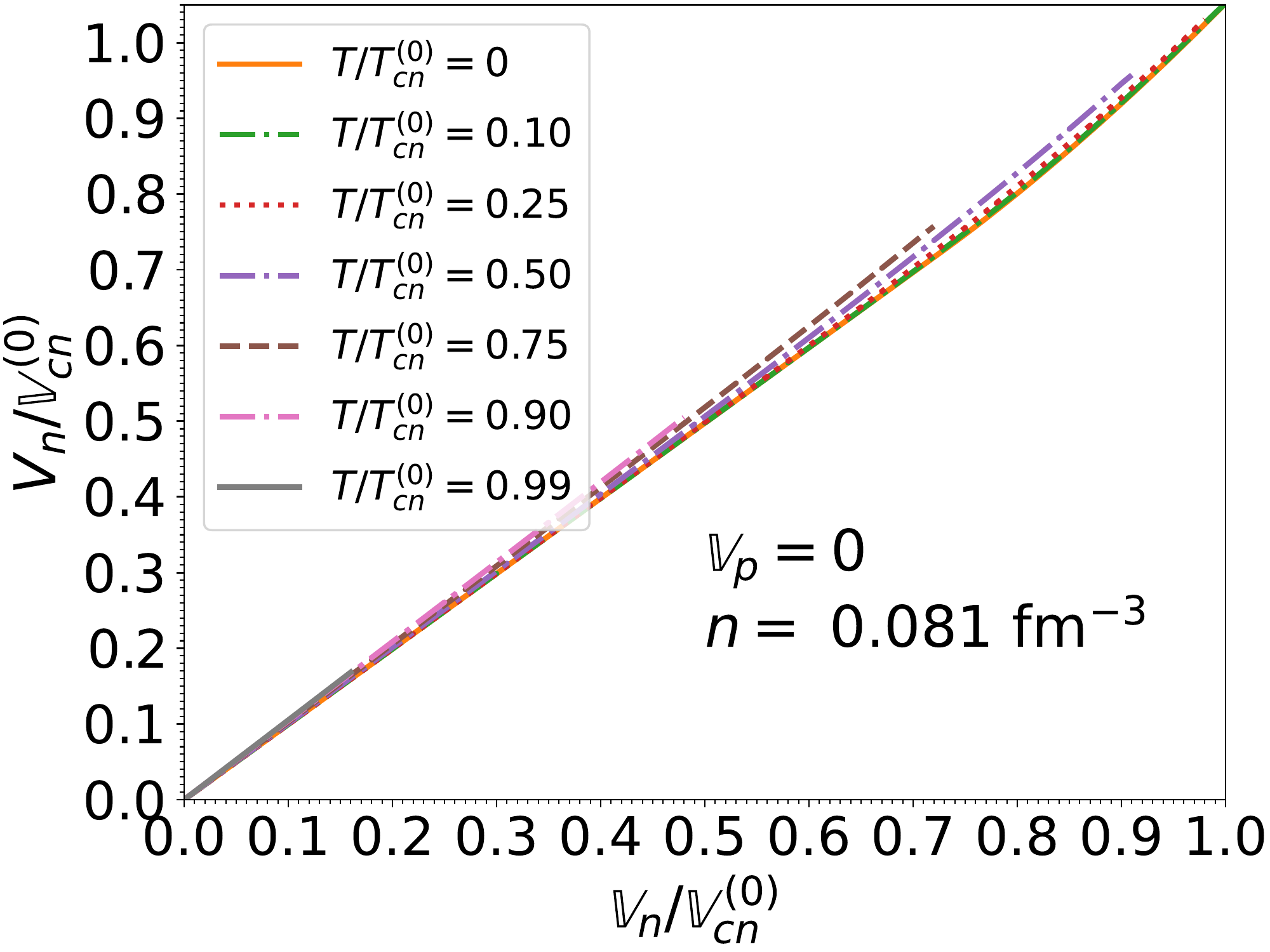}}
\centerline{\includegraphics[width=7cm]{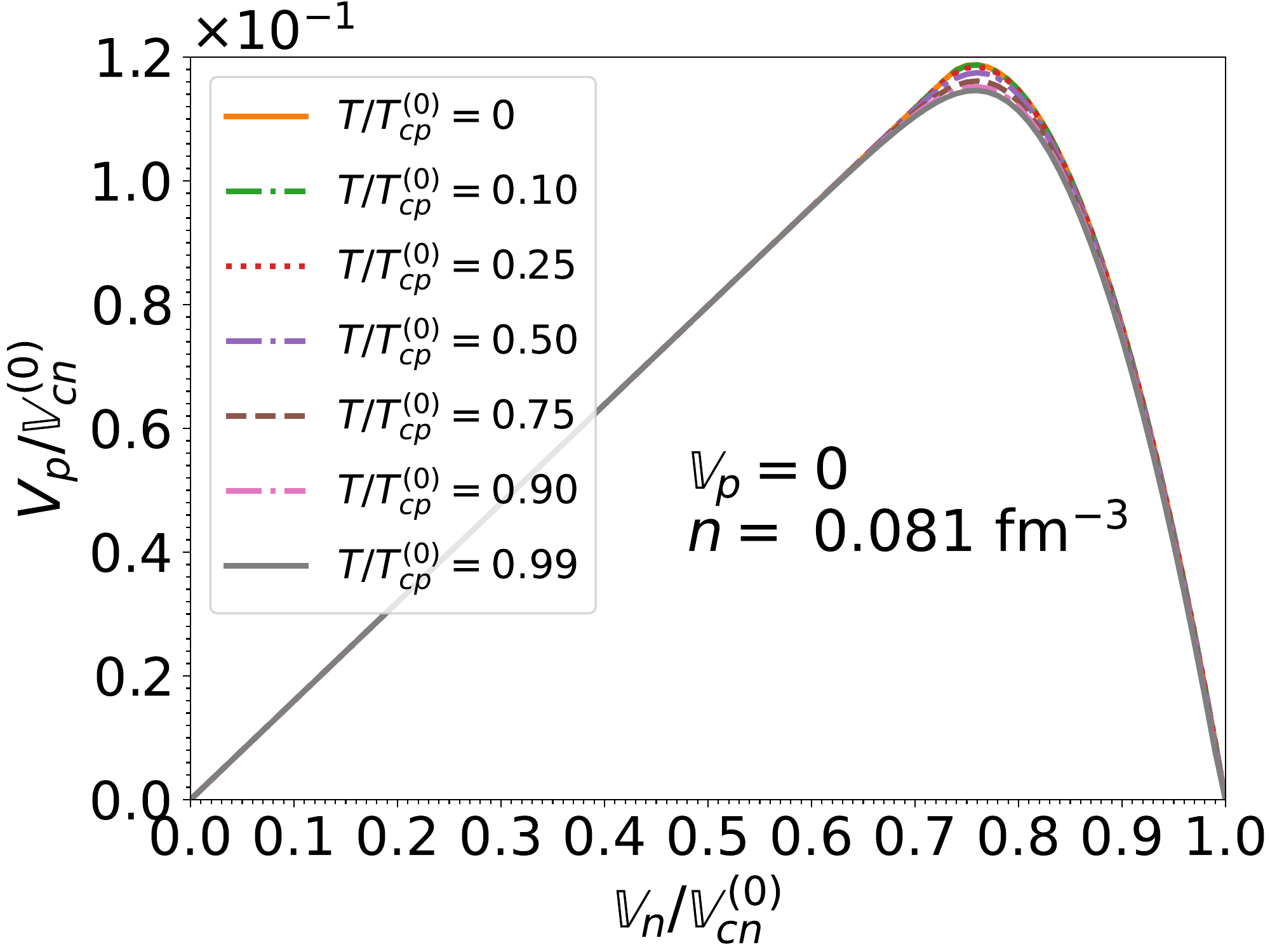}}

\caption{Upper panel: neutron superfluid velocity $V_n$ relative to the corresponding critical velocity as a function of the normalized effective neutron superfluid velocity $\mathbb{V}_n/\mathbb{V}_{cn}^{(0)}$ in $npe\mu$ matter in beta-equilibrium at the crust-core interface for different temperatures. Results were obtained for $\mathbb{V}_p=0$. Lower panel: same for the proton superfluid velocity $V_p$. 
}
\label{fig:ConversionBSk24ncc}
\end{figure}

\begin{figure}[H]
\raggedright
\centerline{\includegraphics[width=7cm]{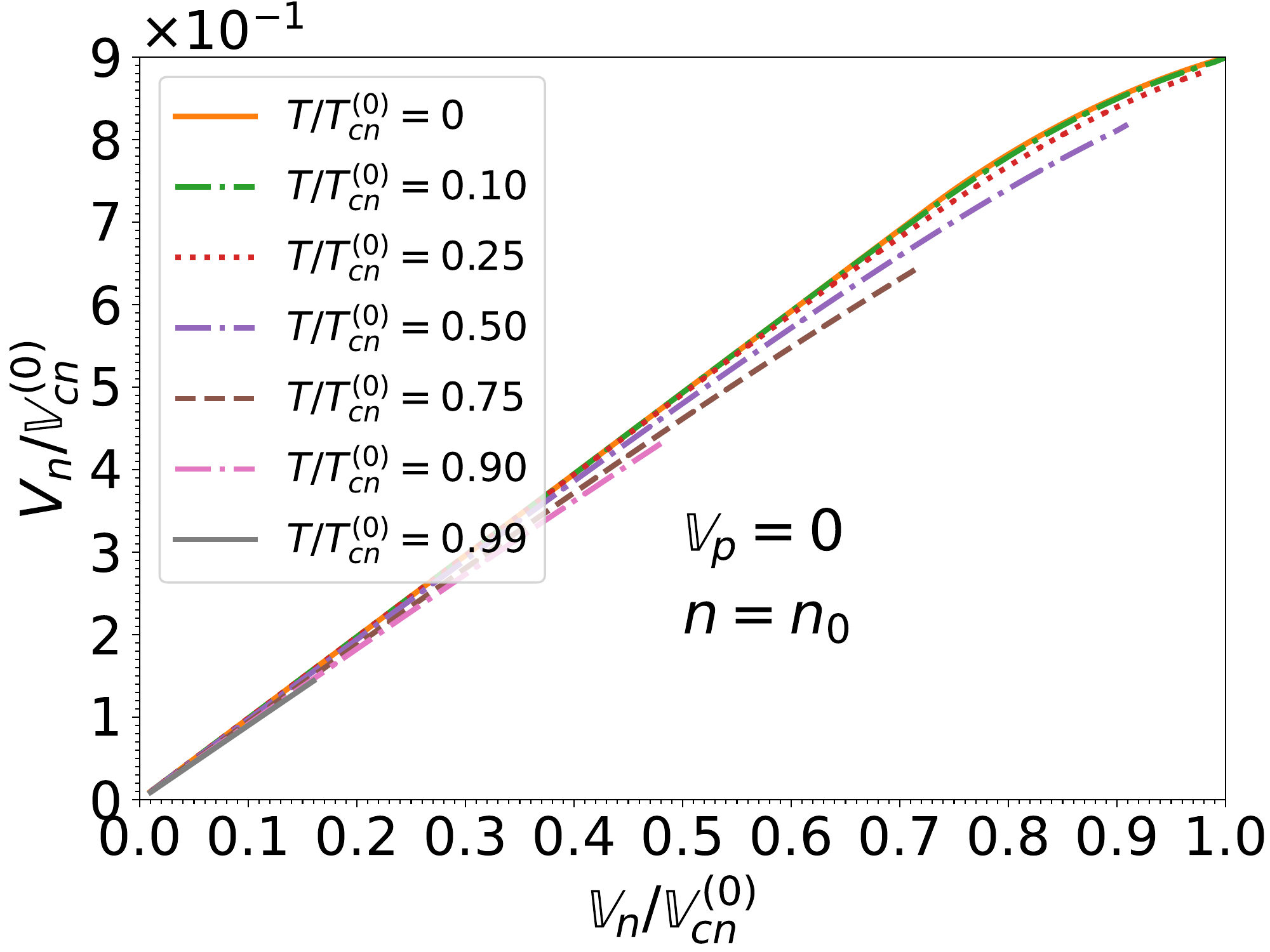}}
\centerline{\includegraphics[width=7cm]{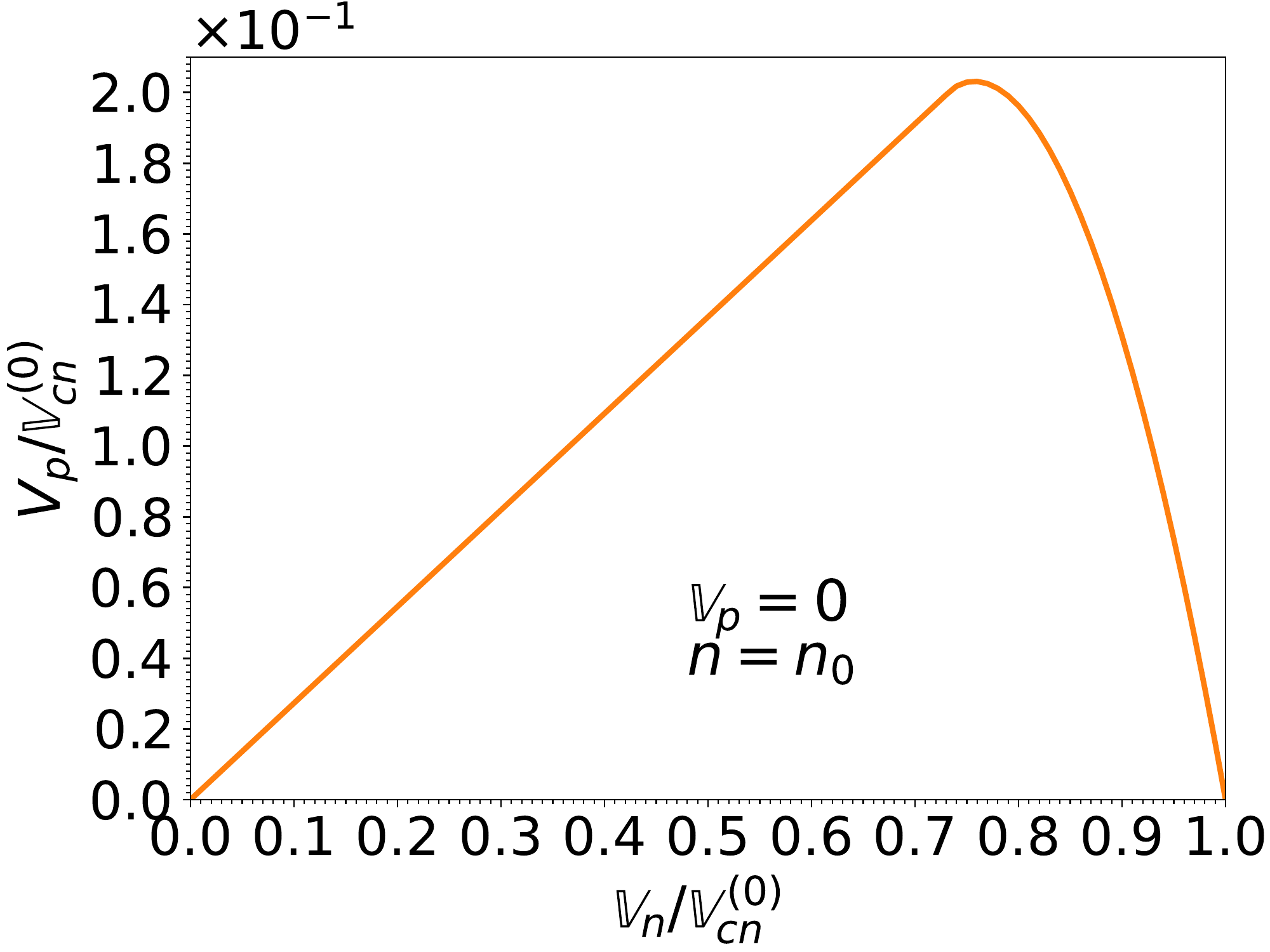}}

\caption{Same as Figure~\ref{fig:ConversionBSk24ncc} at saturation density $n_0$. Please note that for the proton superfluid velocity $V_p$, results for temperatures $T\leq T_{cp}^{(0)}$ are all contained within the thickness of the solid line.}
\label{fig:ConversionBSk24n0}
\end{figure}

Results for the norms, considering $npe\mu$ matter in beta-equilibrium, are displayed in Figures~\ref{fig:ConversionBSk24ncc} and~\ref{fig:ConversionBSk24n0} for two different densities. These superfluid velocities are only defined in the superfluid phase, for effective superfluid velocities and temperatures lower than their associated critical values given by~\eqref{eq:InterpolationCriticalQuantities}. Indeed, in the normal phase, the abnormal densities $\widetilde{n}_q$ vanish identically and the associated superfluid velocities are therefore ill defined. However, this has no physical implications since the superfluid velocities are irrelevant in this case. 

Although the neutron superfluid velocity is roughly equal to the effective superfluid velocity, $V_n\approx \mathbb{V}_n$, the proton superfluid velocity exhibits a more complicated behavior as a function of $\mathbb{V}_n$. From Equation~\eqref{eq:VitesseProton}, we have $V_p\propto \left(1-\mathcal{Y}_n\right)\mathbb{V}_n$. For sufficiently low neutron effective superfluid velocities, $\mathcal{Y}_n \approx 0$ therefore $V_p$ increases linearly with $\mathbb{V}_n$. However, $\mathcal{Y}_n$ increases with $\mathbb{V}_n$ leading to a decrease of $V_p$ (for $\mathbb{V}_n\simeq \mathbb{V}_{Ln}$), which vanishes when $\mathbb{V}_n=\mathbb{V}_{cn}^{(0)}$ corresponding to $\mathcal{Y}_n=1$.

% start a new page without indent 4.6cm

\vspace{-12pt}

% start a new page without indent 4.6cm
%\clearpage

\vspace{-18pt}

\subsection{Entrainment Matrix}
\label{sec:entrainment}

Having computed the pairing gaps, chemical potentials, and functions $\mathcal{Y}_q$ we can now determine the entrainment matrix from Equation~\eqref{eq:EntrainmentMatrix}. To better see the influence of temperature and superflows, results will be compared to those obtained at zero temperature and in the limit of small currents (conditions for which pairing can be ignored) using the following expression that we have previously calculated within the time-dependent Hartree–Fock (TDHF) theory~\cite{ChamelAllard2019}: 
\beqy
\rho_{np}^\mathrm{TDHF}=\rho_{pn}^\mathrm{TDHF}=-\frac{2}{\hbar^2}\rho_n \rho_p \left(\frac{\delta E_\mathrm{nuc}^{j}}{\delta X_0}-\frac{\delta E_\mathrm{nuc}^{j}}{\delta X_1}\right) \, ,
\eeqy
\beqy
\rho_{nn}^\mathrm{TDHF} = \rho_n - \rho_{np}^\mathrm{TDHF}\, ,
\eeqy
\beqy
\rho_{pp}^\mathrm{TDHF} = \rho_p - \rho_{pn}^\mathrm{TDHF}\, .
\eeqy

\begin{figure}[H]

\centerline{\includegraphics[width=7cm]{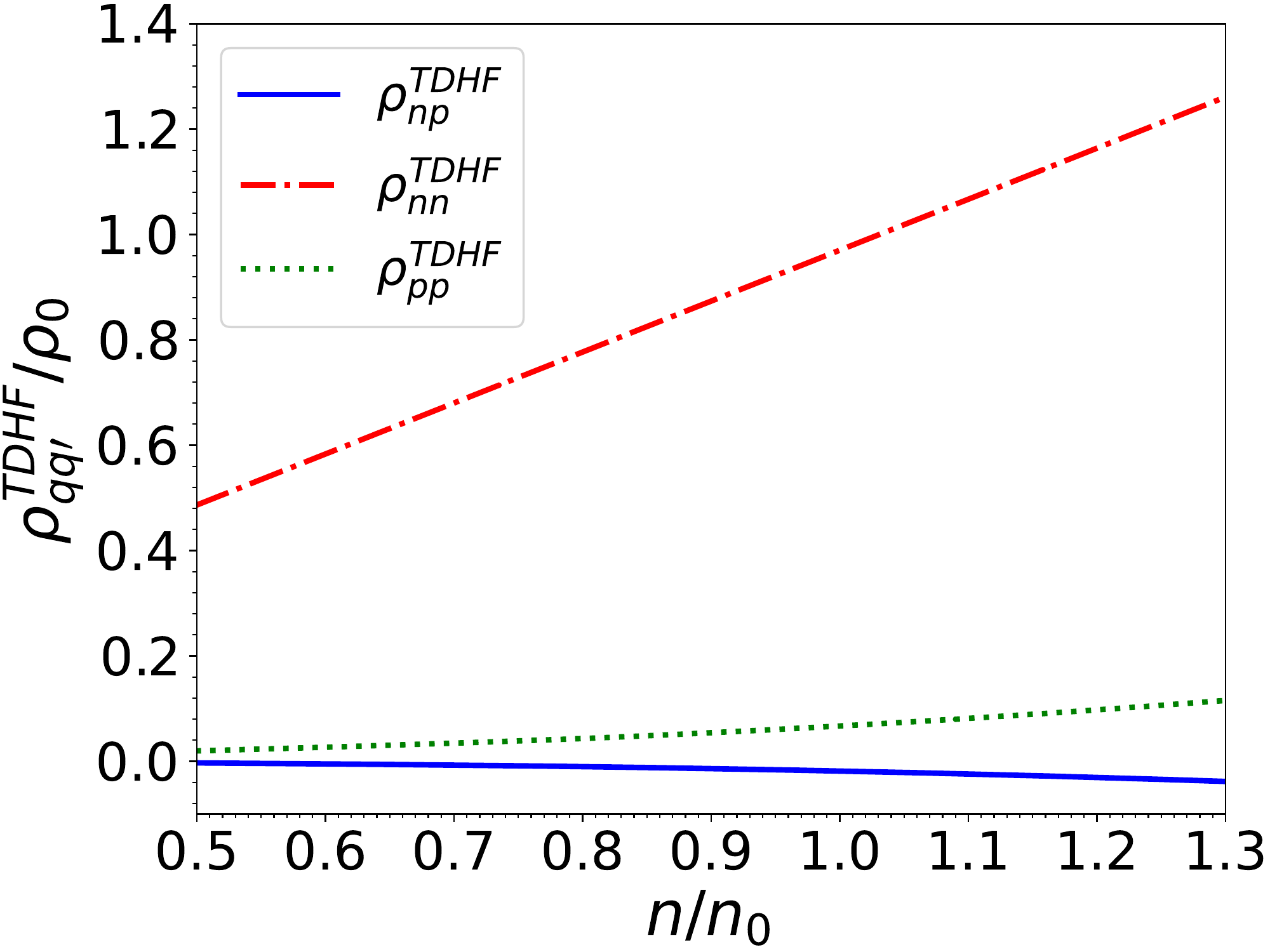}}
\caption{Entrainment matrix elements  (normalized by the saturation density $\rho_0=n_0 m$) for $npe\mu$ matter in beta-equilibrium at zero temperature and in the limit of small currents at baryon densities prevailing in neutron-star cores in units of the saturation density $n_0$. 
}
\label{fig:RhoTDHF}
\end{figure}

\begin{figure}[H]
\raggedright

\centerline{\includegraphics[width=7cm]{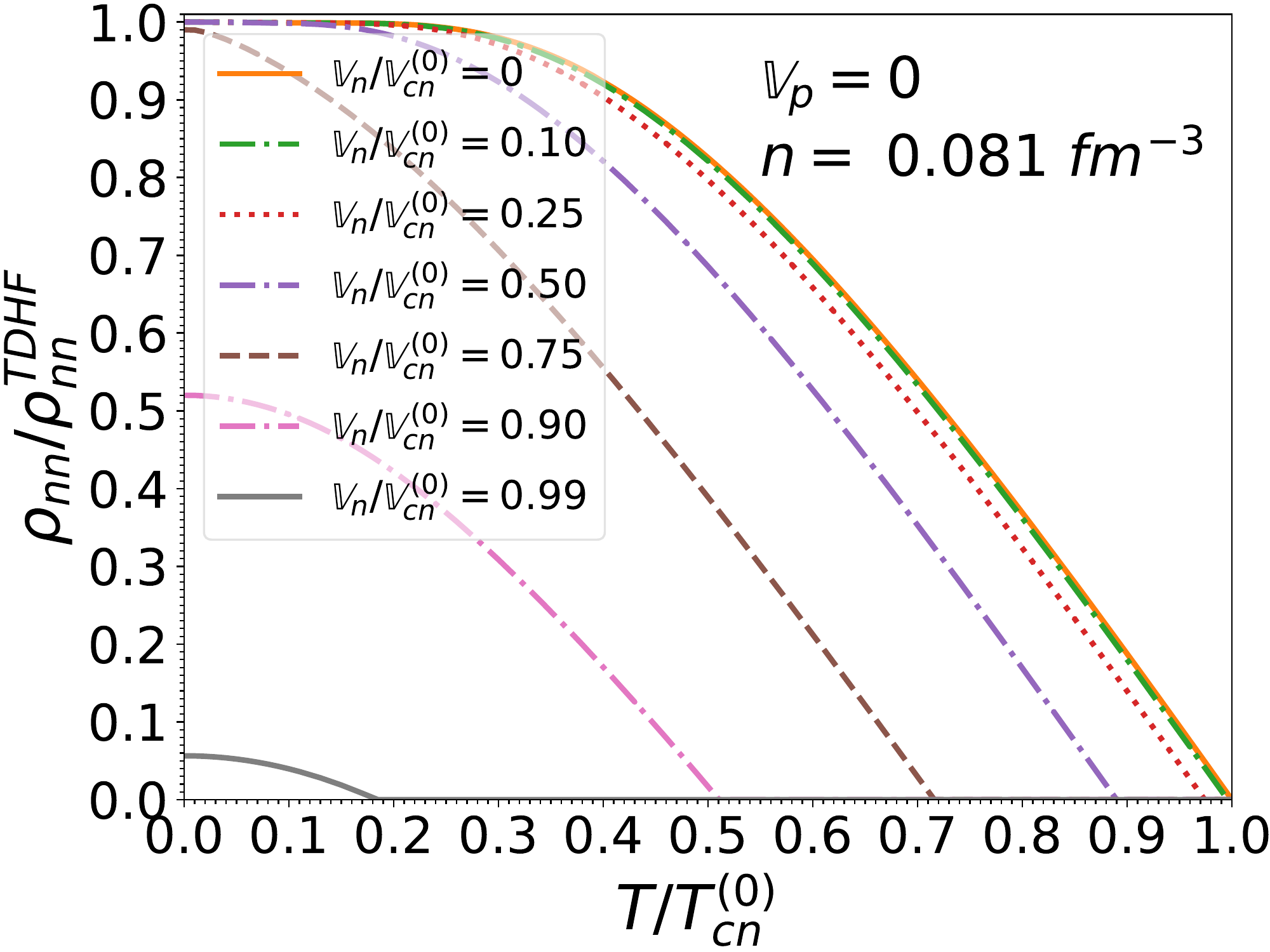}}
\centerline{\includegraphics[width=7cm]{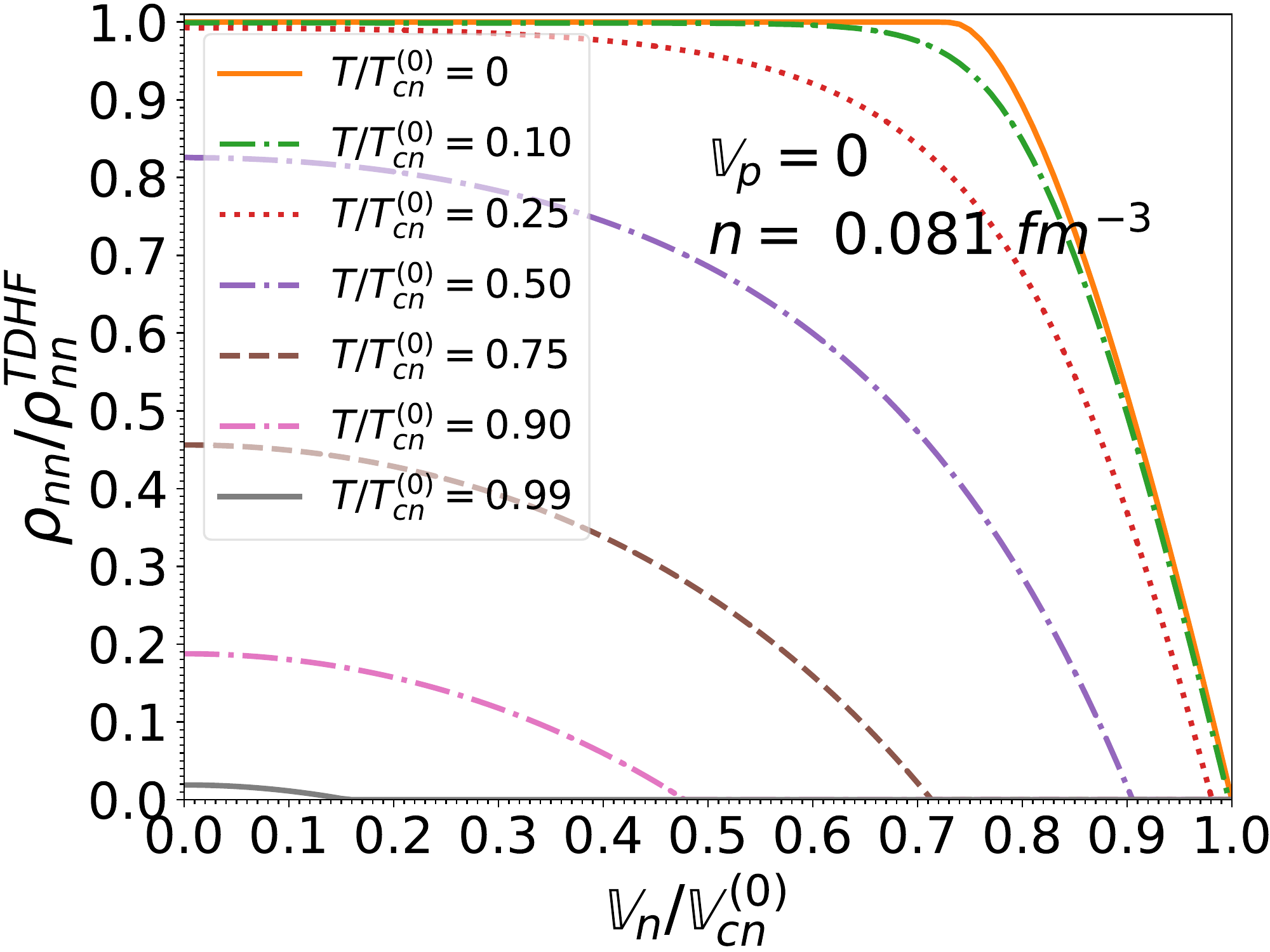}}

\caption{Dimensionless entrainment matrix element $\rho_{nn}/\rho_{nn}^{TDHF}$ as a function of the normalized effective superfluid velocity $\mathbb{V}_n/\mathbb{V}_{cn}^{(0)}$ and the normalized temperature $T/T_{cn}^{(0)}$ for $npe\mu$ matter in beta-equilibrium at the crust-core interface and for $\mathbb{V}_p =0$. }
\label{fig:RhoNNBSk24ncc}
\end{figure}

\begin{figure}[H]

\centerline{\includegraphics[width=7.5cm]{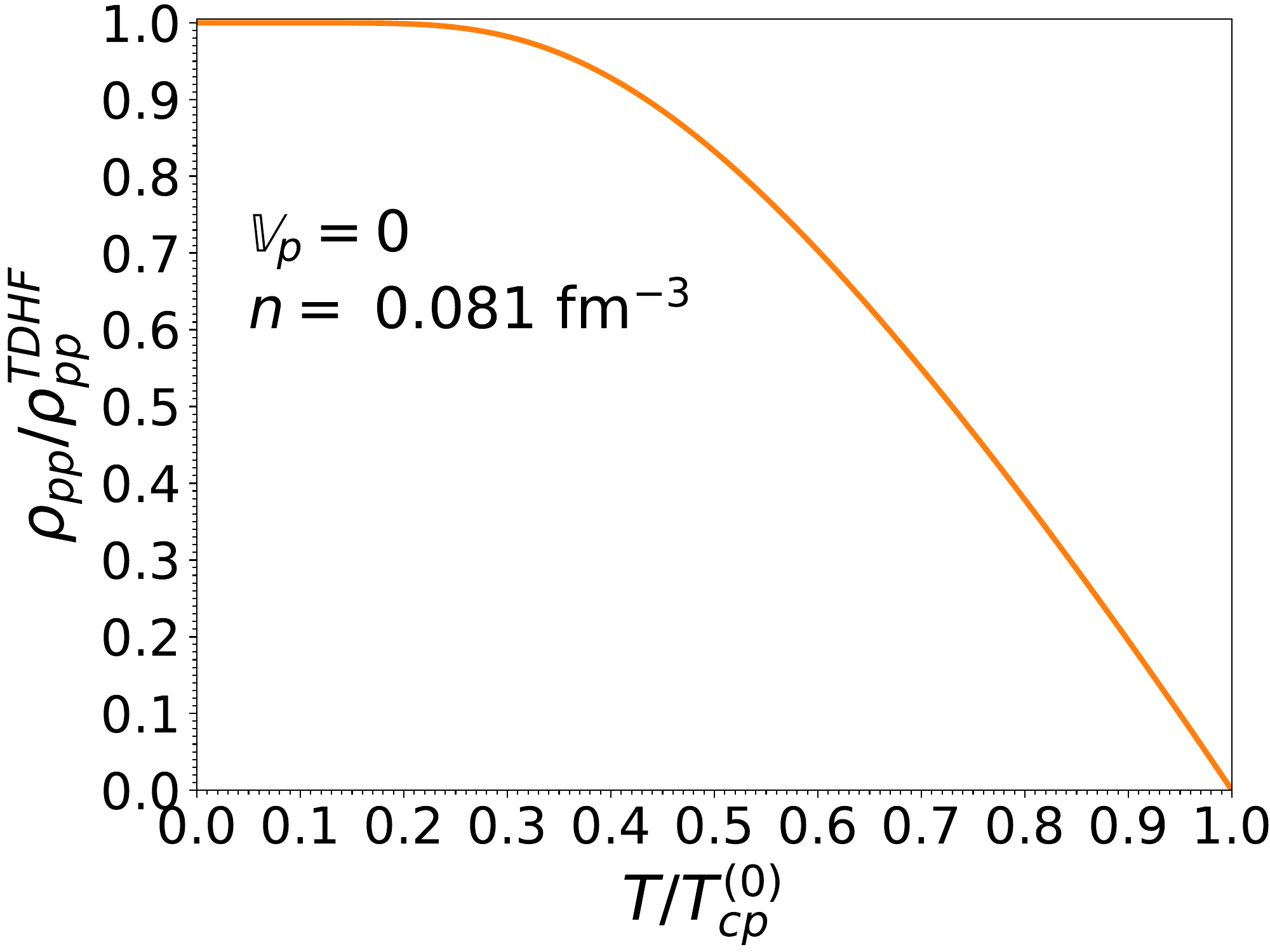}}
\caption{Dimensionless entrainment matrix element $\rho_{pp}/\rho_{pp}^{TDHF}$ as a function of the normalized temperature $T/T_{cp}^{(0)}$, for $npe\mu$ matter in beta-equilibrium at the crust-core interface and for $\mathbb{V}_p=0$. Results obtained for different neutron effective superfluid velocities are indistinguishable. 
}
\label{fig:RhoPPBSk24ncc}
\end{figure}

\begin{figure}[H]
\raggedright
\centerline{\includegraphics[width=7.7cm]{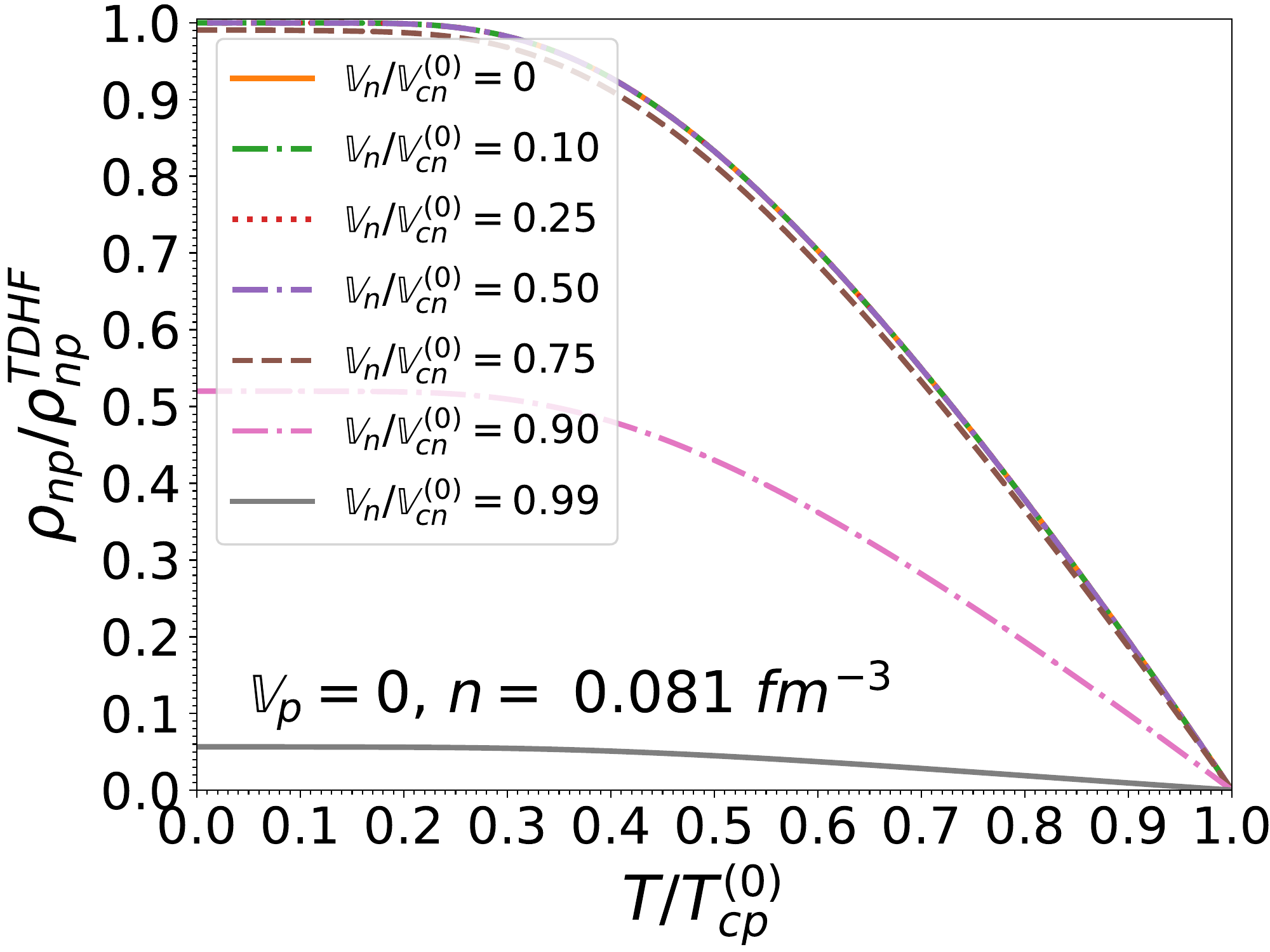}}
\centerline{\includegraphics[width=7.7cm]{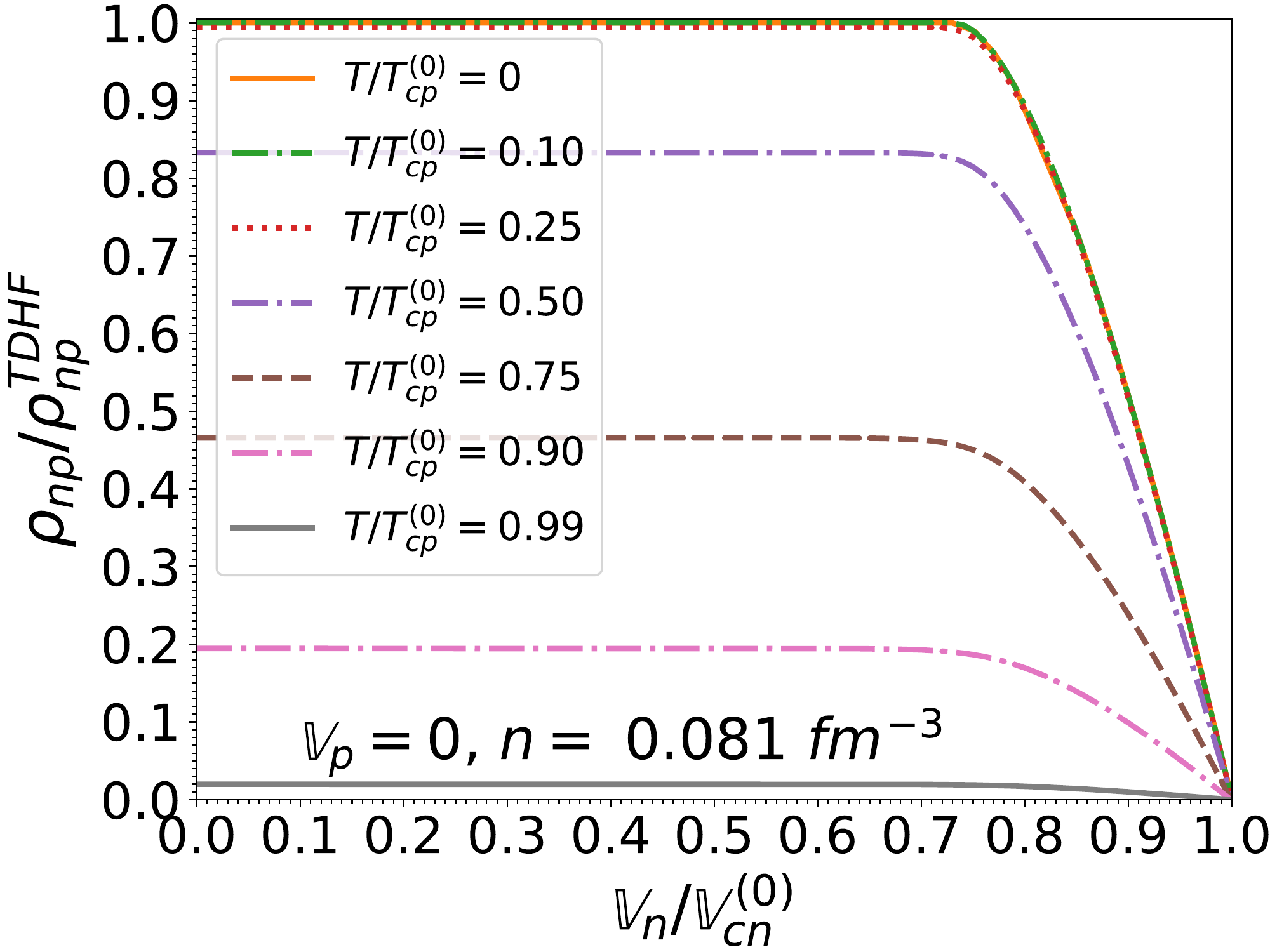}}

\caption{Dimensionless entrainment matrix element $\rho_{np}/\rho_{np}^{TDHF}$ as a function of the normalized effective superfluid velocity $\mathbb{V}_n/\mathbb{V}_{cn}^{(0)}$ and the normalized temperature $T/T_{cp}^{(0)}$, for $npe\mu$ matter in beta-equilibrium at the crust-core interface and for $\mathbb{V}_p=0$.}
\label{fig:RhoNPBSk24ncc}
\end{figure}

\begin{figure}[H]
\raggedright

\centerline{\includegraphics[width=7.7cm]{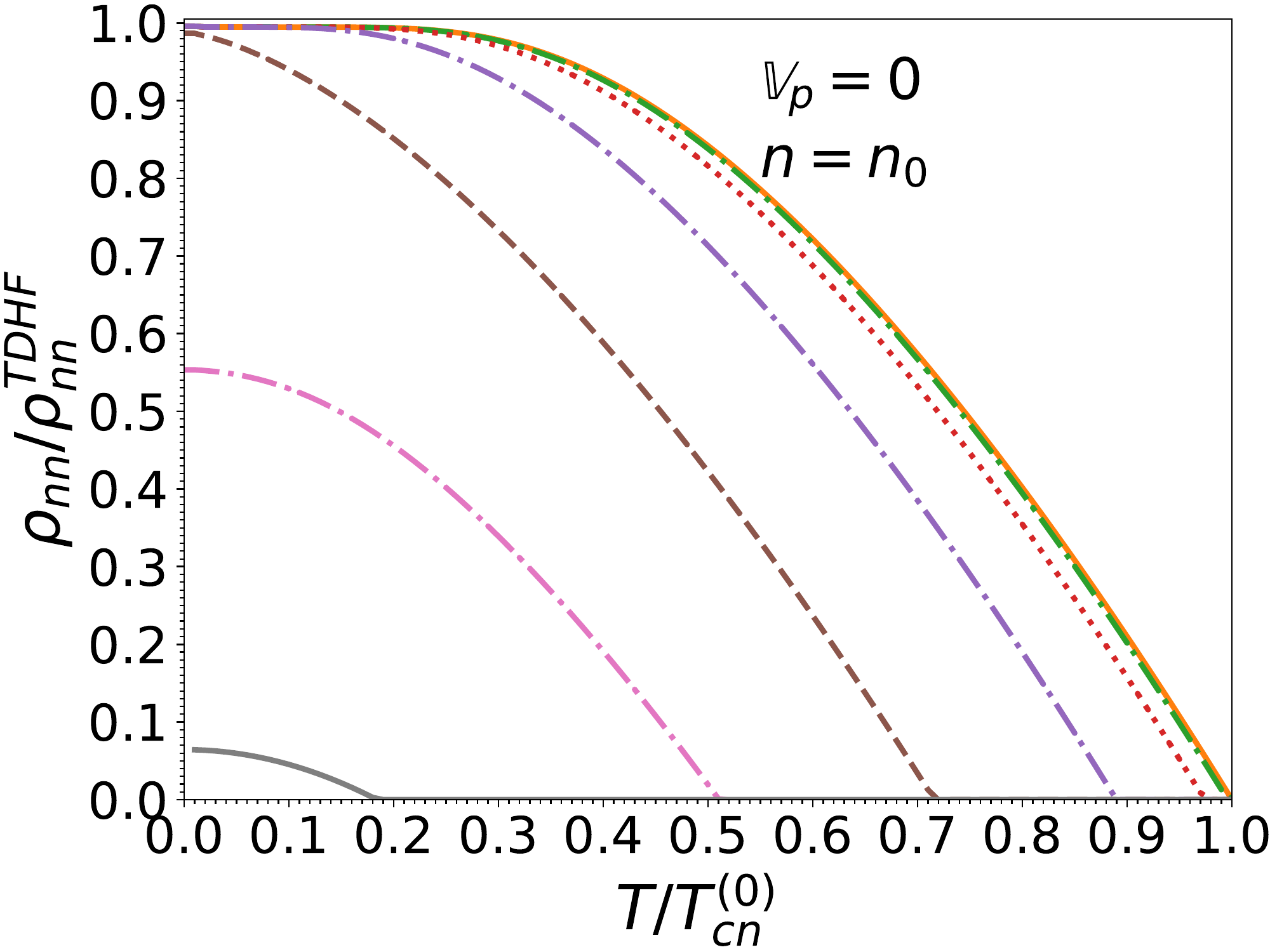}}
\centerline{\includegraphics[width=7.7cm]{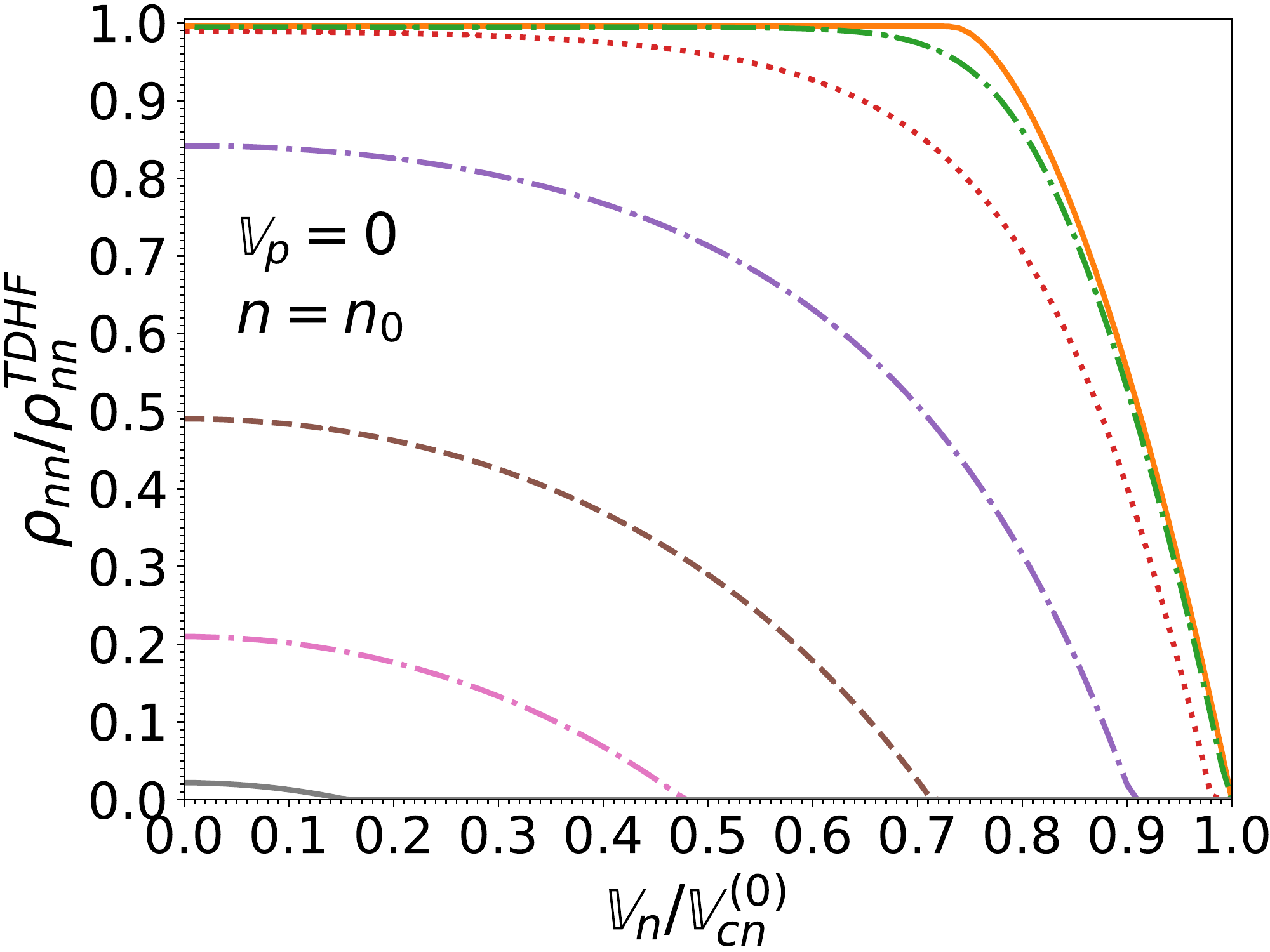}}

\caption{Same
 as Figure \ref{fig:RhoNNBSk24ncc} at the saturation density $n_0$ 
 using the same notation for the meaning of the different curves. 
 }
\label{fig:RhoNNBSk24n0}
\end{figure}

\begin{figure}[H]

\centerline{\includegraphics[width=7.5cm]{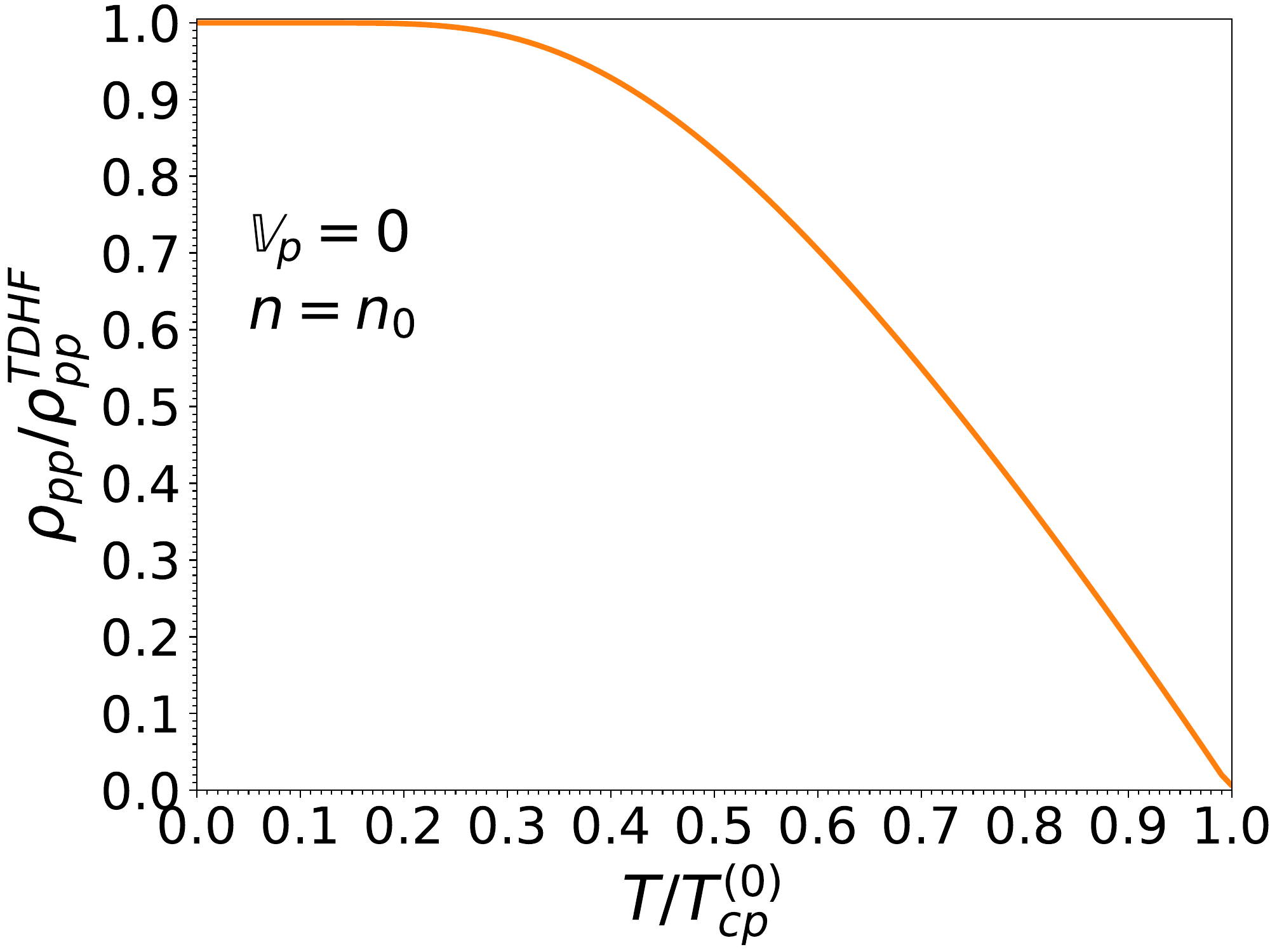}}
\caption{Same as Figure~\ref{fig:RhoPPBSk24ncc} at the saturation density $n_0$.}
\label{fig:RhoPPBSk24n0}
\end{figure}

\begin{figure}[H]
\raggedright

\centerline{\includegraphics[width=7.9cm]{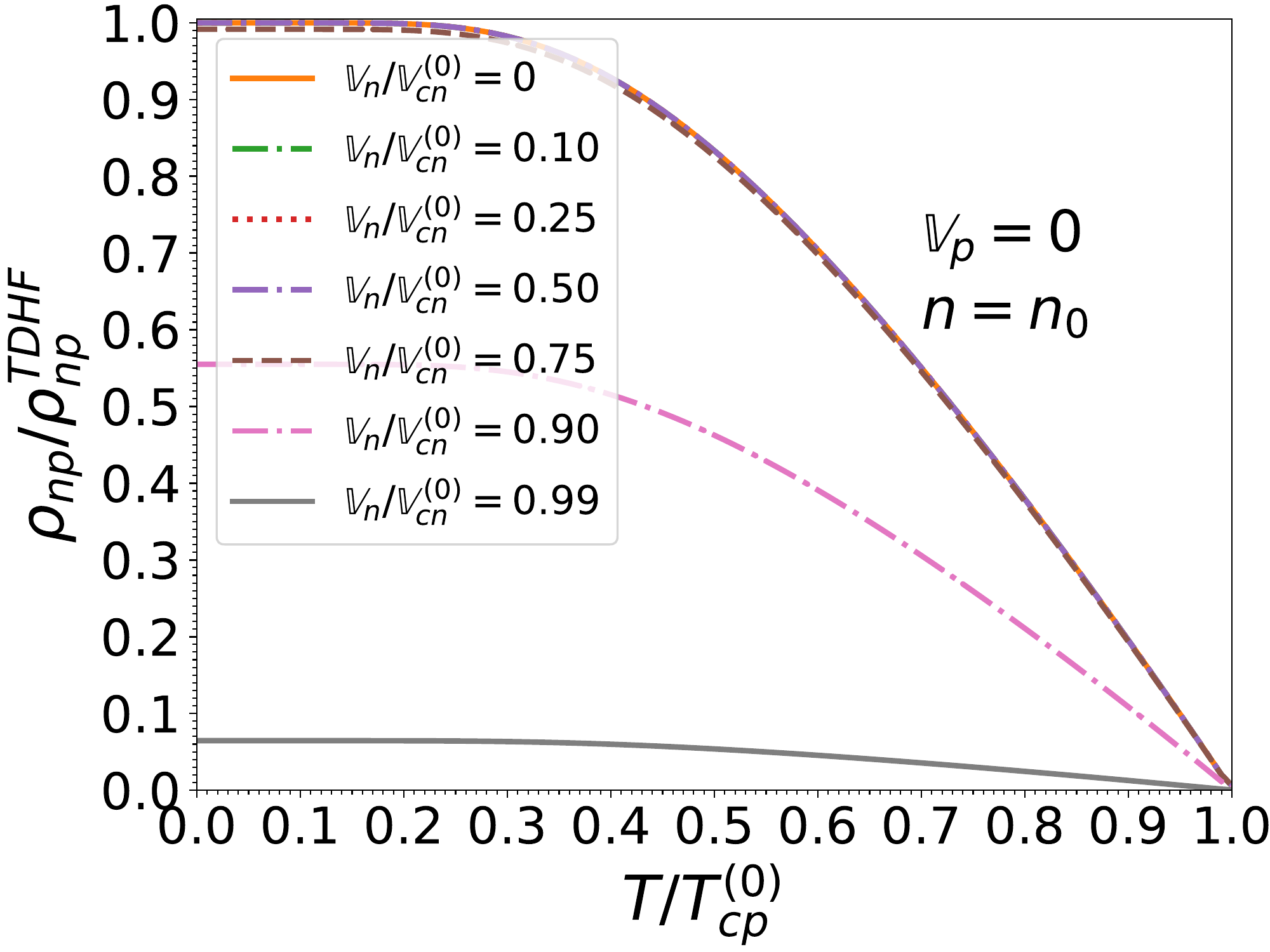}}
\centerline{\includegraphics[width=7.9cm]{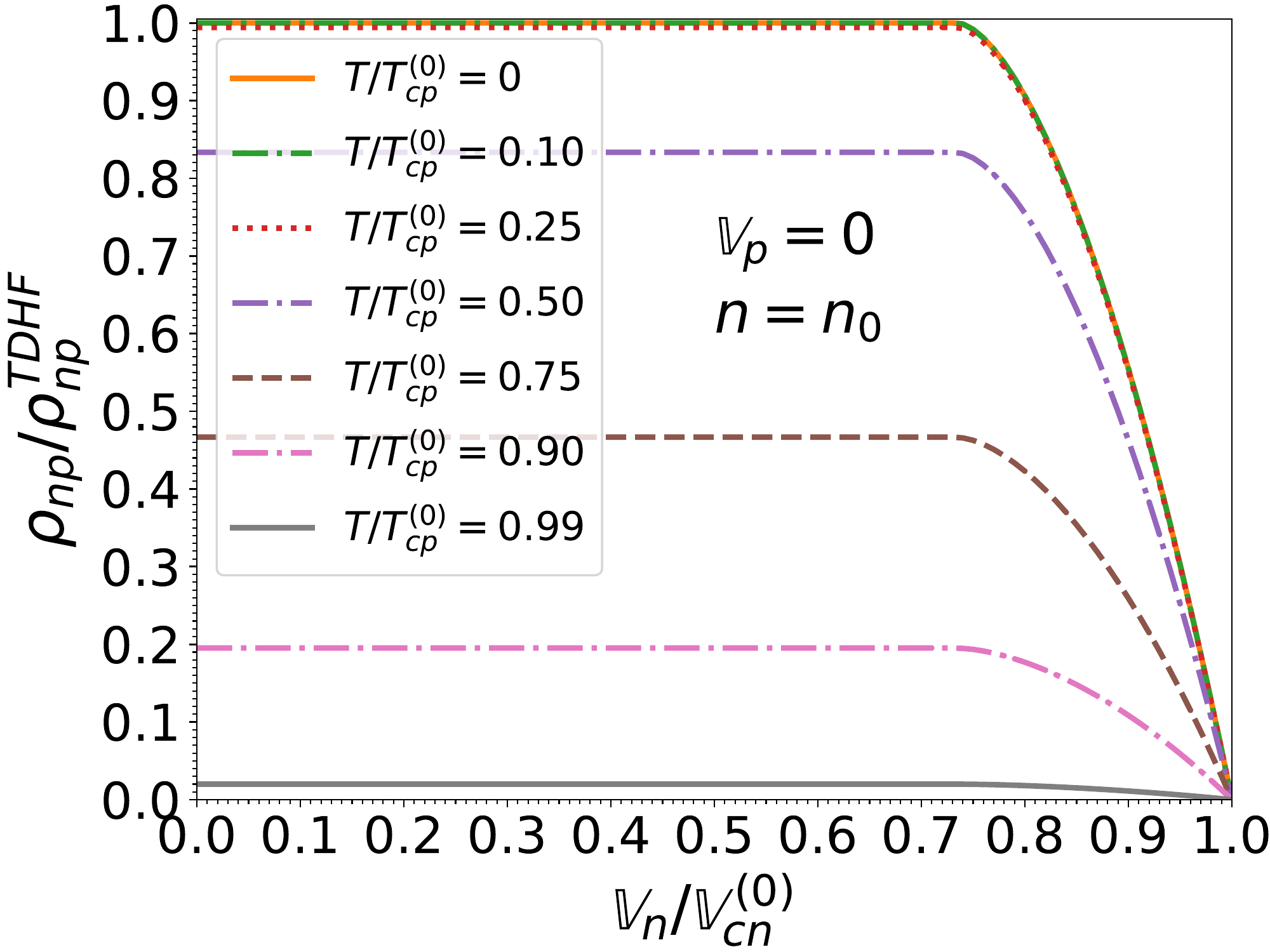}}

\caption{Same as Figure~\ref{fig:RhoNPBSk24ncc} at the saturation density $n_0$. }
\label{fig:RhoNPBSk24n0}
\end{figure}

These matrix elements are shown in Figure~\ref{fig:RhoTDHF} for $npe\mu$ matter in beta-equilibrium at densities relevant 
for the outer core of neutron stars. Results within the TDHFB theory for finite temperatures and different neutron effective superfluid velocities 
(recalling that we set $\mathbb{V}_p=0$ as discussed in Section~\ref{sec:velocities}) are plotted in Figures~\ref{fig:RhoNNBSk24ncc}--\ref{fig:RhoNPBSk24ncc} at the crust-core interface, 
and  in Figures~\ref{fig:RhoNNBSk24n0}--\ref{fig:RhoNPBSk24n0} at the saturation density.

Quite remarkably, the entrainment matrix at $T=0$ remains independent of the neutron effective superfluid velocity provided the latter does not exceed Landau's critical velocity. In other words, the expressions obtained in \cite{ChamelAllard2019}
in the limit of vanishing small effective superfluid velocities are actually exact for any effective superfluid velocity lying below Landau's critical value. This can be traced back to the vanishing of the function $\mathcal{Y}_q$ for $\mathbb{V}_q\leq \mathbb{V}_{Lq}$, as can be seen in Figure~\ref{fig:YqVitesseBSk24ncc}. This also entails that the entrainment matrix does not depend on the pairing gaps in this regime, thus justifying a posteriori our application of the TDHF theory~\cite{ChamelAllard2019} instead of TDHFB~\cite{ChamelAllard2020} since the gaps can thus artificially be set to zero. However, the TDHFB theory is still required for the determination of the actual value for $\mathbb{V}_{Lq}$. 

At finite but sufficiently low temperatures, the entrainment matrix remains weakly dependent on the neutron effective superfluid velocity provided $\mathbb{V}_q\leq \mathbb{V}_{Lq}$. For higher neutron effective superfluid velocities, the entrainment matrix elements $\rho_{nn}$ and $\rho_{np}=\rho_{pn}$ are reduced, even at $T=0$. The element $\rho_{pp}$ is essentially independent of $\mathbb{V}_n$. The influence of $T$ and $\mathbb{V}_n$ becomes increasingly important as these parameters approach their critical value. In particular, $\rho_{nn}$ and $\rho_{np}=\rho_{pn}$ both vanish when $\mathbb{V}_n=\mathbb{V}_{cn}^{(0)}$: neutron superfluidity is destroyed and the neutron mass is thus entirely transported by the normal fluid, as can be seen from Equation~\eqref{eq:NeutronMassCurrent}. On the other hand, protons remain superconducting but are no longer entrained by neutrons: the two species are dynamically uncoupled. The proton mass current~\eqref{eq:ProtonMassCurrent} thus reduces to the familiar expression for a single superfluid.  

\vspace{-6pt}

% start a new page without indent 4.6cm
%\clearpage

\vspace{-12pt}

\vspace{-6pt}
% start a new page without indent 4.6cm
%\clearpage

\vspace{-15pt}

% start a new page without indent 4.6cm

\vspace{-6pt}
% start a new page without indent 4.6cm
%\clearpage

\vspace{-18pt}

\subsection{Chemical Potentials}
\label{sec:chemical}

Knowing the relation between the effective superfluid velocities $\pmb{\mathbb{V}_q}$ and the superfluid velocities $\pmb{V_q}$, given by Equations~\eqref{eq:VitesseNeutron} and~\eqref{eq:VitesseProton} respectively, and using Equation~\eqref{eq:UqPotential} for the potentials $U_q$ together with Equation~\eqref{eq:MomentumDensityHomogeneous} for the momentum densities $\pmb{j_q}$ and Equation~\eqref{eq:KineticDensityHomogeneous} for the kinetic-energy densities $\tau_q$, we can compute the true chemical potentials $\lambda_q$ from Equations~\eqref{eq:LambdaN} and~\eqref{eq:LambdaP}. Results for $\mathbb{V}_p=0$ (as discussed in Section~\ref{sec:velocities}) are plotted in Figure~\ref{fig:LambdaBSk24ncc} for $n=n_\text{cc}$ and in Figure~\ref{fig:LambdaBSk24n0} for $n=n_0$ respectively. Although the chemical potentials $\lambda_q$ are found to be very weakly dependent on the temperature and on the neutron effective superfluid velocity (in the regime for which superfluidity exists), they deviate substantially from their corresponding Fermi energies $\Fermi$ due to the contribution from the potential $U_q$.

% start a new page without indent 4.6cm

\begin{figure}[H]
\raggedright
\centerline{\includegraphics[width=7cm]{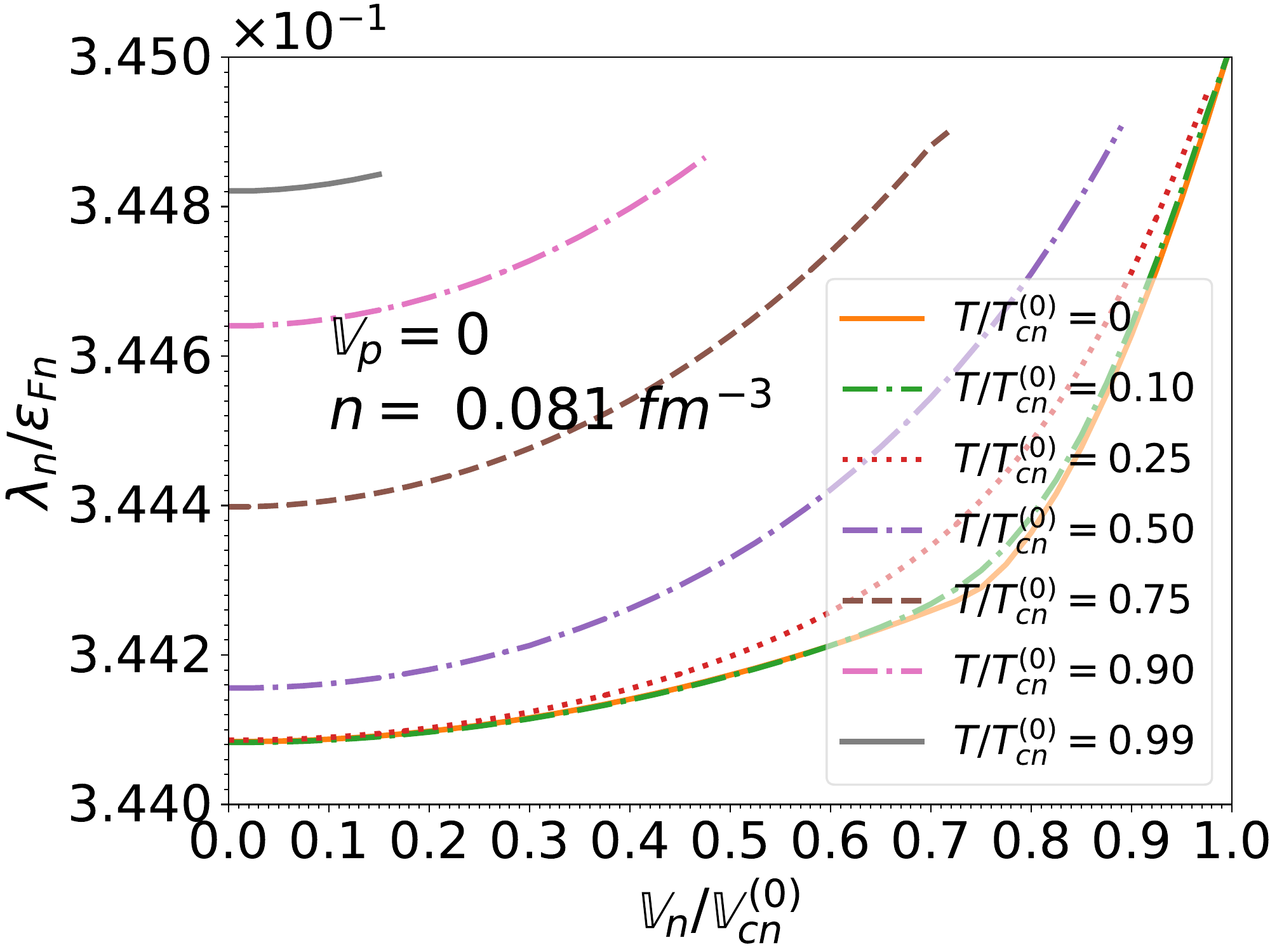}}

\caption{\textit{Cont}.}
\label{fig:petit3leaks3D}
\end{figure}

\begin{figure}[H]\ContinuedFloat
\widefigure
\raggedright

\centerline{\includegraphics[width=7cm]{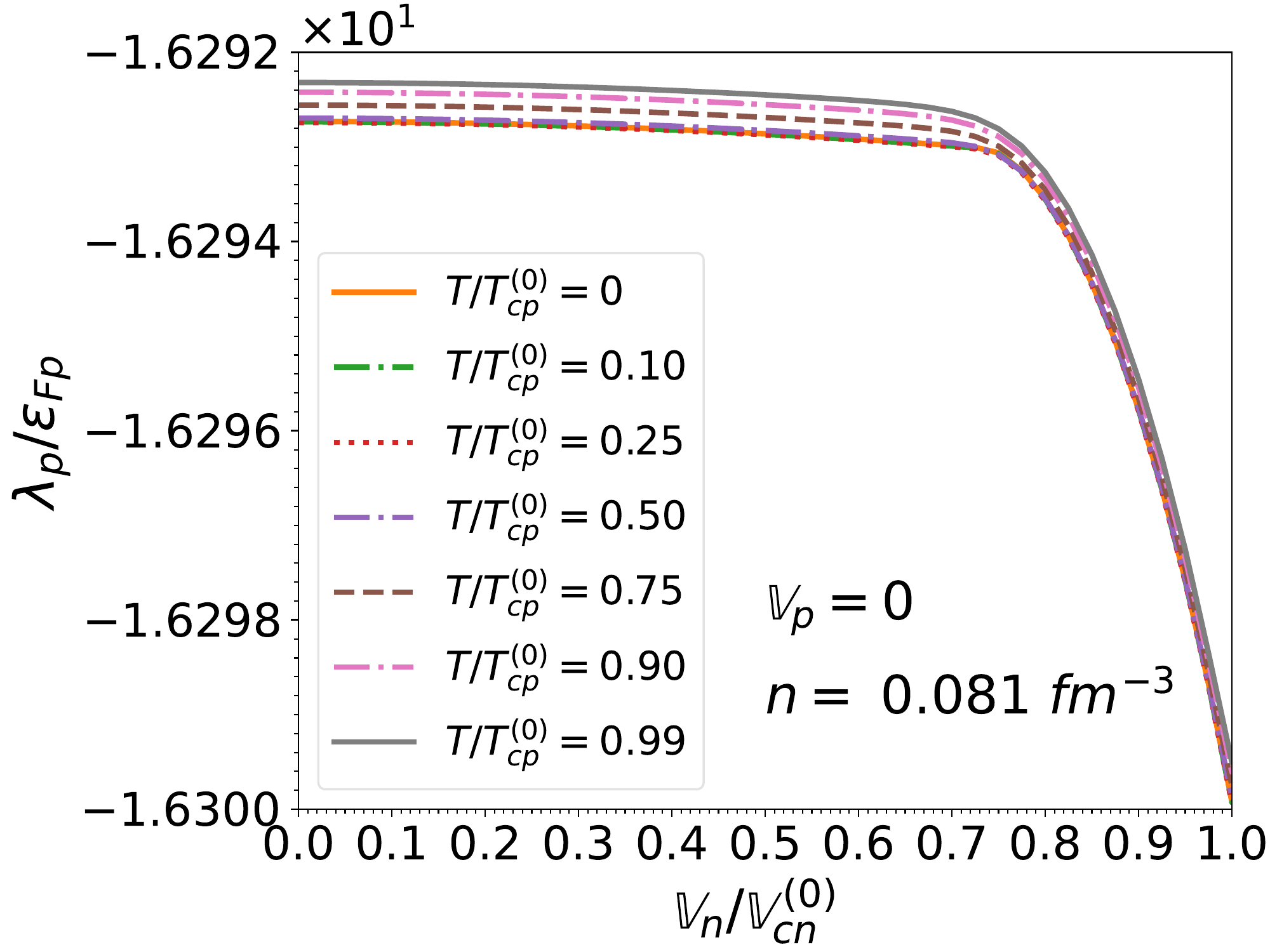}}

\caption{Upper panel: neutron chemical potential $\lambda_n$ normalized to the associated Fermi energy $\varepsilon_{Fn}$ as a function of the normalized neutron effective superfluid velocity $\mathbb{V}_n/\mathbb{V}_{cn}^{(0)}$ and of the normalized temperature $T/T_{cn}^{(0)}$ for $npe\mu$ matter in beta-equilibrium at the crust-core interface and for $\mathbb{V}_p=0$. Lower panel: same for the proton chemical potential $\lambda_p$.}
\label{fig:LambdaBSk24ncc}
\end{figure}
\vspace{-9pt}

% start a new page without indent 4.6cm
%\clearpage

\begin{figure}[H]
\raggedright

\centerline{\includegraphics[width=7cm]{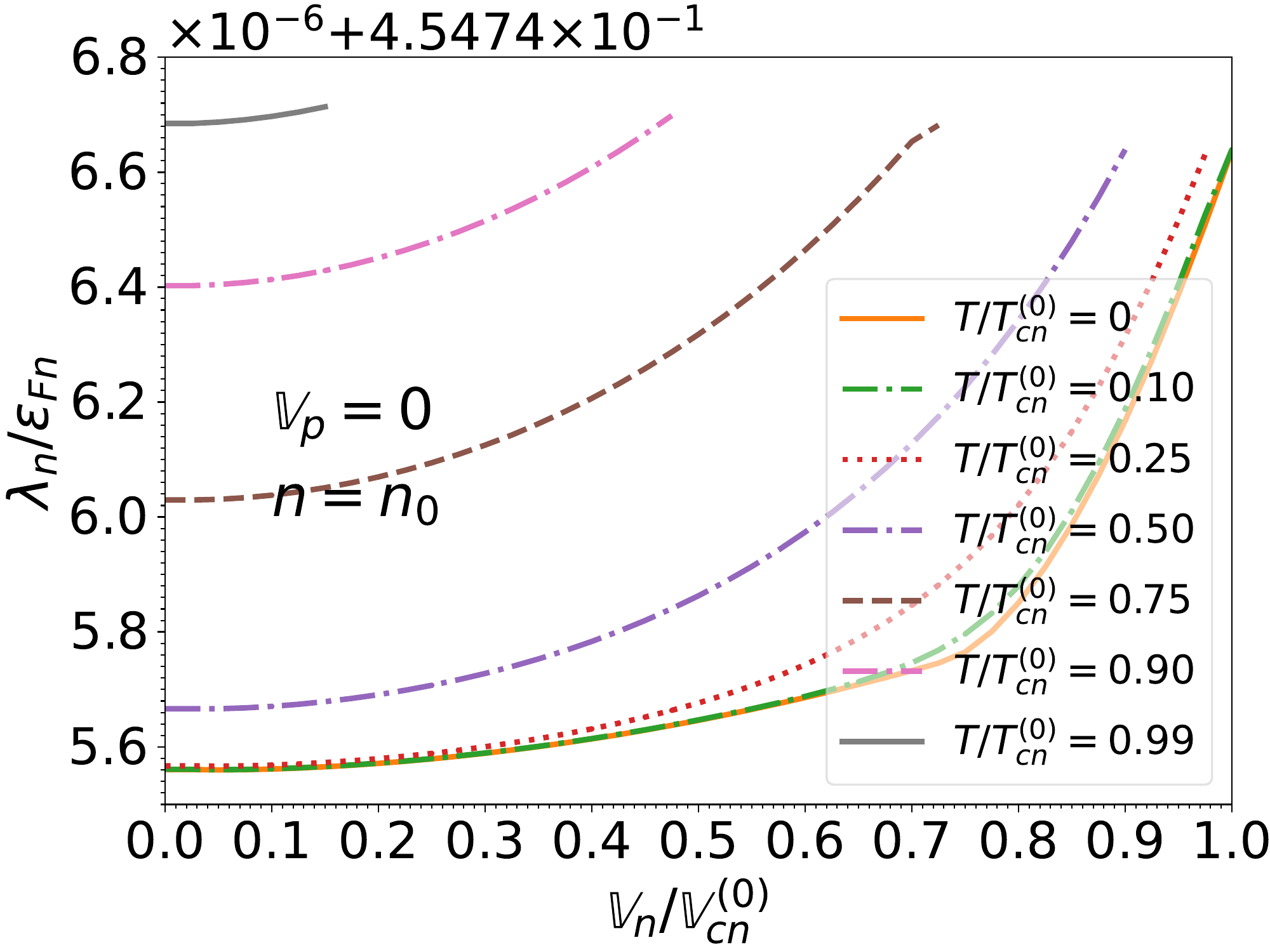}}
\centerline{\includegraphics[width=7cm]{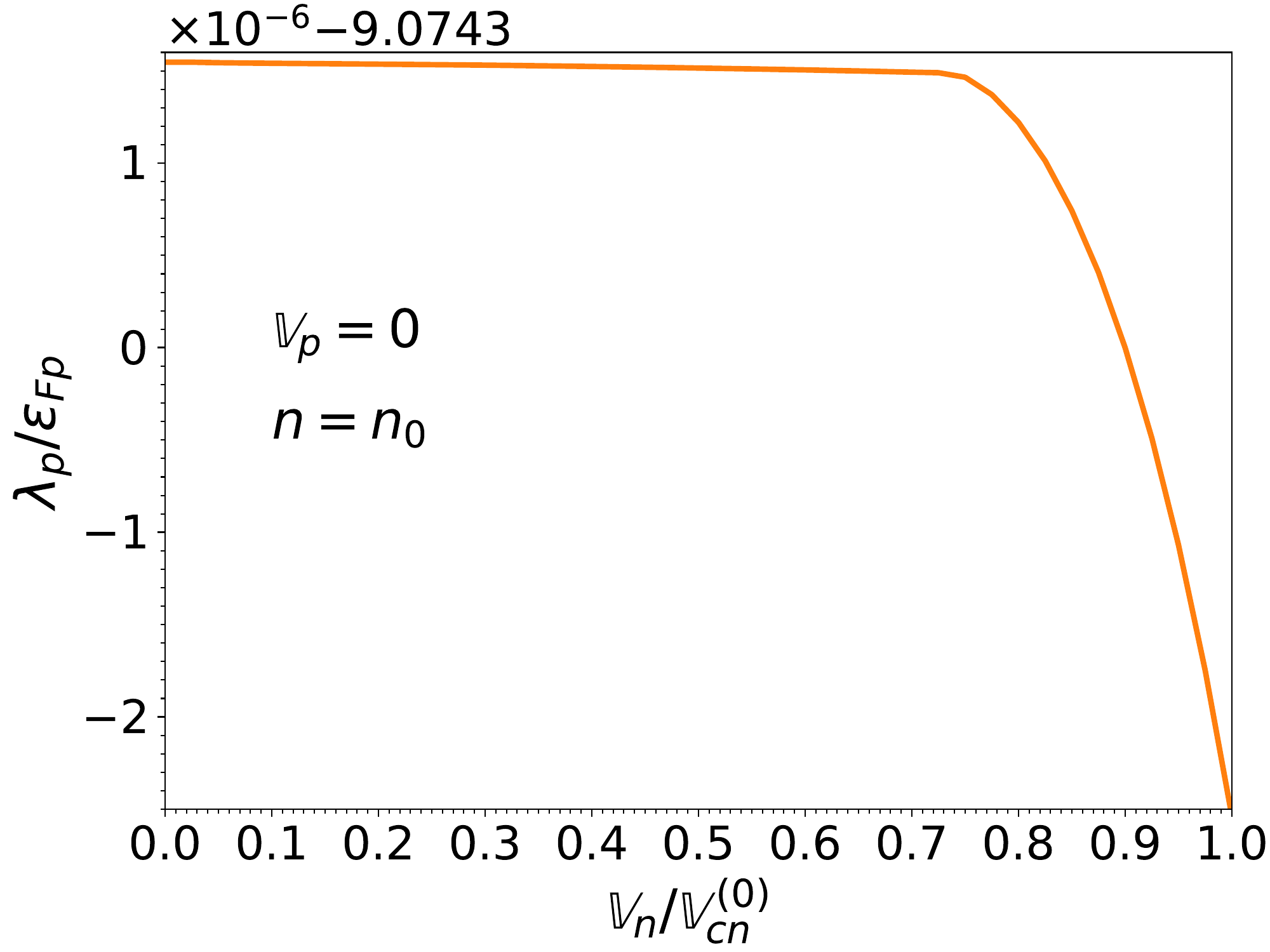}}

\caption{\textls[-15]{Same as Figure~\ref{fig:LambdaBSk24ncc} at the saturation density $n_0$. Please note that for the proton chemical potential $\lambda_p$, results for temperatures (ranging from 0 to $T_{cp}^{(0)}$) are all contained within the thickness of the solid line.}}
\label{fig:LambdaBSk24n0}
\end{figure}
\vspace{-15pt}

\section{Conclusions}
\label{sec:conclusions}

We have studied the neutron–proton superfluid mixture present in the outer core of a neutron star in the framework of the nuclear-energy-density functional theory. In particular, we have calculated consistently the $^1$S$_0$ pairing gaps $\Delta_q$ of each nucleon species $q$, their chemical potentials $\lambda_q$, and the entrainment matrix elements $\rho_{qq'}$ relating the mass currents $\pmb{\rho_q}$ to the so-called superfluid ``velocities'' $\pmb{V_{q}}$ (actually representing superfluid momenta per unit mass) in the normal-fluid rest frame.

To this end, we have solved numerically the self-consistent TDHFB 
equations with the Brussels–Montreal functional BSk24~\cite{goriely2013} for $npe\mu$ matter in beta-equilibrium over the whole range of temperatures and velocities for which nuclear superfluidity can exist using the composition published in \cite{pearson2018}. We have considered the full TDHFB equations without any approximation. In particular, the vector potentials $\pmb{I_q}$ and the contributions from the momentum densities $\hbar\pmb{j_q}$ to the potentials $U_q$ and to the kinetic densities $\tau_q$ have been fully taken into account. We have shown that $\hbar \pmb{j_q}$ represents the total momentum density of a given nucleon species $q$, accounting not only for the superfluid momentum density $\rho_q \pmb{V_q}$ but also for the momentum density carried by the quasiparticles, as shown in Equation~\eqref{eq:total-momentum-density}. Because the true velocity $\pmb{v_q}=\pmb{\rho_q}/\rho_q$ of the nucleon species $q$ in the normal frame is related to the effective superfluid velocity $\pmb{\mathbb{V}_q}$ introduced in Equation~\eqref{eq:EffectiveSuperfluidVelocity} through Equation~\eqref{eq:true-velocity}, $\pmb{\mathbb{V}_q}$ appears as the natural variable to characterize the superflow of the nucleon species $q$. 

The $^{1}$S$_0$ proton pairing gaps $\Delta_p^{(0)}$ at zero temperature and in the absence of flows are found to be significantly smaller than the neutron gaps  $\Delta_n^{(0)}$, unlike those generally considered in neutron-star simulations. Although proton gaps are mainly determined by the empirical interpolation~\eqref{eq:reference-gaps} between the reference gaps in symmetric nuclear matter and pure neutron matter, they turn out to be consistent with recent diagrammatic calculations taking into account medium-polarization effects and considering both two- and three-body interactions~\cite{guo2019}. We have shown that the gaps  $\Delta_q^{(0)}$  are accurately  reproduced by the approximate formula~\eqref{eq:ChamelApprox}. 

The normalized  $^{1}$S$_0$ pairing gaps $\Delta_q/\Delta_q^{(0)}$ 
and the fraction $\mathcal{Y}_q$ of quasiparticles are found to be universal functions of $T/T^{(0)}_{cq}$ and $\mathbb{V}_q/\mathbb{V}_{cq}^{(0)}$, with the critical temperature and critical velocity given by Equations~\eqref{eq:Tc0} and \eqref{eq:Vc0} respectively, in the sense that they depend neither 
on the composition nor on the density, and are the same for both neutrons and protons.  This result can be understood from the fact that $^{1}$S$_0$ nucleon superfluidity in the core of neutron stars is in the (weak-coupling) BCS regime. We have found that the temperature dependence of the normalized pairing gaps in the absence of flows is well fitted by Equation~\eqref{eq:YavovlevGapInterpolation} proposed in \cite{YakovlevandLevenfish1994}. 
We have obtained new accurate interpolating formulas for describing the velocity-dependence of the normalized pairing gaps at zero temperature~\eqref{eq:GapInterpolation}, as well as for the critical temperatures~\eqref{eq:InterpolationCriticalQuantities}. For arbitrary temperatures and velocities, the pairing gaps can be determined with a very good accuracy by solving numerically Equation~\eqref{eq:LandauGapWeak} instead of the full TDHFB equations.

We have found that the approximations reducing the TDHFB equations to Landau's theory provide accurate results for the entrainment matrix $\rho_{qq'}$ provided the 
critical temperatures $T^{(0)}_{cq}$ in the absence of superflow are given. Moreover, the \emph{reduced} chemical 
potentials $\mu_q$ defined by Equation~\eqref{eq:ReducedChemicalPotential}  are well approximated by the corresponding Fermi energies $\Fermi$ given by Equation~\eqref{eq:eF}. However, this conclusion may change depending on the functional, especially if the adopted one predicts stronger pairing. Moreover, numerical solutions of the full TDHFB equations are still required for calculating 
the chemical potentials $\lambda_q$. 

Together with the results published in \cite{pearson2018,shelley2020,pearson2020,mutafchieva2019} for the composition and the equation of state, our calculations provide consistent microscopic inputs for modeling the global structure and dynamics of superfluid neutron stars. Although we have considered the specific functional BSk24 because it has been shown to be in excellent agreement with existing nuclear data and astrophysical observations~\cite{perot2019,thi2021}, we have also derived all the necessary equations to evaluate superfluid properties for any other functional of the Brussels–Montreal type (this includes all the functionals based on standard Skyrme effective nucleon-nucleon interactions). Extension of the TDHFB theory to account for $^{3}$PF$_2$ neutron superfluidity in the inner core of massive neutron stars is left for future studies.

%%%%%%%%%%%%%%%%%%%%%%%%%%%%%%%%%%%%%%%%%%
\vspace{6pt} 

%%%%%%%%%%%%%%%%%%%%%%%%%%%%%%%%%%%%%%%%%%
%% optional
%\supplementary{The following are available online at \linksupplementary{s1}, Figure S1: title, Table S1: title, Video S1: title.}

% Only for the journal Methods and Protocols:
% If you wish to submit a video article, please do so with any other supplementary material.
% \supplementary{The following are available at \linksupplementary{s1}, Figure S1: title, Table S1: title, Video S1: title. A supporting video article is available at doi: link.} 

%%%%%%%%%%%%%%%%%%%%%%%%%%%%%%%%%%%%%%%%%%
\authorcontributions{Conceptualization, N.C. and V.A.; methodology, V.A. and N.C.; software, V.A.; validation, V.A. and N.C.; formal analysis, V.A.; investigation, V.A.; resources, N.C.; data curation, V.A.; writing---original draft preparation, V.A. and N.C.; writing---review and editing, N.C. and V.A.; visualization, V.A.; supervision, N.C.; project administration, N.C.; funding acquisition, N.C. All authors have read and agreed to the published version of the manuscript. }

\funding{This work was financially  supported by the Fonds de la Recherche Scientifique (Belgium) under Grant No. PDR T.004320. This work was also partially supported by the European Cooperation in Science and Technology action (EU) CA16214.}

\institutionalreview{Not applicable.}

\informedconsent{Not applicable.}

\dataavailability{Not applicable.} 

%\acknowledgments{In this section you can acknowledge any support given which is not covered by the author contribution or funding sections. This may include administrative and technical support, or donations in kind (e.g., materials used for experiments).}

\conflictsofinterest{The authors declare no conflict of interest.}

%% Optional
\abbreviations{Abbreviations}{

\noindent 
\begin{tabular}{@{}ll}
TDHFB & time-dependent Hartree–Fock–Bogoliubov\\
TDHF & time-dependent Hartree–Fock\\
NM & (Pure) Neutron matter \\
SM & Symmetric matter \\
\end{tabular}}

%%%%%%%%%%%%%%%%%%%%%%%%%%%%%%%%%%%%%%%%%%
%% Optional
\appendixtitles{yes} % Leave argument "no" if all appendix headings stay EMPTY (then no dot is printed after "Appendix A"). If the appendix sections contain a heading then change the argument to "yes".
\appendixstart
\appendix
\section{Weak-Coupling Approximation}
\label{app:weak-coupling}

Adopting Landau's approximations discussed in Section~\ref{sec:LandauApprox}, the gap equation reads
\beqy\label{eq:LandauGap2}
\breve{\Delta}_q \approx -\frac{1}{2}v^{\pi q}\breve{\mathcal{D}}(0)\int_{x_\text{min}}^{\Cutoff/\Fermi} \text{d}x \frac{\breve{\Delta}_q}{\ExLandau}\frac{\Tbar}{2\Vbar}\log\left[\cosh\left(\frac{\ExLandau}{2\Tbar} + \frac{\Vbar}{\Tbar} \right)\sech\left(\frac{\ExLandau}{2\Tbar} - \frac{\Vbar}{\Tbar} \right)\right] \, ,
\eeqy
where the pairing strength is given by Equation~\eqref{eq:Vpi}, 
the density of single-particle states by Equation~\eqref{eq:DoS-Fermi} and Landau's quasiparticle energy by Equation~\eqref{eq:ExLandau}. The lower bound of the integral consistent with the approximate single-particle energies~\eqref{eq:Landau-energy} is $x_\text{min}=-2$. 

Focusing on the superfluid phase ($T<T_{cq}$) such that $\breve{\Delta}_q\neq 0$, dividing Equation~\eqref{eq:LandauGap2} by $\breve{\Delta}_q$ yields
% start a new page without indent 4.6cm
%\clearpage
%\appendix
%\end{paracol}
%\appendix
%\nointerlineskip
\begin{align}\label{eq:LandauGap3}
&1+\frac{1}{2}v^{\pi q}\breve{\mathcal{D}}(0)\int_{x_\text{min}}^{\Cutoff/\Fermi}  \frac{\text{d}x}{\ExLandau}\nonumber\\
&\approx-\frac{1}{2}v^{\pi q}\breve{\mathcal{D}}(0) \int_{x_\text{min}}^{\Cutoff/\Fermi}\frac{\text{d}x}{\ExLandau}\left\{\frac{\Tbar}{2\Vbar}\log\left[\cosh\left(\frac{\ExLandau}{2\Tbar}+\frac{\Vbar}{\Tbar}\right)\sech\left(\frac{\ExLandau}{2\Tbar}-\frac{\Vbar}{\Tbar}\right)\right]-1\right\}\, .
\end{align}
%\begin{paracol}{2}
%\linenumbers
%\switchcolumn

\textls[-15]{
Let us remark that $\displaystyle \frac{\Tbar}{2\Vbar}\log\left[\cosh\left(\frac{\ExLandau}{2\Tbar} + \frac{\Vbar}{\Tbar} \right)\sech\left(\frac{\ExLandau}{2\Tbar} - \frac{\Vbar}{\Tbar} \right)\right]\rightarrow \displaystyle \tanh\left(\frac{\ExLandau}{2\Tbar}\right)$ in the limit $\Vbar\rightarrow 0$, and is equal to $1$ when 
evaluated at $T=0$ since $\ExLandau>0$. 
Taking the limit of Equation~\eqref{eq:LandauGap2} in the absence of currents and evaluating it at zero temperature,  
we thus obtain the approximate gap equation for $\breve{\Delta}_q^{(0)}$}
\beqy\label{eq:LeinsonGapZero}
\breve{\Delta}_q^{(0)}\approx -\frac{1}{2}v^{\pi q}\breve{\mathcal{D}}(0)\int_{x_\text{min}}^{\Cutoff/\Fermi} \text{d}x\; \breve{\Delta}_q^{(0)}\left[x^2 + \left(\frac{\breve{\Delta}_q^{(0)}}{\Fermi}\right)^2\right]^{-1/2}\, .
\eeqy

We can divide both sides of Equation~\eqref{eq:LeinsonGapZero} by $\breve{\Delta}_q^{(0)}$. Therefore, the left-hand side of Equation~\eqref{eq:LandauGap3} can be expressed as 
% start a new page without indent 4.6cm
%\clearpage
%\end{paracol}
%\appendix
%\nointerlineskip
\begin{align}\label{eq:RHSLeinson}
&1+\frac{1}{2}v^{\pi q}\breve{\mathcal{D}}(0)\int_{x_\text{min}}^{\Cutoff/\Fermi}  \frac{\text{d}x}{\ExLandau}\nonumber\\
&=-\frac{1}{2}v^{\pi q}\breve{\mathcal{D}}(0)\left\{\int_{x_\text{min}}^{\Cutoff/\Fermi}  \text{d}x \,  \left[x^2 + \left(\frac{\breve{\Delta}_q^{(0)}}{\Fermi}\right)^2\right]^{-1/2} -\int_{x_\text{min}}^{\Cutoff/\Fermi} \frac{\text{d}x }{\ExLandau}\right\}\nonumber\\
& = -\frac{1}{2}v^{\pi q}\breve{\mathcal{D}}(0)\left[\asinh\left(\frac{\Cutoff}{\breve{\Delta}_q^{(0)}}\right) - \asinh\left(\frac{x_\text{min}\Fermi}{\breve{\Delta}_q^{(0)}}\right)-\asinh\left(\frac{\Cutoff}{\breve{\Delta}_q}\right) + \asinh\left(\frac{x_\text{min}\Fermi}{\breve{\Delta}_q}\right)\right]\, .
\end{align}
%\begin{paracol}{2}
%\linenumbers
%\switchcolumn

Following \cite{leinson2018}, the weak-coupling approximation $\displaystyle\breve{\Delta}_q,\; \breve{\Delta}_q^{(0)} \ll \Cutoff, \Fermi $ allows us to replace the inverse hyperbolic sine function by its asymptotic form $\displaystyle\asinh(u)\approx \log\left(2u\right)$. Taking the limit 
$\Cutoff/\Fermi\rightarrow +\infty$ and $x_{\text{min}}\rightarrow -\infty$ and eliminating $v^{\pi q}\breve{\mathcal{D}}(0)$  leads to Equation~\eqref{eq:LandauGapWeak}. 
Please note that unlike the original gap Equation~\eqref{eq:LandauGap2}, the limit $\Cutoff/\Fermi\rightarrow +\infty$ can be taken here since the integral does not exhibit any divergence.

%%%%%%%%%%%%%%%%%%%%%%%%%%%%%%%%%%%%%%%%%%
\end{paracol}
\reftitle{References}

%=====================================
% References, variant A: external bibliography
%=====================================

\end{document}